\documentclass[a4paper,11pt]{article}
\pdfoutput=1
\usepackage{cite}
\usepackage{multirow}
\usepackage{jheppub}
\usepackage{hyperref}
\usepackage{enumerate}
\usepackage{relsize}
\usepackage{mathrsfs}
\usepackage{inputenc}
\usepackage{braket}
\usepackage{graphicx}
\usepackage{chemarrow}
\usepackage{mathtools}
\usepackage{caption}
\usepackage{blindtext}
\usepackage{bbold,amsmath,lipsum,amssymb,leftindex}
\usepackage[T1]{fontenc}

\title{Tensionless Strings in a Kalb-Ramond Background}
\author{Aritra Banerjee\,$^a$, Ritankar Chatterjee\,$^b$, Priyadarshini Pandit\,$^b$}
\affiliation{$^a$\,Birla Institute of Technology and Science, Pilani Campus, Rajasthan 333031, India. \\ 
$^b$\,Indian Institute of Technology Kanpur, Kanpur, 208018, India.}
\emailAdd{aritra.banerjee@pilani.bits-pilani.ac.in}
	\emailAdd{ritankar@iitk.ac.in}\emailAdd{ppandit@iitk.ac.in} 

\abstract{We investigate tensionless (or null) bosonic string 
theory with a Kalb-Ramond background turned on. In analogy with the tensile
case, we find that the Kalb-Ramond field has a non-trivial
effect on the spectrum  only when the theory is compactified on an \(\left(S^1\right)^{\otimes d}\) background with
\(d\geq 2\). We discuss the effect of this background field on the tensionless spectrum constructed on three known consistent null string vacua. We elucidate further on the intriguing fate of duality symmetries in these classes of string theories when the background field is turned on.}
	 
\keywords{}
\begin{document}
\maketitle
	\flushbottom

\section{Introduction}

String theory, for past many decades, continues to hold significant prominence as a potential quantum theory of gravity. At its core, string theory relies on an inherent length scale identified as the tension ($T$) of the fundamental string, which determines its behavior. By considering two extreme limits based on this tension, we can understand different aspects of the theory. When $T$ becomes extremely large ($T\to\infty$), string theory converges to the point particle limit, where it effectively reduces to a theory of (super)gravity. Conversely, when $T$ approaches zero ($T\to 0$), string length becomes extremely large, thereby enabling the exploration of ultra high energy regime of the theory. This intriguing ultra high energy null string sector was first analyzed by Schild \cite{Schild:1976vq}. In this limit, the string scattering amplitudes are found to exhibit a remarkably simple behavior \cite{Gross:1987kza, Gross:1987ar}, characterized by an infinite number of linear relations. Comparing with the massless limit of the point particle, all this suggests the presence of a possible higher symmetry structure \cite{Gross:1988ue} within this tensionless regime. Tensionless strings are also expected to exhibit massless higher spin symmetries \cite{Vasiliev:2003cph, Sagnotti:2003qa, Bonelli:2003kh}, resulting in connections to holographic dualities \cite{Sezgin:2002rt, Gaberdiel:2014cha}. 

\medskip
 
Tensionless strings arise in various scenarios, providing valuable insights into the behavior of strings in extreme situations, for example near generic event horizons. It has been demonstrated in \cite{Bagchi:2021ban, Bagchi:2020ats} that a tensile closed string gradually becomes tensionless when it encounters the event horizon of a Schwarzschild black hole. One significant manifestation of tensionless strings occurs near the Hagedorn temperature $\mathcal{T}_H$ \cite{PhysRevD.26.3735, Olesen:1985ej, Atick:1988si}, the very high-temperature limit of string theory. An interesting observation in support of this is the discovery of a closed-to-open string transition when the string tension is tuned to zero \cite{Bagchi:2019cay}.  This observation highlights the significance of studying the tensionless limit of string theory to gain a deeper understanding of how strings perceive generic null surfaces, and perhaps by extension, spacetime singularities.
Furthermore, tensionless or null strings have recently been employed to develop a quantum model of black hole horizons, specifically BTZ black holes in AdS$_3$ \cite{Bagchi:2022iqb}, in a setup reminiscent of the black hole membrane paradigm. By counting the microstates of null strings, it has been possible to obtain insights into the entropy and logarithmic corrections associated with these black holes.
\medskip 

Nevertheless, developments in the formalism front associated to null string theory is still only in a nascent state.
Since the early work of Schild \cite{Schild:1976vq}, there have been several studies on tensionless strings or null strings in flat spacetime \cite{Karlhede:1986wb,Lizzi:1986nv,Gamboa:1989zc,Gamboa:1989px,Gustafsson:1994kr,Lindstrom:2003mg}. In recent years, however, the study of tensionless string theory has been pursued, in a rather streamlined way, through two distinct approaches: the intrinsic approach \cite{Isberg:1993av} and the limiting approach \cite{Bagchi:2013bga, Bagchi:2015nca}. The intrinsic approach involves constructing a worldsheet action that incorporates the degenerate metric structure of tensionless strings, while preserving invariance under a gauge symmetry, namely worldsheet diffeomorphism invariance. However, this gauge symmetry can only be partially fixed, similar to the case of tensile theory. After gauge fixing, the action still remains invariant under a residual gauge symmetry, which remarkably results in the emergence of the BMS$_3$ (3d Bondi-Metzner-Sachs) algebra \cite{Bagchi:2013bga}, replacing the two copies of the Virasoro algebra present in the tensile case. Notably, the BMS algebra represents the asymptotic symmetry algebra of Minkowski spacetime at null infinity \cite{Bondi:1962px,Sachs:1962wk,Ashtekar:1996cd}. And it comes as no surprise that the same algebra appears in the context of symmetries associated to generic null hypersurfaces \cite{Lévy1965,NDS,Henneaux:1979vn}. This algebra has contributed significantly to our understanding of holographic description of flat spacetime \cite{Bagchi:2010zz,Bagchi:2012xr,Bagchi:2012cy,Bagchi:2012yk,Bagchi:2014iea,Bagchi:2016bcd,Asadi:2016plj,Jiang:2017ecm}, black holes \cite{Donnay:2019jiz}, fluid dynamics near high rapidity surfaces \cite{Bagchi:2023ysc, Bagchi:2023rwd, Armas:2023dcz}, physics of Fractons \cite{Bidussi:2021nmp, Figueroa-OFarrill:2023vbj}, and flat bands in condensed matter physics \cite{Bagchi:2022eui,Marsot:2022imf}, to only name a few.
\medskip

The alternative approach, known as the limiting approach, involves taking a specific limit on the worldsheet coordinates from the tensile string theory \cite{Bagchi:2013bga, Bagchi:2015nca}. This limit corresponds to an ultra-relativistic regime or Carrollian limit \cite{Duval:2014lpa} where the worldsheet speed of light tends to zero. In terms of the worldsheet coordinates ($\tau,\sigma$), this limit can be expressed as \{$\tau\to \epsilon\tau, \sigma\to \sigma$\}, with $\epsilon\to 0$.  As expected, under this scaling, the two copies of the Virasoro algebra transforms into the BMS$_3$ algebra, establishing the consistency between the limiting approach and the intrinsic approach \footnote{In fact, the Carrollian Conformal Algebra in two dimensions (CCA$_2$), the Carrollian contraction of two copies of Virasoro, is isomorphic to the BMS$_3$ \cite{Duval:2014lpa}.}. Furthermore, investigations have been carried out to explore the supersymmetric extension of tensionless string theory \cite{Bagchi:2016yyf,Bagchi:2017cte,Bagchi:2018wsn}, both intrinsically and from limits, where the residual gauge symmetry is that of a Super-BMS algebra.
\medskip

Significant advancement has been achieved in recent years in the quantization of tensionless strings as well, leading to the discovery of three consistent quantum theories based on different BMS invariant vacua: \textit{Oscillator, Induced, and Flipped} \cite{Bagchi:2020fpr, Bagchi:2019cay, Bagchi:2021rfw}. The tensionless limit of the conventional tensile quantum theory based on highest weight representation connects to the Induced vacuum, which in turn corresponds to the induced representation of the BMS algebra. 
The Flipped vacuum, which belongs to the highest weight representation of the BMS algebra itself, corresponds to the tensionless limit of a \textit{twisted} string theory \cite{Casali:2016atr}, a cousin of usual tensile string theory. These two theories although classically equivalent, are quantum mechanically very different \cite{Casali:2016atr, Lee:2017utr}. The Oscillator vacuum on the other hand, exhibits intriguing connections to the Induced vacuum through Bogoliubov transformation. The light cone quantization and path integral quantization of all the three tensionless Bosonic theories have also been studied in \cite{Bagchi:2021rfw, Chen:2023esw}.

\medskip
In this paper, we would like to continue understanding the formal structure of a null string theory in better detail. 
As we know, tensile string theory requires 
extra dimensions to be compactified on a manifold \cite{Polchinski:1998rq,Blumenhagen:2013fgp} to write down a four dimensional effective theory. Compactification introduces quantum numbers like winding number and quantized momentum, leading to a spectrum of massless and massive vector and scalar states. 
Recently, this study of compactified string theory has been extended to tensionless regime \cite{Banerjee:2023ekd} by the present set of authors. The compactified spectrum for all three consistent string vacua has been explored in detail. Unsurprisingly, it has also been noted that the impact of a constant Kalb-Ramond field switched on the tensionless string theory target space can only reflect in the spectrum for a theory with multiple dimension compactification (toroidal compactification on $T^d$). This perfectly fits in with the effect of constant $B$ field in compactification of tensile string theory \cite{Polchinski:1998rq,Blumenhagen:2013fgp,Becker:2006dvp}.
\medskip

So it makes perfect sense to study in-depth the impact of a constant $B$ field in the tensionless regime of string theory. Furthermore such a study opens up the possibility of exploring the fate of string dualities on the tensionless spectrum. In the earlier work \cite{Banerjee:2023ekd} we noted that the structure of T-duality along the compactified directions apparently seems to be lost as $\alpha' \to \infty$ in this limit. The present work will also take a close look on whether the presence of a background field can remedy this issue. Consequently we will also look at the generalized symmetry structure of our spectrum in comparison to the tensile counterpart thereof. For different consistent vacua, investigation into the symmetries of mass spectrum will lead to interesting conclusions.
We will also have ample comparisons between tensile and tensionless physics whenever applicable in our discussions.

\subsection*{Plan of the paper}

The paper is organized as follows: In Section \eqref{sec2}, we present an intrinsic formulation of the tensionless bosonic string action with a constant Kalb-Ramond field along with its symmetries, obtained from the Nambu-Goto version of the tensile theory with constant $B$ field. We also revisit the canonical quantization of the tensionless string theory, leading to three distinct quantum theories based on different vacua. The machinery for studying compactification on a $d$-dimensional torus in this case is discussed in Section \eqref{sec3}. Sections \eqref{sec4},\eqref{sec5} and \eqref{sec6} delve into the detailed effects of the constant $B$ field on the level matching conditions, mass spectrum and symmetries thereof for the three quantum theories. Finally, in Section \eqref{sec7}, we summarize our findings and suggest potential future directions. The appendices contain additional computations and supplementary discussions.

\section{Tensionless strings with constant \textit{B} field: Classical analysis} \label{sec2}

We start with a tensile string propagating in a Lorentzian flat target space with an antisymmetric field $B_{\mu\nu}$.
 Note that one can easily write a Nambu Goto version of the usual tensile string action with constant Kalb-Ramond field (see \cite{Gangopadhyay:2006gmx} for example). The Polyakov version of the same can be derived by constructing an interpolating Lagrangian. We start off with the same action \footnote{In \cite{Gangopadhyay:2006gmx}, they have taken the tension of the string to be unity. However, since our actual intention is to take tensionless limit, we need to work with a finite tension $T$.} 
\begin{align}\label{ctb1}
    S=T\int d^2\xi \left[\sqrt{-\text{det} \gamma_{\alpha\beta}}+B_{\mu\nu}\Dot{X}^{\mu}X'^{\nu}\right],
\end{align}
where $\xi^{\alpha}$ ($\sigma, \tau$) are the worldsheet coordinates and $\gamma_{\alpha\beta}$ is the induced metric on the worldsheet given by
\begin{align}  \label{inducedmetric}  \gamma_{\alpha\beta}=\partial_{\alpha}X^{\mu}\partial_{\beta}X^{\nu}\eta_{\mu\nu}.
\end{align}
Here $X^\mu$ denotes the spacetime coordinates, where $\mu\in 0,1,\cdots,D-1$ and $\eta_{\mu\nu}$ is the background flat metric.
The equation of motion of the above action can be calculated as,
\begin{align}
\Dot{\Pi}^{\mu}+K'^{\mu}=0.
\end{align}
Here the expression for $\Pi_{\mu}$ and $K_{\mu}$ are given by
\begin{equation}\label{ctb7}
\begin{split}
    \Pi_{\mu}=\frac{\partial{\mathcal{L}}}{\partial{\Dot{X}^{\mu}}}=T\frac{(\Dot{X}\cdot X')X'_{\mu}-X'^2\Dot{X}_{\mu}}{\sqrt{(\Dot{X}\cdot X')^2-\Dot{X}^2X'^2}}+TB_{\mu\nu}X'^{\nu},\\
    K_{\mu}=\frac{\partial{\mathcal{L}}}{\partial{X'^{\mu}}}=T\frac{(\Dot{X}\cdot X')\Dot{X}_{\mu}-\Dot{X}^2X'_{\mu}}{\sqrt{(\Dot{X}\cdot X')^2-\Dot{X}^2X'^2}}-TB_{\mu\nu}\Dot{X}^{\nu},
    \end{split}
\end{equation}
where ($'$) represents differentiation with respect to $\sigma$ and dot ($\cdot$) represents differentiation with respect to $\tau$. The canonical momentum $\Pi_{\mu}$ satisfies the following constraints
\begin{equation}\label{ctb8}
    \Pi_{\mu}X'^{\mu}=0,~~~~~~~~~~
    (\Pi_{\mu}-TB_{\mu\nu}X'^{\nu})^2+T^2X'^2=0.
\end{equation}
Now, before taking tensionless limit of this action let us have a look at the action in \eqref{ctb1}. In the tensionless limit, this action seemingly vanishes. However, following the method outlined in \cite{Isberg:1993av}, we can go to the Hamiltonian formalism to achieve a form of the Polyakov action with a finite $T\to 0$ limit. In \cite{Gangopadhyay:2006gmx}, a similar method has been used to achieve a form of Polyakov version of the flat space tensile string with constant $B$ field. In the remaining part of the section we will see how does it work for tensionless limit.
\medskip

The canonical Hamiltonian of the action in \eqref{ctb1} can be calculated using \eqref{ctb7} as
\begin{align}
    \Pi_{\mu}\Dot{X}^{\mu}-\mathcal{L}=0.
\end{align}
The result can also be viewed from the reparameterization invariance of the action \eqref{ctb1}. Hence the total Hamiltonian can be written as a linear combination of the constraints given in \eqref{ctb8}.
\begin{align}
    \mathcal{H}_{T}=\rho~\Pi_{\mu}X'^{\mu}+\frac{\lambda}{2}\left\{\left(\Pi_{\mu}-TB_{\mu\nu}X'^{\nu}\right)^2+T^2X'^2 \right\}.
\end{align}
Here $\lambda$ and $\rho$ are Lagrange multipliers. The interpolating Lagrangian is calculated as 
\begin{equation}\label{ctb9}
\begin{split}
    &\mathcal{L}_{I}=\Pi_{\mu}\Dot{X}^{\mu}-\mathcal{H}_{T}\\
    =\Pi_{\mu}\Dot{X}^{\mu}-\rho\Pi_{\mu}X'^{\mu}-\frac{\lambda}{2}& \left[\Pi^2+T^2X'^2-2TB_{\mu\nu}\Pi^{\mu}X'^{\nu}+T^2B_{\mu\nu}B^{\mu}_{\rho}X'^{\nu}X'^{\rho}\right].
    \end{split}
\end{equation}
In this interpolating Lagrangian, $\lambda$ and $\rho$ are treated as scalar fields. Since the canonical momentum $\Pi_{\mu}$ is an auxiliary field, we integrate it out from the interpolating Lagrangian. Finally we obtain the following expression for Lagrangian
\begin{align}
    \mathcal{L}_{I}=\frac{1}{2\lambda}\left[\Dot{X}^2-2\rho(\Dot{X}\cdot X')+(\rho^2-\lambda^2T^2)X'^2+2\lambda TB_{\mu\nu}\Dot{X}^{\mu}X'^{\nu}\right].
\end{align}
We now identify the coefficients of $\Dot{X}^2$, $\Dot{X}\cdot X'$ and $X'^2$ in the form of a metric as
\begin{align}
    g^{\alpha\beta}=\begin{bmatrix}1 & -\rho \\ -\rho & ~~\rho^2-\lambda^2T^2\end{bmatrix},\hspace{5mm}\sqrt{-g}=\sqrt{-\text{det}g_{\alpha\beta}}=\frac{1}{\lambda T}~.
\end{align}
This results in the Polyakov version of the action given by, 
\begin{align}\label{Polyakov}
   S_{I}=\frac{T}{2}\int d\tau d\sigma \left(\sqrt{-g}g^{\alpha\beta}\partial_{\alpha}X^{\mu}\partial_{\beta}X^{\nu}\eta_{\mu\nu}+\epsilon^{\alpha\beta}B_{\mu\nu}\partial_{\alpha}X^{\mu}\partial_{\beta}X^{\nu}\right).
\end{align}
Now taking the tensionless limit, viz. $T\to\epsilon T$, $\epsilon\to 0$ on this action \footnote{Note that this can be equivalently thought of as $\alpha' \to \frac{c'}{\epsilon}$ with a finite $c'$, which we will be using later on.}, we can redefine our worldsheet variables to take care of the degenerate structure \footnote{This kind of limit from tensile string with a background field to its tensionless counterpart has been discussed earlier in a recent work \cite{Gomis:2023eav}. In the notation there, these strings are called M(-1)T strings. Moreover, in \cite{Bagchi:2024unl}, the example of taking ultrarelativistic (Carroll) limit on a theory of two scalars with a $J^1\wedge \bar{J}^2$ marginal deformation has been considered, which boils down to the same physical situation.},
\begin{align}\label{ctb26}
    T\sqrt{-g}g^{\alpha\beta}\to V^{\alpha}V^{\beta}\hspace{3mm} \text{and} \hspace{3mm} B_{\mu\nu}\to \frac{B_{\mu\nu}}{\epsilon},\hspace{5mm} \text{where} \hspace{5mm}V^{\alpha}=\frac{1}{\sqrt{\lambda}}(1,-\rho).
\end{align}
Here, $V^{\alpha}$ is a timelike vielbein density of the worldsheet. Let us define the scaled antisymmetric field $\mathcal{B}_{\mu\nu}$ as
\begin{align}\label{ctb27}
   \mathcal{B}_{\mu\nu}=\frac{1}{\alpha'}B_{\mu\nu},\hspace{5mm}T=\frac{1}{2\pi\alpha'}
\end{align}
The limit on the $B$ field makes sure that $\mathcal{B}_{\mu\nu}$ remains finite. Hence the action of tensionless string with a $B$ field will be
\begin{align}\label{tsc6}
    S=\int d^2\xi \left(V^{\alpha}V^{\beta}\partial_{\alpha}X^{\mu}\partial_{\beta}X^{\nu}\eta_{\mu\nu}+\frac{1}{2\pi}\epsilon^{\alpha\beta}\mathcal{B}_{\mu\nu}\partial_{\alpha}X^{\mu}\partial_{\beta}X^{\nu}\right).
\end{align}
This will be the action we will be mostly concerned with throughout this work. Note that this is nothing but an extension of the so-called ILST action \cite{Isberg:1993av} for tensionless strings.

\subsection{Symmetries of the action}

It can be shown that the action \eqref{tsc6} is invariant under diffeomorphism ($\xi^\alpha\to\xi^{\alpha}+\epsilon^{\alpha}$), under which, the vector density $V^\alpha$ and the scalars $X^{\mu}(\sigma,\tau)$ transform in the following way:
\begin{equation}\label{B3}
\begin{split}
   \delta V^\alpha=-&V^\beta\partial_\beta\epsilon^\alpha+\epsilon^\beta\partial_\beta V^\alpha+\frac{1}{2}(\partial_\beta\epsilon^\beta)V^\alpha\\
    &\delta X^{\mu}(\sigma,\tau)=\epsilon^{\alpha}\partial_{\alpha}X^{\mu}(\sigma,\tau) 
\end{split}
\end{equation}
The action \eqref{tsc6} with $\mathcal{B}_{\mu\nu}=0$ is known to be invariant under the transformation \eqref{B3} (see \cite{Isberg:1993av, Bagchi:2015nca}). We focus on the extra term, and variation of the $B$ field part gives us just a total derivative term
\begin{equation}
\delta(\epsilon^{\alpha\beta}\mathcal{B}_{\mu\nu}\partial_{\alpha}X^{\mu}\partial_{\beta}X^{\nu})=\partial_{\alpha}(\epsilon^{\alpha\beta}\mathcal{B}_{\mu\nu}\epsilon^{\gamma}\partial_{\gamma}X^{\mu}\partial_{\beta}X^{\nu}).
\end{equation}
Hence the worldsheet diffeomorphism leave the full action \eqref{tsc6} invariant as well. This enables us to do gauge fixing analogous to conformal gauge in case of tensile string theory. Since $V^\alpha$ of this action transforms like \eqref{B3}, we can make a gauge choice $V^{\alpha}=(v,0)$, where $v$ is an arbitrary parameter. For our convenience we make the following choice of parameterization for the constant $v$:
\begin{align}\label{tsc4}
    v=\sqrt{\frac{1}{2\pi c'}},\hspace{5mm}\alpha'=\frac{c'}{\epsilon},\hspace{5mm}\epsilon\to 0.
\end{align}
Similar to its tensile counterpart, there is a residual symmetry left over after the above gauge fixing. The gauge fixed action will remain invariant under the following form of $\epsilon^\alpha$
\begin{equation}   \label{B4} \epsilon^\alpha=\left(f'(\sigma)\tau+g(\sigma),f(\sigma)\right).
\end{equation}
Under this set of transformations, a free function $G(\xi^\alpha)$ transforms as:
\begin{equation}
\delta G=\left[f'(\sigma)\tau\partial_\tau+f(\sigma)\partial_\sigma+g(\sigma)\partial_\tau\right]G=\left[L(f)+M(g)\right]G.
\end{equation}
Therefore, the symmetry generators can be defined as:
\begin{equation}
    \begin{split}L(f)=f'&(\sigma)\tau\partial_\tau+f(\sigma)\partial_\sigma=\sum_n a_ne^{in\sigma}(\partial_\sigma+in\tau\partial_\tau)=-i\sum_na_nL_n ,\\
   & M(g)=g(\sigma)\partial_\tau=\sum_n b_n e^{in\sigma}\partial_\tau=-i\sum_n b_nM_n.
   \end{split}
\end{equation}
Here $f$ and $g$ are expanded in Fourier modes as $f=\sum_n a_ne^{in\sigma},~ g=\sum_n b_n e^{in\sigma}$. The symmetry algebra can be written as:
\begin{equation}\label{bmsalgebra}
    [L_m,L_n]=(m-n)L_{m+n},~~ [L_m,M_n]=(m-n)M_{m+n}, ~~[M_m,M_n]=0.
\end{equation}
The above algebra forms the classical part of the $\text{BMS}_3$ algebra, which arises here replacing the two copies of Virasoro (Witt) algebra in the case of tensile strings. Hence, the worldsheet residual gauge symmetry of the tensionless string  theory with a $B$ field remains that of $\text{BMS}_3$. This is what one expects for generic null worldsheet theories \cite{Isberg:1993av,Bagchi:2015nca}. 

\subsection{Energy-Momentum tensor}
We now move on to compute the energy-momentum tensor for the action \eqref{tsc6}. In order to do that, let us recall the definition of energy momentum tensor in terms of vielbeins as in \cite{Hartong:2015xda,Bergshoeff:2017btm, Bagchi:2022eav}:
\begin{align}\label{tsc10}
    T^{\alpha}_{\hspace{1mm}\beta}=\frac{e^{\alpha}_{A}}{2e}\frac{\delta S}{\delta e^{\beta}_{A}}
\end{align}
In the above $e^{\alpha}_{A}$ $(A=0,1)$ are the timelike and spacelike vielbeins repectively.  $e$ is given by the defining relations:
\begin{align}
  e=\det(e^{A}_{\alpha})=\frac{1}{2}\epsilon_{AB}e^{A}_{\alpha}e^{B}_{\beta}\epsilon^{\alpha\beta},\hspace{5mm}e^{A}_{\alpha}e^{\alpha}_{B}=\delta^{A}_{B},\hspace{5mm}e^{A}_{\alpha}e^{\beta}_{A}=\delta^{\beta}_{\alpha} .
\end{align}
Looking at the action \eqref{tsc6} one can identify that $e^{\alpha}_{0}=V^{\alpha}$ for our case. One can also see that there is no spacelike vielbein $e^{\alpha}_{1}$ in the action. Hence, using \eqref{tsc10}, it is easy to find the following energy-momentum tensor: 
\begin{equation}
    T^\alpha_\beta=V^\alpha V^\gamma\partial_\gamma X^\mu\partial_\beta X_\mu-\frac{1}{2}V^\rho V^\gamma \partial_\rho X^\mu\partial_\gamma X_\mu \delta^\alpha _\beta.
\end{equation} 
This EM tensor turns out to be identical to the EM tensor for tensionless string without $B$ field \cite{Isberg:1993av}. We recall that in the tensile counterpart of the theory too, the constant $B$ field part did not affect the EM tensor \cite{Blumenhagen:2013fgp} since it is a boundary term. In the gauge \eqref{tsc4}, the non-trivial components of the (EM) tensor are:
\begin{equation}\label{emtensor}
    T^0_0=-T^1_1=\frac{1}{4\pi c'}\dot{X}^2\equiv T_2(\sigma,\tau),~~~T^0_1=\frac{1}{2\pi c'}\dot{X}\cdot X'\equiv T_1(\sigma,\tau).
\end{equation}
The zeroth component of the Noether current can be integrated to be,
\begin{equation}
    Q=\int d\sigma J^0= \int d\sigma \left[T_1f+T_2(f'\tau+g)\right].
\end{equation}
We now expand $f$ and $g$ in Fourier modes and obtain
\begin{equation}
    Q=\sum_n a_n \int d\sigma \left(T_1+in\tau T_2\right)e^{in\sigma}+\sum_n b_n \int d\sigma T_2e^{in\sigma}=\sum_n a_n L_n +\sum_n b_n M_n.
\end{equation}
Thus, inverting the above relation we can write the expression for EM tensor in terms of generators of BMS algebra as 
\begin{equation}\label{tsc}
    T_1(\tau,\sigma)=\frac{1}{2\pi}\sum_n(L_n-in\tau M_n)e^{-in\sigma},~~~T_2(\tau,\sigma)=\frac{1}{2\pi}\sum_n M_ne^{-in\sigma}.
\end{equation}

\subsection{Equations of motions and mode expansions}\label{secmodeexp}
Now, we calculate the equations of motion for the tensionless string with constant $B$ field. It turns out that, just like the tensile theory, here too, the $B$ field dependent part in the action doesn't have any effect on the equations of motion and we get
\begin{subequations}
    \begin{equation}\label{tsc1} 
\partial_\alpha(V^\alpha V^\beta\partial_\beta X^\mu)=0
\end{equation}
\begin{equation}\label{tsc2}
V^\beta\gamma_{\alpha\beta}=0,
\end{equation}
\end{subequations}
where $\gamma_{\alpha\beta}$ is the induced metric on the worldsheet defined in \eqref{inducedmetric}.
Equations \eqref{tsc1}, \eqref{tsc2} are identical to the equations of motion derived for free tensionless strings in \cite{Isberg:1993av, Bagchi:2015nca}. Equation \eqref{tsc1} is the equation of motion for $X^{\mu}$. Equation \eqref{tsc2} yields two constraints on $X^{\mu}$. In the chosen gauge $V^{\alpha}=(v,0)$, our equation of motion of $X^{\mu}$ becomes
\begin{equation}\label{tsc3}
    \ddot {X}^\mu=0.
\end{equation}
 The equations of motion \eqref{tsc3} with a closed string boundary condition $X^\mu(\tau,\sigma)=X^\mu(\tau, \sigma+2\pi)$, can be solved by the following mode expansion:
\begin{equation}\label{fcb1}
    X^{\mu}(\tau,\sigma)=x^{\mu}+\sqrt{\frac{c'}{2}}B^{\mu}_{0}\tau+i\sqrt{\frac{c'}{2}}\sum_{n\neq 0}\frac{1}{n}(A^{\mu}_{n}-in\tau B^{\mu}_{n})e^{-in\sigma}.
\end{equation}
In the same gauge ($V^{\alpha}=(v,0)$), the constraints \eqref{tsc2} will reduce to
\begin{align}\label{constraints}
   \dot{X}^2=0,\hspace{5mm}\dot{X}\cdot X'=0.
\end{align}
  Following \cite{Isberg:1993av, Bagchi:2015nca}, we identify them as the components of the stress-energy tensor ($T^{0}_{0},T^{0}_{1}$).
We use the mode expansion \eqref{fcb1} on the two constraints \eqref{constraints} and equate the result with the expression for the EM  tensor defined in terms of generators of the residual symmetry algebra in \eqref{tsc}. After some algebra, this readily gives,
\begin{equation}\label{tsc9}
    L_n=\frac{1}{2}\sum_m A_{-m}\cdot B_{m+n},~~~~M_n=\frac{1}{2}\sum_m B_{-m}\cdot B_{m+n}.
\end{equation}
We now determine the algebra satisfied by these oscillator modes ($A,B$). In the case of tensile string theory with a constant $B$ field, we have seen that the only aspect affected by the presence of $B$ field is the canonical momentum \cite{Polchinski:1998rq,Blumenhagen:2013fgp,Becker:2006dvp}. Here too, we are going to see the same. The canonical momentum in our case turns out to be
\begin{align}\label{mom}
    \Pi_{\mu}=\frac{\partial\mathcal{L}}{\partial \dot{X}^{\mu}}=\frac{\dot{X}^{\nu}}{2\pi c'}\eta_{\nu\mu}+\frac{X'^{\nu}}{2\pi}\mathcal{B}_{\mu\nu}. 
\end{align}
In the above equation, we take $\epsilon^{01}=1$. However, the equal time Poisson bracket between $X(\sigma,\tau)$ and $\dot{X}(\sigma',\tau)$ will still be intact. As a result, the Poisson brackets between the oscillators $A$ and $B$ will be same as they are for tensionless strings without $B$ field (obtained, for example, in \cite{Isberg:1993av})
\begin{equation}\label{tsc5}
    \{A_m^\mu,A_n^\nu\}_{P.B}~=~\{B_m^\mu,B_n^\nu\}_{P.B}=0,~~~~ \{A_m^\mu,B_n^\nu\}_{P.B}=-2im\delta_{m+n}\eta^{\mu\nu}.
\end{equation}
The above algebra is not the harmonic oscillator algebra. Hence we define new modes which are linear combination of these $(A,B)$ oscillator modes \cite{Bagchi:2020fpr} satisfying the harmonic oscillator algebra. These new modes can be written as,
\begin{equation}\label{tsc7}
    C_n^\mu=\frac{1}{2}(A_n^\mu+B_n^\nu),~~~\tilde {C}_n^\mu=\frac{1}{2}(-A_{-n}^\mu+B_{-n}^\mu).
\end{equation}
These modes form the oscillator basis of the tensionless string. They satisfy the algebra given by,
\begin{equation}\label{calgebra}
    \{C_m^\mu,C_n^\nu\}=-im\delta_{m+n,0}\eta^{\mu\nu},~~~\{\tilde C_m^\mu,\tilde C_n^\nu\}=-im\delta_{m+n,0}\eta^{\mu\nu},~~~\{C_m^\mu,\tilde C_n^\nu\}=0.
\end{equation}
Now we can write the mode expansion \eqref{fcb1} in terms of these new modes as
\begin{align}\label{fcb2}
    X^{\mu}(\tau,\sigma)=x^{\mu}+2\sqrt{\frac{c'}{2}}C^{\mu}_{0}\tau+i\sqrt{\frac{c'}{2}}\sum_{n\neq0}\frac{1}{n}\left[(C^{\mu}_{n}-\Tilde{C}^{\mu}_{-n})-in\tau(C^{\mu}_{n}+\Tilde{C}^{\mu}_{-n})\right]e^{-in\sigma},
\end{align}
where we have used the periodicity of $X^\mu$.  We can write the mode expansion given above in terms of ``left'' and ``right'' moving modes \cite{Bagchi:2020fpr} as:
\begin{align}\label{lrmodes}
    X^{\mu}_{L}&=\frac{x^{\mu}}{2}+\sqrt{\frac{c'}{2}}C^{\mu}_{0}\tau+i\sqrt{\frac{c'}{2}}\sum_{n\neq0}\frac{1}{n}(C^{\mu}_{n}-in\tau C^{\mu}_{n})e^{-in\sigma}\\
    X^{\mu}_{R}&=\frac{x^{\mu}}{2}+\sqrt{\frac{c'}{2}}\Tilde{C}^{\mu}_{0}\tau+i\sqrt{\frac{c'}{2}}\sum_{n\neq0}\frac{1}{n}(\Tilde{C}^{\mu}_{n}-in\tau \Tilde{C}^{\mu}_{n})e^{in\sigma},
\end{align}
where the zeroth modes are related to the momentum of the tensionless string as
\begin{equation*}
        C_0^\mu=\tilde C_0^\nu=\sqrt{\frac{c'}{2}}k^\mu.
\end{equation*}
Equation \eqref{mom} leads us to the following centre of mass momentum
\begin{align}\label{ctb2}
    \pi_{\mu}=\int_{0}^{2\pi}d\sigma \Pi_{\mu}=k^{\nu}\eta_{\nu\mu}+\frac{1}{2\pi}\Big(X^{\nu}(\sigma+2\pi,\tau)-X^{\nu}(\sigma,\tau)\Big)\mathcal{B}_{\mu\nu},
\end{align}
where $k^{\mu}$ is the centre of mass momentum of the tensionless string without background field. Now, in the case of closed strings in a non-compactified background, we have $X^{\nu}(\sigma+2\pi,\tau)=X^{\nu}(\sigma,\tau)$ and hence there is no modification in the centre of mass momentum. However, in case of compactified background, a winding number ($W$) appears in \eqref{ctb2}. Hence, effect of constant $B$ field can only be observed in case of compactified background spacetime. This prompts us to now move on to quantize the classical bosonic closed tensionless strings with $B$ field, albeit in a compact target space. 

 \subsection{Three possible quantization schemes}
 
After a detailed discussion on the classical picture of tensionless strings with constant Kalb-Ramond field, we now move on to quantize it using the usual method of covariant quantization, analogous to quantization of bosonic tensionless strings without $B$ field. We begin with the tensionless action with $B$ field \eqref{tsc6},
and notice that the equations of motion and constraints obtained from this action in the gauge $V^\alpha=(v,0)$, are identical to the case of tensionless strings without $B$ field. Hence, the commutation relations between the operators $X^\mu$ and its canonical momenta $P^\mu$ also takes the same form:
 \begin{equation}
 \begin{split}
     &~\left[X^\mu(\tau,\sigma),P_\nu(\tau,\sigma')\right]=i\delta (\sigma-\sigma')\delta^\mu_\nu.
     \end{split}
 \end{equation}
 We use the above relations in the mode expansion \eqref{fcb2} to compute the commutators of $C$ oscillators satisfying harmonic oscillator algebra, which rather unsurprisingly reads as,
  \begin{align}\label{TSQR3}
    [C^{\mu}_{m},C^{\nu}_{n}]=[\Tilde{C}^{\mu}_{m},\Tilde{C}^{\nu}_{n}]=m\eta^{\mu\nu}\delta_{m+n},\hspace{5mm}[C^{\mu}_{m},\Tilde{C}^{\nu}_{n}]=0.
    \end{align}
 Now we move on to define putative vacua of the theory using  $C$ oscillators and build a Hilbert space of physical states on it. The spectrum of physical states however demands imposition of constraints on the Hilbert space. There are different ways to apply these constraints, as elucidated in \cite{Bagchi:2020fpr}, resulting in different quantum theories, as will be discussed below.

\subsection*{Imposing quantum constraints}
 Here we focus on different ways of applying constraints on the Hilbert space to pick out the physical state spectrum. Classically we restrict the Hilbert space simply by demanding vanishing of the component of the energy momentum tensor \eqref{tsc2}. However, when we quantize the theory, we promote these components of energy momentum tensor $T_1$ and $T_2$ to operators and demand vanishing of all its matrix element on the physical states,
 \begin{equation}
     \bra{phys'}T_1\ket{phys}=\bra{phys'}T_2\ket{phys}=0.
 \end{equation}
 Using the relation \eqref{tsc}, the above constraints can be rewritten in terms of generators of BMS$_3$ algebra as 
 \begin{equation}\label{constraint}
     \bra{phys'}L_n\ket{phys}=0,~~~\bra{phys'}M_n\ket{phys}=0, ~~\forall n\in \mathbb{Z}.
 \end{equation}
In order to impose the above constraints, we assume the vacuum state to be a physical state at least for $n\neq0$,
\begin{equation}
    \bra{0}L_n\ket{0}=\bra{0}M_n\ket{0}=0,~~~~ \forall n\neq 0.
\end{equation}
For $n=0$, normal ordering ambiguities takes part in filtering out the physical spectrum on the Hilbert space. Based on equation \eqref{constraint}, a total of nine possible ways might exist for imposing constraints on physical states. However, on detailed analysis, it was found that only three out of these nine possibilities align consistently with the BMS$_3$ algebra \cite{Bagchi:2020fpr},
\begin{subequations}\label{B1}
    \begin{align}
        &L_n\ket{phys}\neq0,~~~ M_n\ket{phys}\neq 0 ~~~(\forall~n),\\
        &L_n\ket{phys} \neq0,~~~ M_n\ket{phys}=0 ~~~(\forall~n\neq 0),\\
        &L_n\ket{phys}=~~M_n\ket{phys}=0 ~~~(\forall~n>0).
    \end{align}
\end{subequations}
The three conditions mentioned above give rise to three inequivalent quantum theories based on three distinct vacua namely, \textit{Oscillator, Induced} and \textit{Flipped} vacuum. In the rest of this work, we will gradually examine the impact of having  a constant $B$ field on each of these cases in detail in separate sections.

\section{Compactification on torus $T^d$} \label{sec3}

Before moving on to the spectra, in this section, we study the effects of a constant $B$ field, which generically apply to all the three quantum theories mentioned earlier. We make two choices to start with: the components of metric along the compactified coordinates are Euclidean, i.e.  $G_{IJ}=\delta_{IJ}$ and the only non-zero components of the $B$ field will be along the compactified directions. 
Let us consider the compactification radius of $I$-th compactified direction to be $R_{I}$. In order to compactify $d$ dimensions, we make the following identification\footnote{For compactified indices $I$, repeated indices are not summed over unless explicitly mentioned.} 
\begin{align}\label{chh42}
      X^{I}\sim X^{I}+2\pi \omega^{I}R_{I} \hspace{5mm} I\in\{1,2,....,d\} .
\end{align}
This leads us to the following closed string boundary conditions for compactified coordinates
\begin{align}\label{CTBC1}
     X^{I}(\sigma+2\pi,\tau)=X^{I}(\sigma,\tau)+2\pi \omega^{I}R_{I} \hspace{5mm} I\in\{1,2,....,d\},
\end{align}
where $\omega^I$ are the winding numbers. 
Following \cite{Becker:2006dvp}, we use the following coordinate rescaling on the compactified coordinates
\begin{align}\label{nakgol}
    \widetilde{X}^{I}=\frac{X^{I}}{R_{I}}.
\end{align}
The components of the metric and the antisymmetric $B$ field along the compactified coordinates will change in the following way (Einstein's summation convention not applied in order to avoid confusion, there is no summation over repeated indices)
\begin{equation}\label{BBS1}
\begin{split}
     &G_{IJ}=\frac{\partial X^{K}}{\partial \widetilde{X}^{I}}\frac{\partial X^{L}}{\partial \widetilde{X}^{J}}\delta_{KL}=R_{I}R_{J}\sum_{K=1}^{d}\sum_{L=1}^{d}\delta^{K}_{I}\delta^{L}_{J}\delta_{KL}=R^{2}_{I}\delta_{IJ}\\
     &\mathbb{B}_{IJ}=\frac{\partial X^{K}}{\partial \widetilde{X}^{I}}\frac{\partial X^{L}}{\partial \widetilde{X}^{J}}\mathcal{B}_{KL}=R_{I}R_{J}\sum_{K=1}^{d}\sum_{L=1}^{d}\delta^{K}_{I}\delta^{L}_{J}\mathcal{B}_{KL}=R_{I}R_{J}\mathcal{B}_{IJ}
\end{split}
\end{equation}
The expression of inverse metric $G^{-1}$ along compactified directions becomes
\begin{align}\label{mathagol}
    G^{IJ}=\frac{1}{R^{2}_{I}}\delta^{IJ}.
\end{align}
For the new scaled coordinates $\widetilde{X}^{I}$, the identification \eqref{chh42} looks like
\begin{align}\label{scaled}
    \widetilde{X}^{I}\sim \widetilde{X}^{I}+2\pi \omega^{I}.
\end{align}
 As a result, the new periodic conditions satisfied by $\widetilde{X}^{I}$ will be
    $\widetilde{X}^{I}(\sigma+2\pi,\tau)-\widetilde{X}^{I}(\sigma,\tau)=2\pi \omega^{I}$.
In terms of these new coordinates the relevant part of the gauge fixed action in \eqref{tsc6} becomes
\begin{align}\label{agol}
    \int d^2\xi \left(v^2\dot{\widetilde{X}}^{I}\dot{\widetilde{X}}^{J}G_{IJ}+\epsilon^{\alpha\beta}\mathbb{B}_{IJ}\partial_{\alpha}\widetilde{X}^{I}\partial_{\beta}\widetilde{X}^{J}\right),
\end{align}
where $v$ is the parameter given in \eqref{tsc4}. The canonical momentum along the compactified directions is given by 
\begin{align}
    \widetilde{\Pi}_{I}=\frac{1}{c'}\sum_{J=1}^{d}\dot{\widetilde{X}}^{J}G_{IJ}+\sum_{J=1}^{d}\mathbb{B}_{IJ}\widetilde{X}'^{J}.
    \end{align}
Hence, the centre of mass momentum will be a combination of compactified momenta and winding number, thus including the contribution from the $B$ field:
\begin{align}\label{ctb24}
    \widetilde{\pi}_{I}=\frac{1}{2\pi}\int_{0}^{2\pi}d\sigma \widetilde{\Pi}_{I}=\sum_{J=1}^{d}K^{J}G_{IJ}+  \sum_{J=1}^{d}\mathbb{B}_{IJ}\omega^{J}.
\end{align}
Now since the canonical momentum is the generator of translation, we would like the quantum mechanical wave function containing the factor $\exp\left[{i\sum_{I=1}^{d}\widetilde{\pi}_{I}\widetilde{X}^{I}}\right]$ to be periodic in  $\widetilde{X}^I\sim \widetilde{X}^I+ 2\pi \omega^I$ as well.
As a result we need to impose the extra condition $\omega^{I}\widetilde{\pi}_{I}\in\mathbb{Z}$  on the system. The implication is that $\pi_{I}$ has to be quantized as winding number ($\omega^I$) is an integer:
\begin{equation}\label{ctb12}
\widetilde{\pi}_{I}=k_{I},\hspace{5mm}k_{I}\in\mathbb{Z}.
    \end{equation}
Just like in our earlier work \cite{Banerjee:2023ekd}, here again defining a dimensionless field $Y^{I}$ will make our calculation more convenient. $Y^{I}$ is defined as
\begin{align}\label{chh49}
    \widetilde{X}^{I}=\sqrt{\frac{c'}{2}}Y^{I}.
\end{align} 
Using \eqref{fcb1}, $Y^{I}$ can be expressed in terms of oscillators $A$ and $B$ as 
\begin{align}\label{modeexp}
    Y^{I}=y^{I}+A^{I}_{0}\sigma+B^{I}_{0}\tau+i\sum_{n\neq 0}\frac{1}{n}(A^{I}_{n}-in\tau B^{I}_{n})e^{-in\sigma},
\end{align}
where both zero modes are non-zero in the compactified theory.
From \eqref{scaled}, \eqref{ctb24} and \eqref{ctb12} we obtain the following
form of those:
\begin{align}\label{B2}
  (B_{0})_I=\sqrt{2c'}\Big(k_{I}-\sum_{J=1}^{d}\mathbb{B}_{IJ}\omega^{J}\Big),\hspace{10mm}A^{I}_{0}=\sqrt{\frac{2}{c'}}\omega^{I}.
\end{align}
We can also express $Y^{I}$ in terms of the oscillators $C$ and $\Tilde{C}$ as before:
\begin{equation}\label{ctb10}
\begin{split}
   Y^{I}_{L}&=y^{I}_{L}+K^{I}_{L}(\tau+\sigma)+i\sum_{n\neq0}\frac{1}{n}(C^{\mu}_{n}-in\tau C^{\mu}_{n})e^{-in\sigma},\\
    Y^{I}_{R}&=y^{I}_{R}+K^{I}_{R}(\tau-\sigma)+i\sum_{n\neq0}\frac{1}{n}(\Tilde{C}^{\mu}_{n}-in\tau \Tilde{C}^{\mu}_{n})e^{in\sigma}. 
\end{split}
\end{equation}
Here $(K^{I})_{L,R}$ are the left and right moving momenta of the string along compactified dimensions. Applying \eqref{chh42} and \eqref{ctb12} on \eqref{ctb10} gives us 
\begin{align}\label{ctb15}
    (k_I)_{L,R}=\frac{1}{\sqrt{2}}\bigg\{\sqrt{c'}\Big(k_{I}-\sum_{J=1}^{d}\mathbb{B}_{IJ}\omega^{J}\Big)\pm \frac{1}{\sqrt{c'}}\sum_{J=1}^{d}G_{IJ}\omega^{J}\bigg\},
\end{align}
which is the main input in generating the spectra of the compact theory.
With our basic ingredients in place, we now go forward with the discussion of the three vacua separately in what follows.  

\newpage

\section{Oscillator vacuum}\label{sec4}
In this section, we explore the tensionless theory resulting from the first and the weakest condition in \eqref{B1}. The physical state conditions for a theory constructed upon this vacuum are imposed in pure ``sandwich'' form of stress tensor components:
\begin{align}\label{TSQR1}
    \bra{phys'}L_{n}-a_{L}\delta_{n,0}\ket{phys}= \bra{phys'}M_{n}-a_{M}\delta_{n,0}\ket{phys}=0.
\end{align}
Here, $L_{n}$ and $M_{n}$ are the generators of the worldsheet BMS algebra and $a_{\{L,M\}}$ are the ordering ambiguities associated to the zero modes. Note that these generators can be written in terms of the oscillators $C$ and $\Tilde{C}$ from \eqref{tsc9} as
\begin{equation}\label{B5}
\begin{split}
     L_{n}&=\frac{1}{2}\sum_{m}\left[C_{-m}\cdot C_{m+n}-\Tilde{C}_{-m}\cdot \Tilde{C}_{m-n}\right],\\ 
    M_{n}&=\frac{1}{2}\sum_{m}\left[C_{-m}\cdot C_{m+n}+\Tilde{C}_{-m}\cdot\Tilde{C}_{m-n}+2C_{-m}\cdot\Tilde{C}_{-m-n}\right].
    \end{split}
\end{equation}
The vacuum on which this theory is based is called oscillator vacuum which is consequently defined as
\begin{align}\label{B6}  C^{\mu}_{n}\ket{0,0,k^{\mu},k_I,\omega^I}_{c}~=~\Tilde{C}^{\mu}_{n}\ket{0,0,k^{\mu},k_I,\omega^I}_{c}=0\hspace{5mm}\forall n>0. 
\end{align}
The name reflects the similarity with tensile highest weight string vacuum defined likewise using the oscillators $\alpha_n, \tilde{\alpha}_n$.
To familiarise the reader with the notation used here, note that using a string of $C_{-n}$ and $\Tilde C_{-n}$, we can construct a generic excited state on this Oscillator vacuum as follows,
\begin{align}\label{TSQR11}
    \ket{r,s,k^\mu,k
_I,\omega^I}=\sum_{j}\rho_{j}\Bigg(\prod_{i=1}^{p}C^{a_{i}}_{-m_{i}}\prod_{j=1}^{q}\Tilde{C}^{b_{j}}_{-n_{j}}\Bigg)_{j}\ket{0,0,k^{\mu},k_I,\omega^I}_{c},
\end{align}
where $a_{i}$ is the power of $C_{-m_{i}}$ and $b_j$ is the power of $\Tilde{C}_{-n_{j}}$ oscillators. Here the ``levels'' $r$ and $s$ are given by
\begin{align}\label{TSQR12}    r=\sum_{i}^{p}a_{i}m_{i},\hspace{5mm}s=\sum_{i}^{q}b_{i}n_{i}.
\end{align}
From \eqref{B5}, we obtain the expression of $L_{0}$ in terms of the momenta $k_{I\hspace{.25mm}L,R}$ as
\begin{equation}\label{ctb18}
\begin{split}
    L_{0}=\mathcal{N}-\widetilde{\mathcal{N}}+\frac{1}{2}\sum_{I=1}^{d}\big(K^{I}_{L}&K_{I\hspace{.25mm}L}-K^{I}_{R}K_{I\hspace{.25mm}R}\big)
    =\mathcal{N}-\widetilde{\mathcal{N}}+\sum_{I,J=1}^{d}\Big(k_{I}-\mathbb{B}_{IJ}\omega^{J}\Big)\omega^{I},\\
    &=\mathcal{N}-\widetilde{\mathcal{N}}+\sum_{I=1}^{d}k_{I}\omega^{I}.
\end{split}
\end{equation}
where the operators are defined as
\begin{align}\label{TSQR42}
     \mathcal{N}=\sum_{m>0}C_{-m}\cdot C_{m}~~~~~\text{and}\hspace{5mm} \widetilde{\mathcal{N}}=\sum_{m>0}\Tilde{C}_{-m}\cdot \Tilde{C}_{m}.
\end{align}
Here these $\mathcal{N}$ and $\widetilde{\mathcal{N}}$ are nothing but number operators for the $C$ and $\tilde{C}$. To go from first line to second line in \eqref{ctb18} we have used the fact that $\mathbb{B}_{IJ}$ is antisymmetric in $I,J$ indices and hence the term $\mathbb{B}_{IJ}\omega^{I}\omega^{J}=0$. Similarly, the expression for $M_{0}$ in terms of $(k_I)_{L,R}$ can be obtained from \eqref{B5} as \footnote{Note that there is a cross term in canonical momenta here, which isn't present in the tensile case. This makes sure the canonical basis in $\{ k_I, \omega_I\}$ space takes an antisymmetric structure spanned by $\Big(k_{I}-\mathbb{B}_{IJ}\omega^{J}\Big)$. }:
\begin{equation}
\begin{split}
    &M_{0}=\frac{1}{2}\sum_{I=1}^{d}\big(K^{I}_{L}K_{I\hspace{.25mm}L}+K^{I}_{R}K_{I\hspace{.25mm}R}+2K^{I}_{L}K_{I\hspace{.25mm}R}\big)+c'\sum_{\mu=0}^{25-d} k_{\mu} k^{\mu}+\mathcal{N}+\widetilde{\mathcal{N}}+X+X^{\dagger},\\
    &=c'\sum_{I,J,K,L=1}^{d}\Big(k_{I}-\mathbb{B}_{IJ}\omega^{J}\Big)G^{IL}\Big(k_{L}-\mathbb{B}_{LK}\omega^{K}\Big)+c'\sum_{\mu=0}^{25-d} k_{\mu} k^{\mu}+\mathcal{N}+\widetilde{\mathcal{N}}+X+X^{\dagger}.
\end{split}
\end{equation}
where 
\begin{align}
    m^2=\sum_{\mu=0}^{25-d} k_\mu k^\mu ~~~~~~~~~\text{and}~~~~~~~~X=\sum_{m>0}C_{m}\cdot \Tilde{C}_{m}
\end{align}
 The state $\ket{r,s}$ described in \eqref{TSQR11} is an eigenstate of $\mathcal{N}$ and $\widetilde{\mathcal{N}}$ by definition, with eigenvalues $r$ and $s$ respectively. The $X$ operators are special and need to be dealt with separately under sandwiches as in \cite{Bagchi:2020fpr}. These eigenstates of $\mathcal{N}$ and $\widetilde{\mathcal{N}}$ forms the basis of the tensionless Hilbert space.


\subsection{Level matching condition and mass spectrum}
We now move on to investigate the modification in the level matching condition due to presence of this constant $B$ field. It is known that the level matching condition for a physical state is derived from the physical state condition for $L_{0}$ with two identical states, which in this case is given by 
\begin{align}\label{ctb17}
    \bra{r,s}L_{0}-a_{L}\ket{r,s}=0
\end{align}
It has been shown in \cite{Bagchi:2020fpr, Bagchi:2021rfw} that compliance with the algebra demands $a_{L}=0$ in the present case. Applying \eqref{ctb17} with the expression of $L_{0}$ as given in \eqref{ctb18} on a generic state $\ket{r,s,k^\mu,k_I,\omega^I}$ we end up with the following level matching condition for physical state 
\begin{align}\label{B8}
    s-r=\sum_{I=1}^{d}k_{I}\omega^{I}.
\end{align}
This level matching condition turns out to be identical to the level matching condition for a compactified theory without $B$ field, further in line with expectation coming from tensile theory.
\medskip

The mass spectrum on the other hand is computed from the $M_{0}$ condition \cite{Bagchi:2020fpr}, given in the sandwiched version:
\begin{align}\label{massspectrum}
    \bra{r,s}M_{0}-a_{M}\ket{r,s}=0
\end{align}
We again recall that $a_{M}=2$ for this theory.
Following the analysis in \cite{Bagchi:2020fpr} we can see that the expression of $M_{0}$ in \eqref{massspectrum} assigns the following mass squared matrix to the particular state $\ket{r,s,k^\mu,k_I,\omega^I}$ 
\begin{equation}\label{tsc11}
\begin{split}
     m^2&=\mathlarger{\mathlarger{\sum}}_{I,J,K,L=1}^{d}\Big(k_{I}-\mathbb{B}_{IJ}\omega^{J}\Big)G^{IL}\Big(k_{L}-\mathbb{B}_{LK}\omega^{K}\Big)+\frac{1}{c'}(r+s-2).\\
    &=k^{T}G^{-1}k-2k^{T}G^{-1}\mathbb{B}\omega-\omega^{T}\mathbb{B}G^{-1}\mathbb{B}\omega+\frac{1}{c'}(r+s-2)
\end{split}
\end{equation}
In the last line we have expressed the winding and discrete momentum part of the mass spectrum in matrix notation where $B$ field and inverse metric has been expressed as square matrices $(\mathbb{B}, G)$ and $k$ and $\omega$ are expressed as column matrices
\begin{align}\label{bogol}
    k=\begin{bmatrix}
        k_{1}\\ k_{2}\\ \vdots \\ k_{d}
    \end{bmatrix},\hspace{5mm}
     \omega=\begin{bmatrix}
        \omega_{1}\\ \omega_{2}\\ \vdots \\ \omega_{d}
    \end{bmatrix}.
\end{align}
One should note that although we started from a diagonal metric as given in \eqref{BBS1}, while deriving the mass spectrum \eqref{tsc11} from the action \eqref{agol} we did not use the fact that $G_{IJ}$ is diagonal. Hence, the mass spectrum \eqref{tsc11} is valid for any symmetric $G_{IJ}$. The terms in the mass spectrum \eqref{tsc11} involving the internal momentum $k_{I}$ and winding number $\omega^{I}$ can be rewritten as
\begin{align}\label{ctb20}
  \big[\omega^{T} \hspace{3mm} k^{T}\big]
\hspace{2mm}\mathcal{G}
\begin{bmatrix}
    \omega \\ k
\end{bmatrix},
\end{align}
where $\mathcal{G}$ by definition is given by a $2d\times 2d$ matrix defined along the compactified directions,
\begin{align}\label{Buscher1}
    \mathcal{G}=\begin{bmatrix}
    -\mathbb{B}G^{-1}\mathbb{B} & \mathbb{B}G^{-1}\\-G^{-1}\mathbb{B} & G^{-1}
\end{bmatrix}
\end{align}
If we switch off the $B$ field in \eqref{tsc11} we obtain the mass spectrum 
\begin{align}
    m^2=\sum_{I,J=1}^{d}k_{I}G^{IJ}k_{J}+\frac{1}{c'}(r+s-2).
\end{align}
This is identical to the mass spectrum for tensionless string theory constructed on oscillator vacuum in a multi-dimension compactified target space, albeit with a diagonal metric \cite{Banerjee:2023ekd}. Note that the above was a purely KK momentum dominated spectrum, whereas the inclusion of $B$ field makes sure \eqref{tsc11} we have non-trivial winding contribution as well. This will be an important point in our later discussions. 

\medskip 
One thing to immediately note about the matrix $\mathcal{G}$, often called the generalised metric \cite{Aldazabal:2013sca}, is that it is \textit{not invertible} unlike the corresponding matrix in tensile case \cite{Becker:2006dvp}, which makes all the difference. In a separate section dedicated to the theory constructed on Induced vacuum we will discuss how the corresponding matrix in the tensile case explicitly reduces to  the degenrate $\mathcal{G}$ at tensionless limit. Now the degeneracies in the generalised metric is not uncommon, and in the literature one can find instances (see for example \cite{Morand:2017fnv} and follow ups thereof) where embedding of non-Riemannian target spaces in Double Field Theory (DFT) has been considered. We just would like to point out to the reader, that for our case the compactified target space is still non-degenerate ($\sim \delta^{IJ}$), however the worldsheet is Carrollian, which leads to the degeneration of the whole generalised metric itself. Having said that, it is also fair to point out that a zero determinant generalised metric can also appear from a degenerate compact target space, something we will see in a future section.


\subsection{Symmetries of the mass spectrum}
Let us now look closely at the detailed symmetries of the Tensionless mass spectrum in our case. For the tensile case with constant background metric and antisymmetric field, the symmetries of string mass spectrum compactified on a $d$-Torus can be generalized to the group $O(d,d;\mathbb{Z})$ \cite{Becker:2006dvp,Blumenhagen:2013fgp}. The ${O(d,d;\mathbb{Z})}$ group is a group of transformations matrices ($O$) in $d+d$ dimensions which satisfy the following 
\begin{align}
    O^{T}\begin{bmatrix}
       0 & \mathbb{1}_{d}\\
        \mathbb{1}_{d} &0
   \end{bmatrix}O=\begin{bmatrix}
       0 & \mathbb{1}_{d}\\
        \mathbb{1}_{d} &0
   \end{bmatrix}.
\end{align}
This symmetry of tensile string mass spectrum, written in terms of the generalised metric, has been well-known \cite{Aldazabal:2013sca}.
In what follows, we will carefully analyze the fate of all of these symmetries in our degenerate case.
\subsubsection{Fate of the $O(d,d;\mathbb{Z})$ Symmetry}
In general, the $O(d,d;\mathbb{Z})$ symmetries in the tensile spectrum could be thought of as symmetries of the $O(d,d)$ invariant generalised metric, the tensile cousin of \eqref{Buscher1}. In this metric, momentum and winding modes are $d$ dimensional vectors as written in \eqref{bogol}, which can be embedded into $2d$ dimensional generalised momentum, so that the compactified contribution to the mass formula becomes an inner product of generalised momenta w.r.t. the $O(d,d)$ metric. The symmetry group $O(d,d)$ can further be decomposed as successive products of subgroup symmetries, and we will give a description of these subgroups in the tensionless version of the theory. 

\paragraph{Inversion symmetries}
We start with the part of the mass spectrum in \eqref{ctb20} along with the definition \eqref{Buscher1}. The first question we ask: is it possible to find a transformation for {$\mathcal{G}$} which, along with the duality transformation $\omega\leftrightarrow k$, would make the mass spectrum \eqref{ctb20} invariant? It is not hard to see that if we make $\omega\leftrightarrow k$ transformation on \eqref{ctb20} then it can remain invariant, but only if we make the following transformation on $\mathcal{G}$
\begin{align}\label{Buscher2}
    \mathcal{G}\to\mathcal{G'}\hspace{5mm}\mathcal{G'}=\begin{bmatrix}
    -\mathbb{B}'G'^{-1}\mathbb{B}' & \mathbb{B}'G'^{-1}\\-G'^{-1}\mathbb{B}' & G'^{-1}
\end{bmatrix}=\begin{bmatrix}
   G^{-1}  & -G^{-1}\mathbb{B}\\\mathbb{B}G^{-1} & -\mathbb{B}G^{-1}\mathbb{B}
\end{bmatrix}.
\end{align}
However, there is one catch! Does the transformation \eqref{Buscher2} make sense for our case? Clearly for $\mathbb{B}=0$ it doesn't. At $\mathbb{B}=0$, $\mathcal{G}$ and $\mathcal{G}'$ respectively become 
\begin{align*}
    \mathcal{G}=\begin{bmatrix}
       0_{d\times d} & 0_{d\times d}\\
       0_{d\times d} & G^{-1}
   \end{bmatrix}\hspace{5mm}\mathcal{G}'=\begin{bmatrix}
       G^{-1} & 0_{d\times d}\\
       0_{d\times d} & 0_{d\times d}
   \end{bmatrix}.
\end{align*}
If we try to write this transformation in terms of compactification radii, it would very weirdly look like
\begin{align}\label{thang-gol}
    \frac{1}{R_{i}}\leftrightarrow 0.
\end{align}
Which could be interpreted as the $\alpha' \to \infty$ fate of the $R \to \frac{\alpha'}{R}$ duality prevalent in the tensile case.
In order to make the transformation \eqref{Buscher2} sensible, the $B$ field we switched on ($\mathbb{B}$ as well as $\mathbb{B}'$) must satisfy certain conditions. In fact, \eqref{Buscher2} demands the following equations to be simultaneously satisfied
\begin{subequations}
    \begin{equation}\label{Buscher3}
    G'^{-1}=-\mathbb{B}G^{-1}\mathbb{B}
\end{equation}
\begin{equation}\label{Buscher4}
    -G'^{-1}\mathbb{B}'=\mathbb{B}G^{-1}
\end{equation}
\end{subequations}
Applying \eqref{Buscher3} on \eqref{Buscher4} one gets the following
\begin{align*}
\mathbb{B}G^{-1}\mathbb{B}\mathbb{B}'=\mathbb{B}G^{-1} 
   \implies \mathbb{B}\mathbb{B}' =\mathbb{1}_{d}
\end{align*}
Clearly the transformation in \eqref{Buscher2} is valid only if $\mathbb{B}'=\mathbb{B}^{-1}$. The implication is that the transformation \eqref{Buscher3} can be consistently defined only for a $B$ field that is invertible. Hence the inversion symmetry can be observed only for explicitly invertible $B$ fields in particular subspaces. Although for $\mathbb{B}=0$ the transformation could be trivially defined, as we have just seen, it does not make sense.

\medskip Let us compare this to the known tensile case. In case of tensile string theory, $\mathcal{G}$ is invertible and the symmetry was 
    $\omega\leftrightarrow k,\hspace{1mm}\mathcal{G}\to\mathcal{G}^{-1}$.
Here, however, $\mathcal{G}$ is not invertible and consequently one can see:
\begin{align}
    \mathcal{G}\mathcal{G}'=0.
\end{align}
Hence we should readily expect some of the symmetries of the parent $O(d,d)$ to be broken for null string spectrum.


\paragraph{Basis transformations}

As discussed in \cite{Giveon:1994fu}, one other component of  ${O(d,d;\mathbb{Z})}$ group is basis change matrices, given by
\begin{align}\label{potol}
    g_{C}=\begin{bmatrix}
       C & 0\\
        0 & (C^{T})^{-1}
   \end{bmatrix},   
\end{align}
where $C\in GL(d,\mathbb{Z})$. The effect of this transformation is more or less straightforward. Applying the transformation $g_{C}$ on the tensionless generalised metric, $\mathcal{G}$ we get the following:
\begin{align}\label{panchu}
    \mathcal{G}'&=\begin{bmatrix}
       C & 0\\
        0 & (C^{T})^{-1}
   \end{bmatrix}\begin{bmatrix}
    -\mathbb{B}G^{-1}\mathbb{B} & \mathbb{B}G^{-1}\\-G^{-1}\mathbb{B} & G^{-1}\end{bmatrix}\begin{bmatrix}
       C^{T} & 0\\
        0 & C^{-1}
   \end{bmatrix}
   =\begin{bmatrix}
    -\mathbb{B}'G'^{-1}\mathbb{B}' & \mathbb{B}'G'^{-1}\\-G'^{-1}\mathbb{B}' & G'^{-1}
\end{bmatrix},
   \end{align}
where $\mathbb{B}'$ and $G'$ are respectively given by
\begin{align}\label{ghenchu}
\mathbb{B}'=C\mathbb{B}(C^T)\hspace{5mm}G'=CG(C^T).
\end{align}
The compactified part of the mass spectrum corresponding to the new generalised metric $\mathcal{G'}$ is given by 
\begin{align}
    m^2=[\omega^{T} \hspace{3mm} k^{T}]
\hspace{2mm}\mathcal{G}'
\begin{bmatrix}
    \omega \\ k
\end{bmatrix}=[\omega'^{T} \hspace{3mm} k'^{T}]
\hspace{2mm}\mathcal{G}
\begin{bmatrix}
    \omega' \\ k'
\end{bmatrix},\hspace{5mm}\begin{bmatrix}
    \omega' \\ k'
\end{bmatrix}=\begin{bmatrix}
       C^{T} & 0\\
        0 & C^{-1}
   \end{bmatrix}\begin{bmatrix}
    \omega \\ k
\end{bmatrix}
\end{align}
Hence, a tensionless mass spectrum with metric $G'$ and $B$ field $\mathbb{B}'$ with winding number and internal momenta given by the $2d$ vector $\{\omega,k\}$ is equal to a tensionless mass spectrum with metric $G$ and $B$ field $\mathbb{B}$ with winding number $C^T\omega$ and internal momenta $C^{-1}k$.

\paragraph{Sectorised T-duality}
We have already seen in the preceding sections that for invertible $B$ fields the mass spectrum \eqref{ctb20} is symmetric under the inversion transformation $\omega \leftrightarrow k$ and $\mathcal{G}\leftrightarrow \mathcal{G}'$ where $\mathcal{G}'$ is given by \eqref{Buscher2}. Note that this transformation in terms of matrix notation can be given as: 
\begin{equation}\label{tintin}
    \mathcal{G'} \to A ~\mathcal{G}~A^T ~~~\text{and}~~~~\begin{bmatrix}
       \omega'\\
       k' \end{bmatrix} \longrightarrow A~\begin{bmatrix}
       \omega\\
       k
   \end{bmatrix}, ~~~A=\begin{bmatrix}
       0 & \mathbb{1}_{d}\\
        \mathbb{1}_{d} &0
   \end{bmatrix}.
\end{equation}
sectorised T-duality is a generalisation of the circle inversion duality and acts on a \textit{specifically chosen compact direction}, instead of all directions at once.  In retrospect, the sectorised duality matrices are given by 
\begin{align}\label{hotol}
    g_{D_{i}}=\begin{bmatrix}
       \mathbb{1}_{d}-e_{i} & e_{i}\\
        e_{i} & \mathbb{1}_{d}-e_{i}
   \end{bmatrix},
\end{align}
where $e_{i}$ is a matrix with $1$ in the $ii$ position and all the remaining elements are zeros. The sectorised duality matrix $g_{D_{i}}$, when applied on $\{\omega,k\}$, will interchange a particular $\omega_{i}$ and $k_{i}$, with $i \in \{I\}$, leaving the rest of the $\omega$'s and $k$'s intact. If we apply this matrix on our degenerate  $\mathcal{G}$ as $\mathcal{G}'=g_{D_{i}}\mathcal{G}g_{D_{i}}^{T}$, remarkably it leads us to find that $\mathcal{G}'$ cannot be consistently expressed in terms of a metric and a $B$ field in the form given in \eqref{Buscher1} like we did in case of $g_{C}$. Hence, even though the mass spectrum is still symmetric\footnote{Since $g_{D_{i}}^{T}g_{D_{i}}=\mathbb{1}$, if we apply $\mathcal{G}'=g_{D_{i}}\mathcal{G}g_{D_{i}}^{T}$ and $\begin{bmatrix}
       \omega'\\
       k'
   \end{bmatrix}=g_{D_{i}}\begin{bmatrix}
       \omega\\
       k
   \end{bmatrix}$ then the mass spectrum will still be same.}, the symmetry would not close to a different action for tensionless string with a different $B$ field.  One could say this indicates in general the closure of sectorised T-dual theory is not defined in the same compact moduli space for the tensionless string, owing to the fact that generalised metric is degenerate. In the coming section we will work out the example of $d=2$, where we will see this effect even more explicitly.

   \medskip Note that the inversion transformation as written in \eqref{tintin} is generated by the sectorised duality matrices in the following way
   \begin{align}
       A=\begin{bmatrix}
       0 & \mathbb{1}_{d}\\
        \mathbb{1}_{d} &0
   \end{bmatrix}=\begin{bmatrix}
       \mathbb{1}_{d}-\sum_{i=1}^{d}e_{i} & \sum_{i}^{d}e_{i}\\
        \sum_{i}^{d}e_{i} & \mathbb{1}_{d}-\sum_{i}^{d}e_{i}
   \end{bmatrix}=\prod_{i=1}^{d}g_{D_{i}}.
   \end{align}
   As we have seen, although $g_{D_{i}}$s individually do not lead to any mapping to a new tensionless action with different $G$ and $B$, the inversion symmetry, which is generated by product of all $g_{D_{i}}$s, does lead us to an action with $G'=-\mathbb{B}^{-1}G\mathbb{B}^{-1}$ and $\mathbb{B}'=\mathbb{B}^{-1}$ (provided that $\mathbb{B}$ is invertible along all the $i$th legs).


\paragraph{Discrete shift symmetry}
 Let us recall the level matching condition and  mass spectrum of the physical states constructed on oscillator vacuum. We write them out explicitly in the following suggestive way 
\begin{equation}\label{gadha7}
    \begin{split}
        s&-r=\frac{1}{2}\sum_{I=1}^{d}\big(K^{I}_{L}K_{I\hspace{.25mm}L}-K^{I}_{R}K_{I\hspace{.25mm}R}\big),\\
        m^2=\frac{1}{2c'}&\sum_{I=1}^{d}(K^{I}_{L}+K^{I}_{R})(K_{I\hspace{.25mm}L}+K_{I\hspace{.25mm}R})+\frac{1}{c'}(r+s-2).
    \end{split}
\end{equation}
Let us reiterate that $(K_{I})_{L,R}$ are as obtained in \eqref{ctb15}
\begin{align*}
      (K_{I})_{\hspace{.25mm}L,R}=\frac{1}{\sqrt{2}}\bigg\{\sqrt{c'}\Big(k_{I}-\sum_{J=1}^{d}\mathbb{B}_{IJ}\omega^{J}\Big)\pm \frac{1}{\sqrt{c'}}\sum_{J=1}^{d}G_{IJ}\omega^{J}\bigg\}.
\end{align*}
It can be easily seen that $(K_{I})_{\hspace{.25mm}L,R}$ are invariant under the following discrete transformations 
\begin{align}\label{ctb21}
    \mathbb{B}_{IJ}\to \mathbb{B}_{IJ}+\mathcal{N}_{IJ},\hspace{8mm} k_{I}\to k_{I}+\mathcal{N}_{IJ}\omega^{J},\hspace{8mm} \omega^I\to \omega^I, \hspace{8mm}\mathcal{N}_{IJ}=-\mathcal{N}_{JI}, 
\end{align}
where $\mathcal{N}_{IJ}\in \mathbb{Z}$. This can be expressed  in matrix notation as:
\begin{align}\label{alu}
    g_{N}=\begin{bmatrix}
        \mathbb{1}_d & 0\\
        N_{IJ} & \mathbb{1}_d
    \end{bmatrix}
\end{align}
If we apply the $g_{N}$ transformation on the generalised metric, then $\mathcal{G}'=g_{N}\mathcal{G}g_{N}^{T}$ can be consistently expressed in terms of $G'_{IJ}=G_{IJ}$ and $\mathbb{B}'_{IJ}=\mathbb{B}_{IJ}+\mathcal{N}_{IJ}$ in the form given in \eqref{Buscher1}.
Since the part of the mass spectrum \eqref{gadha7} depending on $\{\mathbb{B}_{IJ},k_{I},\omega^{I}\}$ comes entirely from $(k_{I})_{L,R}$, it is apparent that the tensionless mass spectrum \eqref{ctb20}, like the tensile counterpart, is invariant under the transformations \eqref{ctb21}. 

\subsection{Concrete examples of compactification}

\subsubsection*{ Even dimension compactified ($d=2$)}
Now, we look into the mass spectrum of tensionless string with constant $B$ field in a spacetime with only two of the dimensions compactified on a two torus. The reader may compare the contents in this section with the mass spectrum symmetries of tensile strings with a $B$ field, having two compactified target space directions, as discussed in Appendix \eqref{apA}.  When we compactify only two dimensions and demand that the constant $B$ field has non-zero components only along those two compactified directions, we immediately arrive at a unique choice of the $B$ field which is 
\begin{align}\label{chhagol}
    \mathcal{B}_{IJ}=b\hspace{1mm}\epsilon_{IJ},\hspace{5mm}I,J\in \{1,2\}
\end{align}
where $\epsilon_{IJ}$ is the antisymmetric tensor in two dimensions. We choose $\epsilon_{12}=1$ here. After the coordinate scaling defined in \eqref{BBS1}, the rescaled $\mathbb{B}_{IJ}$ becomes 
\begin{equation}\label{gadheswar}
    \mathbb{B}_{IJ}=b R_{I}R_{J}\epsilon_{IJ}.
\end{equation}
The above $\mathbb{B}_{IJ}$ is invertible, and hence, we can expect a sensible inversion symmetry to exist for the resulting mass spectrum. We find that for the $B$ field \eqref{chhagol}, equation \eqref{tsc11} gets simplified to 
\begin{align}\label{pagol}
    m_2^2=\frac{k_{1}^2}{R_{1}^2}+\frac{k_{2}^2}{R_{2}^2}+2b\Bigg(\frac{R_{1}}{R_{2}}k_2\omega_1-\frac{R_{2}}{R_{1}}k_1\omega_2\Bigg)+b^2(R_1^2\omega_1^2+R_2^2\omega_2^2)+\frac{1}{c'}(r+s-2).
\end{align} 
In the above, we have made a slight abuse of notation. In order to make the expression less messy we chose to write \{$\omega_1, \omega_2$\} instead of \{$\omega^1, \omega^2$\}. Hereafter in the explicit form of mass spectrum (i.e. not in the matrix form), we continue to follow this notation. We have seen that the mass spectrum in \eqref{tsc11} is symmetric under the transformations:
\begin{align}\label{B14}    k_{I}\leftrightarrow\omega^{I},\hspace{5mm}R_I\leftrightarrow \frac{1}{bR_I},\hspace{5mm}b\to-b\hspace{5mm}I\in\{1,2\}
\end{align}
It can be seen that for $d=2$, $G^{-1}$ and $B$ field given in \eqref{mathagol} and \eqref{pagol} respectively, the transformation defined on the generalised metric $\mathcal{G}$ defined in \eqref{Buscher2} will boil down to \eqref{B14}. Hence we see that indeed the mass-spectrum displays a sensible inversion symmetry, which very evidently would not happen when the $B$ field is turned off. This non-trivial effect of the antisymmetric field in reinstating a version of T-duality remains a very crucial point in our discussion.

\medskip Now, we want to see whether this mass spectrum displays the full $O(2,2;\mathbb{Z})$ symmetry, in line with our discussion before. It can be easily seen that the discrete shift  symmetry is preserved for the mass spectrum: for $d=2$, the transformation \eqref{ctb21} in terms of $b$, $R_{1}$ and $R_{2}$ becomes the following
\begin{align}\label{babui}
    b\to b+\frac{n}{R_{1}R_{2}}\hspace{5mm}k_{I}\to k_{I}+n\epsilon_{IJ}\omega^J\hspace{5mm}n\in\mathbb{Z}.
\end{align}
One can further check that the mass spectrum is symmetric under the basis change transformation where the transformation of $G$ and $\mathbb{B}$ will be dictated by \eqref{panchu} and \eqref{ghenchu}. 
\medskip

Since we discussed that sectorised dualities are non-trivial in this case, we can try to find the impact of $g_{D_{i}}$ on $\mathcal{G}$. Let us first express it for $d=2$ and $\mathbb{B}_{IJ}$ as written in \eqref{gadheswar} in the form of a $4\times 4$ matrix:
\begin{align}
 \mathcal{G}=\begin{bmatrix}
        b^2R^{2}_{1} & 0 & 0 & \frac{bR_{1}}{R_{2}}
       \\  0 & b^2R^{2}_{2} & -\frac{bR_{2}}{R_{1}} & 0\\
       0 & -\frac{bR_{2}}{R_{1}} & \frac{1}{R^{2}_{1}} & 0\\
       \frac{bR_{1}}{R_{2}} & 0 & 0 & \frac{1}{R^{2}_{2}}
    \end{bmatrix}
    \end{align}
    Let us consider the sectorised transformation $g_{D_{2}}$. It will explicitly interchange $k_{2}$ and $\omega_{2}$ and matrix form of this is given by
    \begin{align}
        \begin{bmatrix}
        1 & 0 & 0 & 0
       \\  0 & 0 & 0 & 1\\
       0 & 0 & 1 & 0\\
       0 & 1 & 0 & 0
    \end{bmatrix}
    \end{align}
Now we would like to study how $\mathcal{G}$ evolves for the $g_{D_{2}}$ transformation. It is easy to arrive at:
\begin{align}\label{opodartho7}
   \mathcal{G}'= g_{D_{2}}\mathcal{G}g_{D_{2}}^T=\begin{bmatrix}
        b^2R^{2}_{1} & \frac{bR_{1}}{R_{2}} & 0 & 0
       \\  \frac{bR_{1}}{R_{2}} & \frac{1}{R^{2}_{2}} & 0 & 0\\
       0 & 0 & \frac{1}{R^{2}_{1}} & -\frac{bR_{2}}{R_{1}}\\
       0 & 0 & -\frac{bR_{2}}{R_{1}} & b^2R^{2}_{2}
    \end{bmatrix}
   \end{align}
If we recall the expression of $\mathcal{G}$ in terms of $G$ and $\mathbb{B}$ from \eqref{Buscher1}, we see that the off-diagonal blocks seem to suggest that $\mathbb{B'}=0$, however, the diagonal blocks suggest that $\mathbb{B'}\neq0$. Hence $\mathcal{G}'$ cannot be put into the form of a generalised metric derived solely from any tensionless string action. Although the mass spectrum is still invariant under this transformation, this transformation cannot be called a symmetry of tensionless action since it does not lead to any mapping between two different tensionless theories. Note that this property is very unique to the tensionless string sigma model, and hints at the same mass spectrum being shared by some other theory, \textit{not covered by the structure} of \eqref{Buscher1}. We will comment more on this when we study the limiting spectrum in the Induced vacuum section.

\medskip
It is important to note that although the transformations $g_{D_{1}}$ and $g_{D_{2}}$ individually do not lead us to any new tensionless action, the inversion symmetry as given in \eqref{B14} is generated by both of them combined, as the field is invertible along these two compact legs of the target space \footnote{In fact in two dimensional compact space, the sectorised T-duality group in the tensile string case comprises of $\{\mathbb{I},g_{D_1},g_{D_2},g_{D_1}g_{D_2}\}$, while the analogue in the null case reads $\{\mathbb{I},g_{D_1}g_{D_2}\}$ i.e. only even duality transforms work. This can be thought of as a breaking of the symmetry $\mathbb{Z}_2\times\mathbb{Z}_2 \to \mathbb{Z}_2$.}. Moreover as we have seen, it connects the theory to an action with a different $B$ field. Again, remember that this depends crucially on having an invertible $B$ field along both of the compact directions. 
\subsubsection*{Example of odd dimensions compactified ($d=3$)}
Next, to understand these dualities better, we find the mass spectrum of tensionless strings in a target spacetime with three dimensions compactified with constant $B$ fields and see whether we can find similar forms of symmetry in the spectrum as we found in the $2d$ case. Now since we have more than one directional option to put the legs of our two-form field on, the dynamics is surely more interesting.  The most general constant $3\times 3$ $B$ field is given by
\begin{align}
  \mathbb{B}_{IJ}= \begin{bmatrix}
        0 && aR_1R_2  && bR_1R_3 \\
        -aR_1R_2 && 0 && cR_2R_3 \\
        -bR_1R_3  && -cR_2R_3  && 0\\
    \end{bmatrix}
\end{align}
The mass spectrum obtained from the above $B$ field is given by
\begin{equation}\label{hatgol}
\begin{split}
    m_3^2=&\frac{k_{1}^2}{R_{1}^2}+\frac{k_{2}^2}{R_{2}^2}+\frac{k_{3}^2}{R_{3}^2}+2a\left(\frac{R_{1}}{R_{2}}k_2\omega_1-\frac{R_{2}}{R_{1}}k_1\omega_2\right)+2b\left(\frac{R_{1}}{R_{3}}k_3\omega_1-\frac{R_{3}}{R_{1}}k_1\omega_3\right)\\&+2c\left(\frac{R_{2}}{R_{3}}k_3\omega_2-\frac{R_{3}}{R_{2}}k_2\omega_3\right)+\left(a R_1\omega_1-cR_3\omega_3\right)^2+(a R_2 \omega_2+b R_3 \omega_3)^2\\&
    +(bR_1\omega_1+c R_2\omega_2)^2+\frac{1}{c'}(r+s-2).
    \end{split}
\end{equation}
Here $a,b,c$ are constants and $R_I$ are the relevant compact radii.
Clearly this mass spectrum won't be symmetric under the duality transformation of the type given in \eqref{B14} if all of these fields are turned on. In fact if we interchange $k_{I}$ and $\omega_{I}$ in the mass spectrum \eqref{hatgol}, there is no transformation on $R_{I}$ which will take us from quadratic terms of $k$'s to the
quadratic terms of $\omega$'s. The reason is that the quadratic terms in $\omega$ involves three cross terms like $\omega_{K}\omega_{J}$, and there is no such cross terms in the quadratic terms of $k$'s. The only way we can get rid of all three cross terms in the quadratic terms of $\omega$'s is to set at least two of the three parameters $a,b,c$ to zero, i.e. working with one of the following three $B$ fields with two legs among the three compact directions:
\begin{align}\label{level}
\begin{split}
    \mathbb{B}_{IJ}^1= &\begin{bmatrix}
        0 && aR_1R_2  && 0 \\
        -aR_1R_2 && 0 && 0 \\
        0  && 0  && 0\\
    \end{bmatrix},\hspace{5mm} \mathbb{B}_{IJ}^2=\begin{bmatrix}
        0 && 0  && bR_1R_3 \\
        0 && 0 && 0 \\
        -bR_1R_3  && 0  && 0\\
     \end{bmatrix},\\&\hspace{15mm}\mathbb{B}_{IJ}^3=\begin{bmatrix}
        0 && 0  && 0 \\
        0 && 0 && cR_2R_3 \\
        0  && -cR_2R_3  && 0\\
    \end{bmatrix}.
    \end{split}
\end{align}
However, whenever we try to work on one of the above three $B$ fields we end up getting a mass spectrum where in order to define a T-dual transformation of the type \eqref{B14}, we need to make a nonsensical transformation of type \eqref{thang-gol} on one of the compactified radius. For example, if we work with the first $B$ field where $a\neq 0, b=c=0$, the quadratic terms of $\omega$ will be
\begin{align}
    a^2(R^{2}_{1}\omega^{2}_{1}+R^{2}_{2}\omega^{2}_{2}).
    \end{align}
 The absence of the $\omega^{2}_{3}R^{2}_{3}$ term in the mass spectrum now implies the transformation of type \eqref{thang-gol} on $R_{3}$. Hence, for $d=3$, we cannot define the full T-duality, i.e. the full inversion symmetry involving all compact directions.
 
 \medskip
 
 It can be easily seen that the discrete shift  symmetry is preserved for the mass spectrum \eqref{hatgol} for $d=3$, where the transformation \eqref{ctb21} in terms of $a,b,c, R_{1},R_{2}$ and $R_{3}$ becomes the following
\begin{align}\label{babui}
    a\to a+\frac{l}{R_{1}R_{2}},\hspace{3mm}b\to b+\frac{m}{R_{1}R_{3}}, \hspace{3mm}c\to c+\frac{n}{R_{2}R_{3}}\hspace{3mm}k_{I}\to k_{I}+\mathcal{N}_{IJ}\omega^J\hspace{3mm}l,m,n\in\mathbb{Z}
\end{align}
where $\mathcal{N}_{IJ}$ is given by a $3\times 3$ matrix:
\begin{align}
    \mathcal{N}_{IJ}= \begin{bmatrix}
        0 && l  && m \\
        -l && 0 && n \\
        -m  && -n  && 0\\
    \end{bmatrix}.
\end{align}
Now, sectorised duality transformations, as defined in \eqref{hotol}, are not respected by the mass spectrum \eqref{hatgol}. However, if we use any one of the $\mathbb{B}_{IJ}^i$ for $(i=1,2,3)$ as given in \eqref{level}, then we observe that the mass spectrum will be symmetric under the following combination of transformations respectively:
    \begin{center}
\begin{tabular}{ |c|c| } 
\hline
 $B$ field & Even sectorised duality \\
\hline
$\mathbb{B}_{IJ}^1 $ & $g_{D_1}g_{D_2}$  \\ 
\hline
 $\mathbb{B}_{IJ}^2$ & $g_{D_1}g_{D_3}$   \\ 
\hline
$\mathbb{B}_{IJ}^3$ & $g_{D_2}g_{D_3}$   \\ 
\hline
\end{tabular}
\captionof{table}{Configuration of symmetries for $B$ field along pairwise legs of the compact manifold. \label{Tab} }
\end{center}
This indeed mirrors our discussions in the $2d$ compact case, where such even sectorised dualities were evident along the $B$ field directions.
Let us consider $\mathcal{G}$ defined in \eqref{Buscher1} constructed using $\mathbb{B}_{IJ}^1$ and $G_{IJ}=R_I^2\delta_{IJ}$, which corresponds to the mass spectrum \eqref{hatgol} with $b=c=0$. It can be shown that the transformation $g_{D_1}g_{D_{2}}$ connects this one to a different action involving $\mathbb{B}',G'$ given below 
\begin{align}\label{new}
    \mathbb{B}_{IJ}'^1= &\begin{bmatrix}
        0 && -\frac{1}{aR_1R_2}  && 0 \\
        \frac{1}{aR_1R_2} && 0 && 0 \\
        0  && 0  && 0\\
    \end{bmatrix},\hspace{5mm} G_{IJ}'^1=\begin{bmatrix}
        \frac{1}{a^2R_1^2} && 0  && 0 \\
        0 && \frac{1}{a^2R_2^2} && 0 \\
        0  && 0  && R_3^2\\
     \end{bmatrix}
\end{align}
For detailed calculation, the reader is referred to appendix \eqref{factorised}. For $\mathbb{B}_{IJ}^2$ and $\mathbb{B}_{IJ}^3$, the transformations listed in Table \eqref{Tab} is given by
\[  \mathbb{B}_{IJ}^2\; \xrightarrow{g_{D_1}g_{D_{3}}}{}\;\mathbb{B}_{IJ}'^2=\begin{bmatrix}
        0 && 0  && -\frac{1}{bR_1R_3} \\
        0 && 0 && 0 \\
        \frac{1}{bR_1R_3}  && 0  && 0 \\
    \end{bmatrix},~~ G_{IJ}^2\; \xrightarrow{g_{D_1}g_{D_{3}}}{}\;G_{IJ}'^2=\begin{bmatrix}
        \frac{1}{b^2R_1^2} && 0  && 0 \\
        0 && R_2^2 && 0 \\
        0  && 0  && \frac{1}{b^2R_3^2} \\
    \end{bmatrix} \]
    \[ \mathbb{B}_{IJ}^3\; \xrightarrow{g_{D_2}g_{D_{3}}}{}\;\mathbb{B}_{IJ}'^3=\begin{bmatrix}
        0 && 0  && 0 \\
        0 && 0 &&  -\frac{1}{cR_2R_3} \\
        0  && \frac{1}{cR_2R_3}  && 0\\
    \end{bmatrix},~~~    G_{IJ}^3\; \xrightarrow{g_{D_2}g_{D_{3}}}{}\;G_{IJ}'^3=\begin{bmatrix}
        R_1^2 && 0  && 0 \\
        0 && \frac{1}{c^2R_2^2} && 0 \\
        0  && 0  && \frac{1}{c^2R_3^2} \\
    \end{bmatrix}\]
Looking at the above three transformation of $B$ fields we observe one common feature in all of them. All $\mathbb{B}_{IJ}^i$'s listed in \eqref{level} acts like invertible two-form $B$ fields in a 2 dimensional sub-space ($T^2$ $\subset$ $T^3$). The transformations listed in the table are restricted, in a pairwise manner, to those sub-spaces only and in those sub-spaces, they transform those $B$'s to their respective inverses.

\medskip

We also tried sectorised duality transformations on linear combinations of the two-form fields and found that the mass spectrum constructed on different $B$ fields will be symmetric under different combinations of $g_{D_{i}}$'s. They are given below in Table \eqref{combination}.
 \begin{center}
\begin{tabular}{ |c|c| } 
\hline
 $B$ field & Even set of  sectorised dualities \\
\hline
$l_{1}\mathbb{B}_{IJ}^1+l_{2}\mathbb{B}_{IJ}^2 $ & $g_{D_1}g_{D_2}$, $g_{D_1}g_{D_3}$, $g_{D_3}g_{D_1}g_{D_2}g_{D_3}$, $g_{D_1}g_{D_2}g_{D_3}g_{D_2}$ \\ 
\hline
 $l_{2}\mathbb{B}_{IJ}^2+l_{3}\mathbb{B}_{IJ}^3$ & $g_{D_1}g_{D_3}$, $g_{D_2}g_{D_3}$, $g_{D_1}g_{D_2}g_{D_3}g_{D_1}$, $g_{D_1}g_{D_2}g_{D_3}g_{D_2}$  \\ 
\hline
$l_{1}\mathbb{B}_{IJ}^1+l_{3}\mathbb{B}_{IJ}^3$ & $g_{D_1}g_{D_2}$, $g_{D_2}g_{D_3}$, $g_{D_3}g_{D_1}g_{D_2}g_{D_3}$, $g_{D_1}g_{D_2}g_{D_3}g_{D_1}$   \\ 
\hline
\end{tabular}
\captionof{table}{Configurations of linear combinations of $B$ field along the compact manifold. Here $l_i$'s are constant parameters. \label{combination}}
\end{center}
Based on our observation of the effects of inversion transformation and sectorised dualities, both for $T^2$ and $T^3$ compactifications, we can now make some generalised comments on the impact of sectorised duality in any number of compact dimensions. Operating sectorised duality an odd number of times never lead us to any mass spectrum of tensionless strings which can be constructed from some different null string theory compatible with a $\mathbb{B}'$ and $G'$. However for appropriate $\mathbb{B}$'s applying sectorised duality only even number of times gives tensionless mass spectrum with different $\mathbb{B}'$ and $G'$. 
\medskip

For even dimension compactification, the inversion transformation itself becomes an example of even number of sectorised duality transformation, where we need an invertible $B$ field to make it work. However when we compactify odd number of dimensions, the full inversion symmetry consists of odd number of sectorised duality transformations, which by virtue of the discussion above, does not work for any $B$ field. But still in odd dimensions, if we chose a (or a collection of) $B$ field(s) which acts like an invertible $B$ field in an even (two) dimensional subspace, then within that even dimensional subspace, operating sectorised duality
will invert that $B$ field consistently. The degeneracy of the $\mathcal{G}$ thus significantly constrains the allowed forms of symmetries.

\subsection{States arising due to compactification}

In this section we look into the new states arising due to compactification. We study how the presence of a constant $B$ field affects the masses of those states. We also study the nature of the particles corresponding to those states. In order to visualise the effects of Kalb- Ramond field more clearly, we will work out the example of two compactified dimensions.
\subsubsection*{Vacuum}
\begin{figure}[h]
    \centering
    \includegraphics{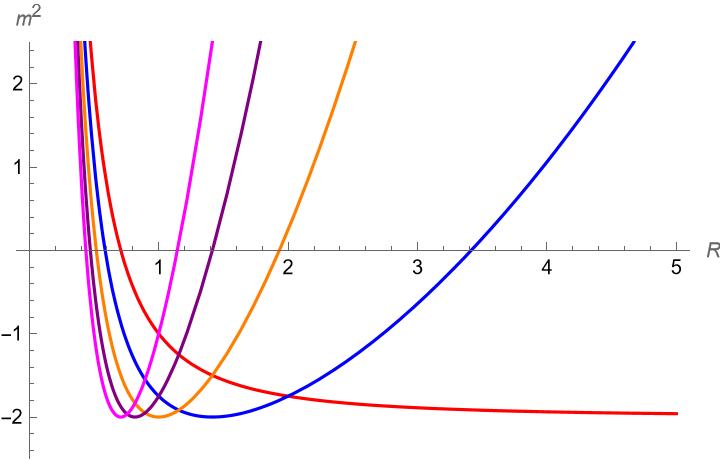}
    \caption{Heuristic plot of $m_2^2$ in \eqref{twomass} vs Radius ($R$) for different values of constant $B$ field. Red is for $b=0$, blue for $b=0.5$, orange for $b=1$, purple for $b=1.5$,
magenta is for $b=2$. Here we have considered $\omega_2=0, \omega_1=1, k_1=0~ \text{and}~ k_2=1$ with $c'=1$. Minima for these curves signifies the position of tachyon. Note that the effective tachyonic region reduces as a the $B$ field is dialled to higher values.}
    \label{massplot}
\end{figure}

In our earlier work \cite{Banerjee:2023ekd} it was found that when we consider oscillator vacuum in a target space with one dimension compactified, the vacuum state $\ket{0,0,k^{\mu},K,W}$ can be physical only if $KW=0$, i.e. either $K$ or $W$ is zero, which is the consequence of the level matching condition. However, when more than one dimensions are compactified, we can have oscillator vacuum $\ket{0,0,k^{\mu},k_{I},\omega^{I}}$ with both $k_{I}\neq 0$ and $\omega^{I}\neq 0$. The only restriction coming from the level matching condition \eqref{B8} is that the internal momentum on the torus $T^{d}$ is always orthogonal to the winding number vector i.e. 
\begin{align}
    \sum_{I=1}^{d}k_{I}\omega^{I}=0.
\end{align}
Mass of these vacuum states are given by
\begin{align*}
    m_{vac,d}^2=k^{T}G^{-1}k-2k^{T}G^{-1}\mathbb{B}\omega-\omega^{T}\mathbb{B}G^{-1}\mathbb{B}\omega-\frac{2}{c'}.
\end{align*}
For the special case of $d=2$, the mass-squared of the vacuum will be given by
 \begin{equation}
 \begin{split}
      m_{vac,2}^2&=\frac{k_{1}^2}{R_{1}^2}+\frac{k_{2}^2}{R_{2}^2}+2b\Bigg(\frac{R_{1}}{R_{2}}k_2\omega_1-\frac{R_{2}}{R_{1}}k_1\omega_2\Bigg)+b^2(R_1^2\omega_1^2+R_2^2\omega_2^2)-\frac{2}{c'}\\
      &= \left(\frac{k_1}{R_{1}}-bR_{2}\omega_2\right)^2+\left(\frac{k_2}{R_{2}}+bR_{1}\omega_1\right)^2-\frac{2}{c'}.
 \end{split}
 \end{equation}
Here we will see that the presence of the $B$ field severely constraints the range of parametric values of \{$k_{I},\omega^{I},R_{I}$\} for which mass of the vacuum becomes tachyonic. Let's do a ballpark calculation: For sake of simplicity we take $R_{1}=R_{2}=R$. The mass of the state $\ket{0,0,k^{\mu},k_{I},\omega^{I}}$ in that case becomes
\begin{align}\label{twomass}
    m_{vac,2}^2=\frac{1}{R^2}(k_{1}^2+k_{2}^2)+2b(k_{2}\omega_{1}-k_{1}\omega_{2})+b^{2}R^{2}(\omega_{1}^2+\omega_{2}^2)-\frac{2}{c'}.
\end{align}
The constraint satisfied by \{$k_{I},\omega^{I}$\} for this case, as per level matching, is 
\begin{equation}
    k_{1}\omega_{1}+k_{2}\omega_{2}=0.
\end{equation}
The following Table \eqref{Tab1} provides us with the details of $m^2$ of vacuum for given values of $k_1, k_2, \omega_1 ~\text{and}~ \omega_2$ so that the above constraint is satisfied. For the sake of the reader, a heuristic plot for the same is provided in fig \eqref{massplot}.
\medskip

 We also can get tachyons for the states \{$r=1,s=0$\} and \{$r=0,s=1$\}, and for detailed analysis the reader is referred to Appendix \eqref{states}. However, we observe that there would be no tachyons for the higher level states \{$r\geq1,s\geq1$\}.
The expression for mass formula for the state \{$r=1,s=1$\} can be expressed as: 
\begin{equation}
    m^2=\left(\frac{k_1}{R}+bR\omega_2\right)^2+\left(\frac{k_2}{R}+bR\omega_1\right)^2
\end{equation}
which is always $\geq 0$. This indicates the absence of tachyons at and above level two spectrum. However massless condition can be achieved in this case, with
$k_1=-b\omega_2 R^2$ and $k_2=b\omega_1R^2$. If we switch off the $B$ field from the expression of mass of the ground state, then as expected, the mass will be independent of the winding number given by
\begin{align}
    m_{B=0}^2=\frac{1}{R^2}\sum_{I,J=1}^{d}k_{I}g^{IJ}k_{J}-\frac{2}{c'}.
\end{align}
This perfectly matches with the result derived in \cite{Banerjee:2023ekd}.

\begin{center}
\small
\begin{tabular}{ |c|c|c|c| }
\hline
$k_{1},\omega_{1}$ & $k_{2},\omega_{2}$ & $m^2$ & Condition for $m^2<0$ \\
\hline
\multirow{3}{6em}{$k_{1}=\omega_{1}=0$} & $k_{2}=\omega_{2}=0$ & $m^2=-\frac{2}{c'}$ & Always  \\ 

& $k_{2}=0,\omega_{2}\neq 0$ & $m^2=b^2R^2\omega^2_{2}-\frac{2}{c'}$ &  $\omega_{2}<\frac{1}{bR}\sqrt{\frac{2}{c'}}$\\

& $k_{2}\neq0,\omega_{2}= 0$ & $m^2=\frac{k_{2}^2}{R^2}-\frac{2}{c'}$ &  $k_{2}<R\sqrt{\frac{2}{c'}}$\\ 

\hline
\multirow{3}{7em}{$k_{1}=0,~\omega_{1}\neq0$} & $k_{2}=\omega_{2}=0$ & $m^2=b^2R^2\omega_1^2-\frac{2}{c'}$ &  $\omega_{1}<\frac{1}{bR}\sqrt{\frac{2}{c'}}$\\ 

& $k_{2}=0,\omega_{2}\neq 0$ & $m^2=b^2R^2(\omega_1^2+\omega^2_{2})-\frac{2}{c'}$ & $\omega_1^2+\omega^2_{2}<\frac{2}{c'b^2R^2}$\\

& $k_{2}\neq0,\omega_{2}= 0$ & $m^2=\left(\frac{k_{2}}{R}-bR\omega_1\right)^2-\frac{2}{c'}$ & $\left(\frac{k_{2}}{R}-bR\omega_1\right)^2<\frac{2}{c'}$\\ 
\hline
\multirow{3}{7em}{$k_{1}\neq0,~\omega_{1}=0$} & $k_{2}=\omega_{2}=0$ & $m^2=\frac{k_1^2}{R^2}-\frac{2}{c'}$ & $k_{1}<R\sqrt{\frac{2}{c'}}$\\ 

& $k_{2}=0,\omega_{2}\neq 0$ & $m^2=\left(\frac{k_{1}}{R}+bR\omega_2\right)^2-\frac{2}{c'}$ & $\left(\frac{k_{2}}{R}+bR\omega_1\right)^2<\frac{2}{c'}$\\

& $k_{2}\neq0,\omega_{2}= 0$ & $m^2=\frac{k_1^2+k_2^2}{R^2}-\frac{2}{c'}$ & $k_1^2+k_2^2<\frac{2R^2}{c'}$\\ 
\hline
\multirow{3}{7em}{$k_{1}\neq0,~\omega_{1}\neq0$} & &  & $k_{2}=\omega_{1}bR^2$ \\ & $k_{2}\neq0,~\omega_{2}\neq0$ & $m^2=(\omega_1^2+\omega^2_{2})\left(\frac{k_{2}}{R\omega_{1}}-bR\right)^2$ & or \\
& & $-\frac{2}{c'}$ & $(\omega_1^2+\omega^2_{2})\left(\frac{k_{2}}{R\omega_{1}}-bR\right)^2<\frac{2}{c'}$ \\
\hline
\end{tabular}
\captionof{table}{Mass square eigenvalues for vacuum state $r=s=0$ for given values of $k_I,\omega^I$. Note that the notation has been simplified here. Conditions for getting tachyonic states are shown separately. \label{Tab1}}
\end{center}

\subsubsection*{Vector states}

Let us consider the following vector states built on a vacuum $\ket{0,0,k^{\mu},k_{I},\omega^{I}}$.
\begin{align}
 \ket{V^{\mu\hspace{.5mm}I}_{k_{I},\omega^{I}}}=C^{\mu}_{-1}\Tilde{C}^{I}_{-1} \ket{0,0,k^{\mu},k_{I},\omega^{I}}\hspace{5mm}\ket{\Tilde{V}^{\mu\hspace{.5mm}I}_{k_{I},\omega^{I}}}=\Tilde{C}^{\mu}_{-1}C^{I}_{-1} \ket{0,0,k^{\mu},k_{I},\omega^{I}}
\end{align}
As we have seen, this state will be physical if $k_{I}\omega^{I}=0$. Looking at \eqref{ctb20} one can see that for $k_{I}=\omega^{I}=0$, this state will be massless. \eqref{ctb20} also makes it clear that if we switch off the $B$ field, then for $k_{I}=0$, these states will be massless for any  \{$\omega^{I}$\}. 

\subsubsection*{Scalar states}

We can also construct $d^2$ number of scalar states on a ground state $\ket{0,0,k^{\mu},k_{I},\omega^{I}}$ as below
\begin{align}
    \ket{\Phi^{IJ}}=C^{I}_{-1}\Tilde{C}^{J}_{-1} \ket{0,0,k^{\mu},k_{I},\omega^{I}}
\end{align}
For $B=0$, the states with zero internal momenta $k_{I}$ will be massless for all \{$\omega^{I}$\}.

\subsection{Take home summary}
Let us summarize the findings of this section for the sake of the reader, since the lessons will be important going forward.
\begin{itemize}
    \item Presence of constant $B$ field does not affect the level matching condition just like tensile theory (see \eqref{B8}).
    
    \item The mass spectrum \eqref{tsc11} however gets contribution from $B$ field for compact target space. Without $B$ field, the mass spectrum is independent of winding number $\omega^I$. Physically, this meant that due to the large string length (much greater than compactification radii)\footnote{It is a known result in string theory that when compactification radius becomes much shorter than the string length, the mass spectrum becomes purely dependent on the momentum modes only. In the tensionless sector, similar situation arises by keeping the compactification radii intact and making the string infinitely long.}, one could wind the string along compact coordinates without requiring extra energy. However, the existence of a topological $B$ field term along the compactified directions results into creating an equilibrium, where one needs a finite amount of energy proportional to the $B$ field. \footnote{It is important to recall here that for string, the role of $B$ field is analogous to the role played by vector potential $A_\mu$ for point particles in electromagnetism.}.
    
    \item The generalised metric $\mathcal{G}$, on which the mass spectrum is based, turns out to be degenerate. This points towards a degenerate version of Double Field Theory when one considers the similar limit on the inverse metric. 
    \item The new mass spectrum \eqref{tsc11} satisfies a partially broken $O(d,d;\mathbb{Z})$ symmetry, where only even number of sectorised T-dualities acting along the compact legs of the $B$ field preserves the structure of a null $\mathcal{G}$. 
    \item Looking at the effects of inversion transformation and sectorised dualities on $T^2$ and $T^3$ explicitly, we can say that only even number of operations of sectorised duality can lead us to mass spectrum of purely tensionless strings constructed from different $\mathbb{B}'$ and $G'$.
    \item  From the above observation we can say that when we compactify even number of dimensions where the full inversion transformation consists of even number of sectorised duality operations, we can get an invertible $B$ field to make it work.
    \item However, for odd dimension compactification, only a particular choice (or a collection) of $B$ field(s), which acts like an invertible $B$ field in an even dimensional subspace, operating sectorised duality in all those even subspaces will invert the $B$ field. 
    \item As shown in our earlier work \cite{Banerjee:2023ekd}, we obtain infinite number of new physical states for each level for multi-dimension compactification. The presence of $B$ field in the theory constraints the range of parametric values for which mass of the vacuum becomes tachyonic. 
\end{itemize}
\newpage 

\section{Induced vacuum} \label{sec5}
This section deals with the study of effect of constant $B$ field on the mass spectrum for Induced vacuum of the tensionless string theory. This vacuum is intriguing, since intrinsically derived quantum mechanical results can be arrived at by taking ultrarelativistic limits on the highest weight states of tensile string theory. In what follows, we will see whether the same works with a constant $B$ field on both regimes of the theory. Our basic setup remains the same as in the last section, only the action of the oscillators will change. 

\subsection{Intrinsic mass spectrum and level matching}

  We begin by reminding the reader the mode expansion of the dimensionless field $Y^{I}$ \eqref{chh49} in terms of oscillators $A$'s and $B$'s as in \eqref{modeexp}
\begin{align*}
    Y^{I}=y^{I}+A^{I}_{0}\sigma+B^{I}_{0}\tau+i\sum_{n\neq 0}\frac{1}{n}\big(A^{I}_{n}-in\tau B^{I}_{n}\big)e^{-in\sigma}.
\end{align*}
Here $A^{I}_{0}$ and $B^{I}_{0}$ are given in \eqref{B2}. It is better to work in this language for the Induced vacuum, as the tensile string states will directly reduce to states constructed on this vacuum in the tensionless limit.  We also use here the equations defined in \eqref{ctb24} and \eqref{ctb12}. Using \eqref{B1} one can see the action of all $M_n,~n\neq0$ annihilates the vacuum, which is a physical state. This in turn signfies that all $B_n,~n\neq 0$ annihilates the physical vacuum \cite{Bagchi:2020fpr}, which becomes the defining relation.
\medskip

To focus on the zero modes, we apply the physical state condition $M_{0}$ with $a_M=0$ (as $B$'s here commute among each other, there is no ordering ambiguity) on the vacuum $\ket{0,0,k^{\mu},k_{I},\omega^{I}}$ to get,
\begin{align}
      &\hspace{1.3cm}\sum_{m}B_{-m}\cdot B_{m}\ket{0,0,k^{\mu},k_{I},\omega^{I}}_{I}=0 \nonumber\\ 
    &\implies \Big( \sum_{m\neq 0}B_{-m}\cdot B_{m}+B^{2}_{0}\Big)\ket{0,0,k^{\mu},k_{I},\omega^{I}}_{I}=0\nonumber\\ 
    \implies &B^{2}_{0}\ket{0,0,k^{\mu},k_{I},\omega^{I}}_{I}=2c'\Bigg(k^2+\sum_{I=1}^{d}K^{I}K_{I}\Bigg)\ket{0,0,k^{\mu},k_{I},\omega^{I}}_{I}=0.
\end{align}
Hence, using \eqref{ctb24} the mass squared eigenvalue of the state $\ket{0,0,k^{\mu},k_{I},\omega^{I}}_{I}$ is given by
\begin{align}
   m^2=\sum_{I=1}^{d}K_{I}G^{IJ}K_{J}=\sum_{I,J,K=1}^{d}\Big(k_{I}-\mathbb{B}_{IK}\omega^{K}\Big)G^{IJ}\Big(k_{J}-\mathbb{B}_{JL}\omega^{L}\Big)
   \end{align}
Using the matrix notation introduced earlier in \eqref{tsc11} and \eqref{bogol} we rewrite the expression of this mass spectrum as
   \begin{align}\label{ctb25}
m^2=k^{T}G^{-1}k-2k^{T}G^{-1}\mathbb{B}\omega-\omega^{T}\mathbb{B}G^{-1}\mathbb{B}\omega.
    \end{align}
We immediately notice two important structures in this mass spectrum. Firstly, the $(\omega, k)$ lattice structure in case of induced vacuum theory is identical to that of the oscillator vacuum theory. Secondly, we recall that in case of tensionless string without $B$ field \cite{Banerjee:2023ekd}, the mass spectrum did not have any contribution from the winding number. In fact, in that case the mass spectrum is just rest energy contributed by the internal momenta. However, for the constant $B$ field case, we do have finite contribution in the mass spectrum from the winding sector. Switching off the $B$ field in \eqref{ctb25} would lead us to the following mass spectrum
\begin{align}
     m^2=\sum_{I,J=1}^{d}k_{I}G^{IJ}k_{J}=\sum_{I}\frac{k^{2}_{I}}{R^{2}_{I}}
\end{align}
If we set all the compactification radii to be equal then this  mass spectrum will reduce to the one for induced vacuum derived in our earlier work \cite{Banerjee:2023ekd}. This purely momentum dominated spectrum can also be reached via a direct limit on the tensile mass spectra. 
\medskip 

On the other hand, the physical state constraint coming from $L_{n}$ will be given by
\begin{equation}
\begin{split}
    &\leftindex_I{\bra{0,0,k^{\mu},k_{I},\omega^{I}}}L_{n} \ket{0,0,k^{\mu},k_{I},\omega^{I}}_{I}=0\hspace{5mm}\forall n\\
\implies&\leftindex_I{\bra{0,0,k^{\mu},k_{I},\omega^{I}}}\Big(A^{\mu}_{n}B_{\mu\hspace{1mm}0}+\sum_{I}A^{I}_{n}B_{I\hspace{1mm}0}\Big)\ket{0,0,k^{\mu},k_{I},\omega^{I}}_{I}=0\hspace{5mm}\forall n.
\end{split}
\end{equation}
For $n=0$, we recall that the normal ordering ambiguity for $L_{0}$ i.e. $a_{L}$ vanishes as shown in \cite{Bagchi:2021rfw,Chen:2023esw}, and that leads us to the following:
\begin{equation}
\begin{split}
     &\leftindex_I{\bra{0,0,k^{\mu},k_{I},\omega^{I}}}\Big(A^{\mu}_{0}B_{\mu\hspace{1mm}0}+\sum_{I}A^{I}_{0}B_{I\hspace{1mm}0}\Big)\ket{0,0,k^{\mu},k_{I},\omega^{I}}_{I}=0\\
    \implies & \sum_{I}k_{I}\omega^{I}\leftindex_I{\braket{0,0,k^{\mu},k_{I},\omega^{I}|0, 0,k^{\mu},k_{I},\omega^{I}}}_{I}=0\\
      \implies & \sum_{I}k_{I}\omega^{I}=0.
\end{split}
\end{equation}
This can be thought of as the reminiscent of the level matching condition of the tensile strings with multiple dimensions compactified.
\subsection{Limit from tensile string}
\subsubsection{Tensionless limit of tensile perturbative states}
As we know from \cite{Bagchi:2015nca} and \cite{Bagchi:2020fpr}, the quantised tensionless string theory built on induced vacuum naturally emerges from the tensionless limit of the tensile string theory. In \cite{Bagchi:2019cay}, it has been shown that when we take direct tensionless limit on the tensile perturbative physical states, at $\epsilon\to 0$ they condense explicitly on the induced vacuum. We briefly review that before discussing the fate of tensile states for our case with the $B$ field.

\medskip
Let us recall that for non-compactified tensile string the perturbative states are level matched. Hence we consider the following excited state created via the action of creation-annihilation operators:
\begin{align}
    \ket{\zeta_{n}}=\rho_{\mu\nu}\alpha^{\mu}_{-n}\Tilde{\alpha}^{\mu}_{-n}\ket{0}_{\alpha},
\end{align}
where $\rho_{\mu\nu}$ is a polarisation tensor. Around tensionless limit i.e. $\epsilon=0$, the tensile string vacuum can be perturbatively expanded in terms of $\epsilon$ in following way
:
\begin{align}\label{chh17}
\ket{0}_{\alpha}=\ket{0}_{I}+\epsilon\ket{I_{1}}+\epsilon^2\ket{I_{2}}+\cdots
\end{align}
We recall the definition of tensile string vacua which is 
\begin{align}\label{chh13} 
    \alpha_{n}\ket{0}_{\alpha}=\Tilde{\alpha}_{n}\ket{0}_{\alpha}=0\hspace{5mm}\forall n>0.
\end{align}
We also recall how the oscillators $\alpha_{n}$ and $\Tilde{\alpha}_{n}$ are connected to tensionless oscillators $A_{n}$ and $B_{n}$\footnote{See Appendix \eqref{apB} for details of the limiting computation of modes.}
\begin{align}\label{chh188}
     \alpha_{n}=\frac{1}{2}\Big[\sqrt{\epsilon} A_{n}+\frac{1}{\sqrt{\epsilon}}B_{n}\Big],\hspace{5mm}\Tilde{\alpha}_{n}=\frac{1}{2}\Big[-\sqrt{\epsilon} A_{-n}+\frac{1}{\sqrt{\epsilon}}B_{-n}\Big].
\end{align}
Applying \eqref{chh188} and \eqref{chh17} to the definition \eqref{chh13} one gets the following
\begin{align}\label{chh16}
   \ket{\zeta_{n}}=\frac{1}{\epsilon}\rho_{\mu\nu}\Big(B^{\mu}_{-n}+\epsilon A^{\mu}_{-n}\Big)\Big(B^{\nu}_{n}-\epsilon A^{\nu}_{n}\Big)\Big(\ket{0}_{I}+\epsilon\ket{I_{1}}+\epsilon^2\ket{I_{2}}+\cdots\Big).
\end{align}
Using the algebra satisfied by $A$ and $B$ one can see that the new perturbative state in the limit:
\begin{align}
     \ket{\zeta_{n}}=\Xi\ket{0}_{I},\hspace{5mm} \Xi=2n\eta^{\mu\nu}\rho_{\mu\nu}. 
\end{align}
In this section we will study the fate of the perturbative physical states of the tensile string theory with background Kalb-Ramond field and see whether they reduce to the physical states of tensionless string theory built on induced vacuum.
We recall that for tensile string theory in a target spacetime with $d$ compactified dimensions, the level matching condition \footnote{It is important to recall here that in tensile theory too, the constant $B$ field does not have any effect on the level matching condition.} is modified to:
\begin{align}
\widetilde N-N=\sum_{I}k_{I}\omega^{I}=l.  
\end{align}
\begin{figure}[ht]
    \includegraphics[scale=0.8]{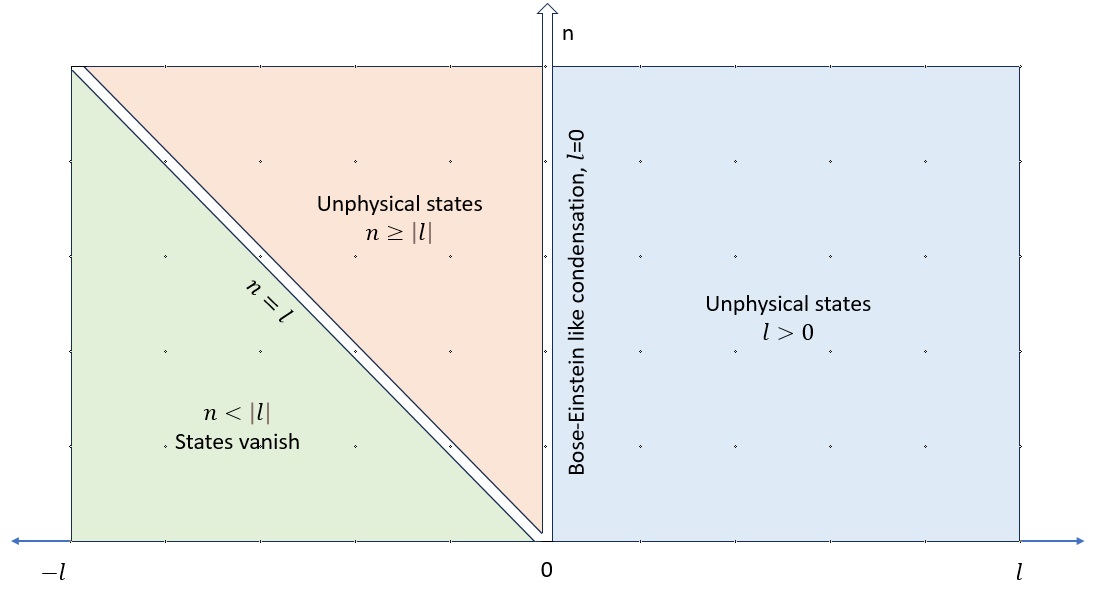}
    \caption{Fate of physical states in the tensionless limit in the Induced case.}
    \label{fig1}
\end{figure}
Let us consider a tensile physical perturbative string state satisfying the above level matching condition 
\begin{align}
    \ket{\zeta_{n}}=\rho_{\mu\nu}\alpha^{\mu}_{-n}\Tilde{\alpha}^{\mu}_{-n-l}\ket{0,0,k^{\mu},k_{I},\omega^{I}}_{\alpha}.
\end{align}
In our previous work \cite{Banerjee:2023ekd} we considered a  target spacetime with only one dimension compactified on a circle. There we considered the tensile state \begin{align*}
    \ket{\xi_{n}}=\rho_{\mu\nu}\alpha^{\mu}_{-n}\Tilde{\alpha}^{\mu}_{-n-KW}\ket{0,0,k^{\mu},K,W}_{\alpha},
\end{align*} and we examined its fate under the explicit tensionless limit using the limiting procedure outlined above. To remind the reader, our findings were as given below
\begin{itemize}
    \item For $KW=0$, $\ket{\xi_{n}}$ condense at the induced vacuum $\ket{0,0,k^{\mu},K,W}_{I}$. Note that this condensation is nothing but the Bose-Einstein like condensation described in \cite{Bagchi:2019cay}.
    \item For $KW>0$, $\ket{\xi_{n}}$ go to unphysical state.
    \item For $KW<0$ and $n\geq|KW|$, $\ket{\xi_{n}}$ goes to unphysical state again.
\item For $KW<0$ and $n<|KW|$, $\ket{\xi_{n}}$ altogether vanishes at tensionless limit.
\end{itemize}
The generalisation to a target spacetime with $d$ dimensions compactified, as in the present case, is straightforward\footnote{The steps of the calculation are identical to the same of one-dimension compactified and hence, not reproduced here. One can see \cite{Banerjee:2023ekd} for details of the calculation.}.
\begin{itemize}
    \item For $l=0$, $\ket{\zeta_{n}}$ condenses at $\ket{0,0,k^{\mu},k_{I},\omega^{I}}_{I}$.
    \item For $l>0$, $\ket{\zeta_{n}}$ tends to unphysical state.
    \item For $l<0$, $n\geq |l|$ too, $\ket{\zeta_{n}}$ tends to unphysical state.
    \item For $l<0$ and $n<|l|$ , $\ket{\zeta_{n}}$ vanishes at tensionless limit.
\end{itemize}
Hence, if a perturbative tensile physical state still remains a physical state in the tensionless limit, it will always condense on the induced vacuum, or else become unphysical or even vanish. A visualisation of the fate of the states is shown in figure \eqref{fig1}.
\medskip

For completeness, let us also comment on the fate of non-perturbative string states. As discussed in \cite{Bagchi:2020fpr}, the physical state conditions for the induced vacuum tensionless string theory allows the following non-perturbative states in the spectrum of physical states. 
\begin{equation}
    \ket{\phi}=\text{exp}\Bigg(i\sum_{n,m}\omega_{n}A_{n-m}\cdot B_{m}\Bigg)\ket{0,0,k^{\mu},k_{I},\omega^{I}}_{I}
\end{equation}
In the light of the fact that both $A_{0}$ and $B_{0}$ commute with all $A_{n}$'s and $B_{n}$'s one can see that $\ket{\phi}$ must have the same momentum and winding number as the vacuum state it has been constructed on
\begin{equation}
   B^{\mu}_{0}\ket{\phi}=k^{\mu}\ket{\phi}\hspace{5mm}B_{I\hspace{1mm}0}\ket{\phi}=k_{I}\ket{\phi}\hspace{5mm}A^{I}_{0}\ket{\phi}=\sqrt{\frac{2}{c'}}\omega^{I}\ket{\phi}.
\end{equation}
Consequently mass of this state also has to be same as the vacuum state it has been constructed upon, i.e.
\begin{align}
   m^2\ket{\phi}=(k^{T}G^{-1}k-2k^{T}G^{-1}\mathbb{B}\omega-\omega^{T}\mathbb{B}G^{-1}\mathbb{B}\omega)\ket{\phi}.
\end{align}
\subsubsection{Limit on mass spectra and the generalised metric}
Now we will discuss what happens to the mass spectrum in the explicit tensionless limit. We already learnt that a subset of the tensile physical perturbative states condense to the induced vacuum in the tensionless limit and the rest simply become unphysical in the new theory. Hence, if we want the limit from tensile string theory to be consistent with the theory intrinsically constructed on the induced vacuum, it is important to show that the mass of induced vacuum in \eqref{ctb25} is identical to the tensionless limit of the mass formula for particular tensile states which remain physical (viz. $l=0$) in the tensionless limit. 

\medskip
Let us consider the Polyakov action of the tensile string with a constant $B$ field \eqref{Polyakov} in a compactified spacetime with $d$ number of compact dimensions, where $B_{\mu\nu}\neq0$ only along the compact dimensions. We recall that the mass spectrum for the tensile physical perturbative states is given by (see for example, \cite{Becker:2006dvp,Blumenhagen:2013fgp})
\begin{align}\label{ctb28}
    m^2=k^{T}G^{-1}k+\frac{1}{\alpha'^2}\omega^{T}(G-bG^{-1}b)\omega-\frac{2}{\alpha'}k^{T}bG^{-1}\omega+\frac{1}{\alpha'}(N_{L}+N_{R}-2).
\end{align}
In the above equation, the metric along compact directions $G_{IJ}=R_{I}^{2}\delta_{IJ}$ and the antisymmetric tensor $b_{IJ}$ is related to the tensor $B_{IJ}$ in exactly the same way as in \eqref{BBS1}, i.e.
\begin{equation}\label{ramgorur}
    b_{IJ}=\frac{\partial X^{K}}{\partial \widetilde{X}^{I}}\frac{\partial X^{L}}{\partial \widetilde{X}^{J}}B_{KL}=R_{I}R_{J}\sum_{K=1}^{d}\sum_{L=1}^{d}\delta^{K}_{I}\delta^{L}_{J}B_{KL}=R_{I}R_{J}B_{IJ}.
\end{equation}
We also have the level matching condition in terms of tensile number operators $N_{R}-N_{L}=\sum_{I=1}^{d}k_{I}\omega^{I}$.
From \eqref{ctb27} and \eqref{BBS1}, one can see that
\begin{align}
    \frac{1}{\alpha'}b_{IJ}=\mathbb{B}_{IJ}.
\end{align}
Writing the part of mass spectrum involving momenta and winding, $\{k_{I},\omega^{I}\}$ of \eqref{ctb28} in the matrix notation we used earlier, we get
\begin{align}\label{tensile}
 \Big[\omega^{T} \hspace{4mm} k^{T}\Big]\begin{bmatrix}
    \cfrac{1}{\alpha'^2}\left(G-b G^{-1}b\right) & ~~\cfrac{1}{\alpha'}~bG^{-1}\\-\cfrac{1}{\alpha'}~G^{-1}b & G^{-1}
\end{bmatrix}
\hspace{2mm}
\begin{bmatrix}
    \omega \\ \\ k
\end{bmatrix},
\end{align}
 Taking tensionless limit corresponds to taking $\alpha'\to\frac{c'}{\epsilon}$ with $\epsilon\to 0$, as explained earlier. We recall the limit taken on $B$ field in tensionless limit as given in \eqref{ctb26} and the definition of $\mathcal{B}_{\mu\nu}$ in \eqref{ctb27}. Together, these limits on $\alpha'$ and $ b_{IJ}$ keeps the term $\mathbb{B}_{IJ}$ intact as we take $\epsilon\to 0$. As a result, the terms in the mass spectrum \eqref{ctb28} involving $\frac{b}{\alpha'}$ survive the tensionless limit. However, the terms only involving $\frac{1}{\alpha'}$ becomes suppressed due to the limit $\alpha'\to\frac{\alpha'}{\epsilon}$. Consequently, at the tensionless limit, \eqref{ctb28} will reduce to \eqref{ctb25}. Therefore, the limit from tensile mass spectrum (i.e., limiting approach) turns out to be consistent with the intrinsically derived mass spectrum.
\medskip

Let us now focus on the tensile mass spectrum as given in \eqref{tensile}. The generalised momentum here belongs to a $2d$ dimensional spacetime where the $2d\times 2d$ dimensional matrix in \eqref{tensile} acts as inverse metric. The metric in the spacetime becomes inverse of the matrix, given by
\begin{align}\label{DFTmetric}
 \begin{bmatrix}
  G^{-1}   & ~~-\cfrac{1}{\alpha'}~G^{-1}b \\ \cfrac{1}{\alpha'}~bG^{-1} & \cfrac{1}{\alpha'^2}\left(G-b G^{-1}b\right)
\end{bmatrix},
\end{align}
In this space a manifestly $O(d,d)$ symmetric  field theory can be defined, often called  as double field theory (DFT) (see \cite{Aldazabal:2013sca}). Taking a consistent tensionless limit ($\alpha'\to\infty$) on the inverse generalised metric in \eqref{tensile} we get:
\begin{align}
     \mathcal{G}=\begin{bmatrix}
    -\mathbb{B}G^{-1}\mathbb{B} & \mathbb{B}G^{-1}\\-G^{-1}\mathbb{B} & G^{-1}
    \end{bmatrix}
\end{align}
It turns out that the new metric is indeed the degenerate one we presented in \eqref{Buscher1}, and a similar limit on \eqref{DFTmetric} will give us the degenerate DFT metric. The degeneracy at tensionless limit becomes even more manifest when we work with the special case of $b=0$. For that case, at $\alpha'\to\infty$, we can easily see how the entire winding sector gets suppressed, a degeneracy of order $d$ appears, and the mass spectrum depends only on the compactified momenta.
\medskip

A well known DFT result is that when the compactification scale becomes much smaller than the string length scale, the DFT only depends on dual/momentum coordinates \cite{Aldazabal:2013sca}. Since the tensionless limit happens to be the ultra-relativistic limit on the worldsheet level, the string here becomes extremely long, even much larger than the compactification radius. Hence, taking tensionless limit can be considered to be an alternative way to achieve the ``compactification scale $\ll$ string length scale'' scenario for Double Field Theory.

 \subsubsection{Narain Lattice at the limit}
 The mass spectrum obtained in \eqref{ctb28} comes from the following left and right moving momentum for the tensile string
 \begin{align}\label{abol}
    (P_I)_{\hspace{.25mm}L,R}=\frac{1}{\sqrt{2}}\bigg\{\sqrt{\alpha'}\Big(k_{I}-\frac{1}{\alpha'}\sum_{J=1}^{d} b_{IJ}\omega^{J}\Big)\pm \frac{1}{\sqrt{\alpha'}}\sum_{J=1}^{d}G_{IJ}\omega^{J}\bigg\}.
\end{align}
As is well known, the vector $P=\big((P_I)_{\hspace{.25mm}L},(P_I)_{\hspace{.25mm}R}\big)$ belongs to a $2d$-dimensional even self-dual lattice of Lorentzian signature $\left((+1)^d,(-1)^d)\right)$, also called Narain lattice \cite{Becker:2006dvp}. The Lorentzian length of the vector $P$ on this lattice space is given by
\begin{equation}
P^2=(P_I)_{\hspace{.25mm}L}G^{IJ}(P_I)_{\hspace{.25mm}L}-(P_I)_{\hspace{.25mm}R}G^{IJ}(P_I)_{\hspace{.25mm}R}
\end{equation}
In matrix notation we can write this as
\begin{align}
    P^2=\Big[(P_I)_{\hspace{.25mm}L}\hspace{2mm}(P_I)_{\hspace{.25mm}R}\Big]\begin{bmatrix}
        G^{IJ} & 0 \\ 0 & -G^{IJ}
    \end{bmatrix}\begin{bmatrix}
     (P_I)_{\hspace{.25mm}L} \\ (P_I)_{\hspace{.25mm}R}
\end{bmatrix}.
\end{align}
We can also write this in following way by rotating the momentum basis of the lattice
\begin{align}
    P^2=\frac{1}{2}\Big[\big((P_I)_{\hspace{.25mm}L}+(P_I)_{\hspace{.25mm}R}\big)\hspace{3mm}\big((P_I)_{\hspace{.25mm}L}-(P_I)_{\hspace{.25mm}R}\big)\Big]\begin{bmatrix}
       0  & G^{IJ} \\ G^{IJ} & 0
    \end{bmatrix}\begin{bmatrix}
     \big((P_I)_{\hspace{.25mm}L}+(P_I)_{\hspace{.25mm}R}\big) \vspace{0.1cm}\\ \big((P_I)_{\hspace{.25mm}L}-(P_I)_{\hspace{.25mm}R}\big)
\end{bmatrix}.
\end{align}
Taking tensionless limit ($\alpha'\to c'/\epsilon$) on $ (P_I)_{\hspace{.25mm}L,R}$ as written in \eqref{ctb15}, one can clearly see that the basis vectors are inhomogeneously scaled up and down by $\epsilon$,
\begin{align}
    \big((P_I)_{\hspace{.25mm}L}+(P_I)_{\hspace{.25mm}R}\big)=\frac{1}{\sqrt{\epsilon}}\big((k_I)_{\hspace{.25mm}L}+(k_I)_{\hspace{.25mm}R}\big),\hspace{5mm}\big((P_I)_{\hspace{.25mm}L}-(P_I)_{\hspace{.25mm}R}\big)=\sqrt{\epsilon}\big((k_I)_{\hspace{.25mm}L}-(k_I)_{\hspace{.25mm}R}\big).
\end{align}
Evidently, at tensionless limit $\epsilon\to 0$, one can see that 
\begin{align}
   P^2=\frac{1}{2}\left[\big((k_I)_{\hspace{.25mm}L}+(k_I)_{\hspace{.25mm}R}\big)\hspace{3mm}\big((k_I)_{\hspace{.25mm}L}-(k_I)_{\hspace{.25mm}R}\big)\right]\begin{bmatrix}
       0  & \frac{\sqrt{\epsilon}}{\sqrt{\epsilon}}G^{IJ} \\ \frac{\sqrt{\epsilon}}{\sqrt{\epsilon}}G^{IJ} & 0
    \end{bmatrix}\begin{bmatrix}
     \big((k_I)_{\hspace{.25mm}L}+(k_I)_{\hspace{.25mm}R}\big) \vspace{0.1cm}\\ \big((k_I)_{\hspace{.25mm}L}-(k_I)_{\hspace{.25mm}R}\big)
\end{bmatrix}.
\end{align}
This implies that at $\epsilon\to0$, although the left-right basis vectors are scaled, the volume of the Narain lattice remains effectively intact, keeping the length of the vector unchanged. This also makes sure that we have the same level matching conditions as in tensile string theory.

\subsection{Concrete example: Two dimensional compact space}\label{bhishmalochan}
Now, we look into the mass spectrum of tensile string with constant $B$ field in a spacetime with only two of the dimensions compactified. We choose $\mathbb{B}$ in the following way
\begin{align}    \mathbb{B}_{IJ}=b\hspace{1mm}\epsilon_{IJ},\hspace{5mm}I,J\in \{1,2\}
\end{align}
where $\epsilon_{IJ}$ is the antisymmetric tensor in two dimensions. Choosing $\epsilon_{12}=1$, we find that for $d=2$, equation \eqref{ctb25} gets simplified to 
\begin{align}\label{gobhut}
   m^2=\frac{k_{1}^2}{R_{1}^2}+\frac{k_{2}^2}{R_{2}^2}+2b\Bigg(\frac{R_{1}}{R_{2}}k_2\omega_1-\frac{R_{2}}{R_{1}}k_1\omega_2\Bigg)+b^2(R_1^2\omega_1^2+R_2^2\omega_2^2)
\end{align}
Just like in the case of oscillator vacuum, this mass spectrum is symmetric under the following set of transformations:
\begin{align}
    k_{I}\leftrightarrow\omega^{I},\hspace{5mm}R_I\leftrightarrow \frac{1}{bR_I},\hspace{5mm}b\to-b\hspace{5mm}I\in\{1,2\}.
\end{align}
Now, let us consider the tensile counterpart of this theory, i.e. a tensile string with a $B$ field compactified on a two-torus. In that case the mass spectrum is 
 \begin{align}\label{beltola}
    m^2=\frac{k_{1}^2}{R_{1}^2}+\frac{k_{2}^2}{R_{2}^2}+\frac{2b}{\alpha'}\Bigg(\frac{R_{1}}{R_{2}}k_2\omega_1-\frac{R_{2}}{R_{1}}k_1\omega_2\Bigg)+\frac{(1+b^2)}{\alpha'^2}(R_1^2\omega_1^2+R_2^2\omega_2^2)+\frac{1}{\alpha'}(N_{L}+N_{R}-2).
\end{align}
As we can clearly see, direct tensionless limit ($\alpha' \to \frac{\alpha'}{\epsilon}$, $b \to \frac{b}{\epsilon}$, $\epsilon \to 0$) taken on this mass spectrum will lead us to the intrinsically derived mass-spectrum in \eqref{gobhut}.

\medskip
Since \eqref{gobhut} is identical to the compact part of the mass spectrum we found in oscillator vacuum (see \eqref{pagol}), the symmetries satisfied (and broken) by this mass  spectrum too, is the same, which means all the discussions in the example of oscillator vacuum compactified on $T^2$ is applicable here as well. Let us recall the transformation of generalised metric under sectorised duality transformation $g_{D_{2}}$ as shown in \eqref{opodartho6}, where we saw that the transformed generalised metric is no longer corresponding to the tensionless sector.
In order to make more sense of this transformation let us look back at the similar scenario in the tensile case where the same sectorised duality transformation is applied on the generalised metric constructed from Polyakov action with $G_{IJ}=R^2_{I}\delta_{IJ}$ and $B$ field $B_{IJ}=B\epsilon_{IJ}.$ It can be shown that the inverse generalised metric for tensile string after applying $g_{D_{2}}$ can be rewritten as below \footnote{For detailed calculations for tensile string calculation see Appendix \eqref{apA}. We use the subscript $\alpha'$ to distinguish the tensile case.}  
\begin{align}
     \mathcal{G}'^{-1}_{\alpha'}= g_{D_{2}}\mathcal{G}^{-1}_{\alpha'}g_{D_{2}}^T=   \begin{bmatrix}
        (1+b^2)R^{2}_{1} & \frac{bR_{1}}{R_{2}} & 0 & 0
       \\  \frac{bR_{1}}{R_{2}} & \frac{1}{R^{2}_{2}} & 0 & 0\\
       0 & 0 & \frac{1}{R^{2}_{1}} & -\frac{bR_{2}}{R_{1}}\\
       0 & 0 & -\frac{bR_{2}}{R_{1}} & (1+b^2)R^{2}_{2}.
    \end{bmatrix}
    \end{align}
It can be easily seen that the two diagonal blocks are inverse of each other. Hence if we consider the following metric on a torus $T^2$
\begin{align}\label{opodartho5}
            (G_{\alpha'})_{IJ}=\begin{bmatrix}
                (1+b^2)R^{2}_{1} & \frac{bR_{1}}{R_{2}}\\\frac{bR_{1}}{R_{2}} & \frac{1}{R^{2}_{2}}
            \end{bmatrix}
        \end{align}
Then the transformed inverse generalised metric can be rewritten as below
\begin{align}\label{opodartho6}
    \mathcal{G}'^{-1}_{\alpha'}= \begin{bmatrix}
        G & 0\\
        0 & G^{-1}
    \end{bmatrix}.
\end{align}
Looking at the expression of tensile generalised metric one can see that this new generalised metric correspond to a Polyakov action with $G_{IJ}$ given in \eqref{opodartho5} and $B_{IJ}=0$. Now, in comparison with this, let us rewrite the diagonal blocks in the $\mathcal{G}'$ from \eqref{opodartho7}:
\begin{align}
    (G_{c'})_{IJ}=\begin{bmatrix}
         b^2R^{2}_{1} & \frac{bR_{1}}{R_{2}}\\ 
         \frac{bR_{1}}{R_{2}} & \frac{1}{R^{2}_{2}}
    \end{bmatrix}\hspace{5mm}(G'_{c'})_{IJ}=\begin{bmatrix}
        \frac{1}{R^{2}_{1}} & -\frac{bR_{2}}{R_{1}}\\
        -\frac{bR_{2}}{R_{1}} & b^2R^{2}_{2}
    \end{bmatrix}.
\end{align}
One can see that $det G_{c'}= det G'_{c'}=0$, and also that  $G_{c'}G'_{c'}=0$. The implication is that $G_{c'}$ and $G'_{c'}$ can be considered to be the metric and inverse metric of a \textit{degenerate} torus. In fact one can see that the degenerate torus having metric $(G_{c'})_{IJ}$ comes from the following equivalent limit of the non-degenerate torus having metric $G_{\alpha'}$ in \eqref{opodartho5}\footnote{i.e. one cycle of the two torus is ``pinched''.}:
\begin{align}
    R_{1}\to\epsilon R_{1},\hspace{3mm}b\to\frac{b}{\epsilon}\hspace{3mm}R_{2}\to R_{2}.
\end{align}
Consequently, $\mathcal{G'}$ in \eqref{opodartho7} can be thought of to be constructed from a different limit of string  theory where the tensile string is compactified on a degenerate torus without any $B$ field. This seems to indicate that sectorised duality connects tensionless strings on a non-degenrate compact manifold to a different corner of string theory. We have found similar connections between tensionless limit and degenerate torus limit for compactification on $T^3$ as well. However, a general proof of such connection for compactification on $T^d$ is beyond the scope of the current work.

\medskip
We must mention here, this kind of connections between different corners of string theory is far from new. It is known for a long time that non-relativistic closed string theory is connected to DLCQ limit of string theory \cite{Gomis:2000bd}. Later many such connections between different limits of string theories has been explored. For detailed and updated picture of this duality web the reader is referred to \cite{Gomis:2023eav,Blair:2023noj}. Although the connections we have found in the current work probably do not correspond to any existing duality connection, one could conjecture that this might lead to an extension of the duality web.

 \newpage 

\section{Flipped vacuum}\label{sec6}
\subsection{Detour: Effect of constant $B$ field on twisted tensile string}
As we know from earlier works \cite{Bagchi:2020fpr}, the quantum tensionless theory constructed on the flipped vacuum can be arrived at by taking direct tensionless limit on tensile version of \textit{twisted} string theory. Hence, before studying the effect of a constant $B$ field on tensionless limit of the theory, we should first learn the effect of constant $B$ field on a tensile twisted string theory. This analysis itself is novel, and can be considered as generalization of \cite{Casali:2017mss, Lee:2017crr}, where $d$ ($>1$) number of dimensions are compactified and a constant $B$ field is present in the target spacetime. 

\subsubsection*{Classical theory and mode expansions}

Classically, this twisted string theory is identical to the usual string theory with $B$ field and hence the action is same as the Polyakov action with a $B$ field:
\begin{align}
    S=\frac{T}{2}\int d^{2}\xi \Big(\sqrt{-g}g^{\alpha\beta}\partial_{\alpha}X^{\mu}\partial_{\beta}X^{\nu}\eta_{\mu\nu}+\epsilon^{\alpha\beta}\partial_{\alpha}X^{\mu}\partial_{\beta}X^{\nu}B_{\mu\nu}\Big).
\end{align}
After conformal gauge fixing, i.e. putting $g^{\alpha\beta} = \eta^{\alpha\beta}$,
the equations of motion, constraints, the commutation relations and the symmetries of this action can be found to be the same as that for the action without $B$ field. The equation of motion along with the closed string boundary conditions (\eqref{CTBC1} for compact coordinates and $X^{\nu}(\sigma+2\pi,\tau)=X^{\nu}(\sigma,\tau)$ for non-compact coordinates) leads us to the following familiar mode expansions: 
\begin{equation*}
    \begin{split}
        &X^{\mu}=X^{\mu}_{L}+X^{\mu}_{R}\\
        X^{\mu}_{L}=x^{\mu}+\sqrt{\frac{\alpha'}{2}}&\alpha^{\mu}_{0}(\tau+\sigma)+i\sqrt{\frac{\alpha'}{2}}\sum_{n\neq 0}\frac{1}{n}\alpha^{\mu}_{n}e^{-in(\tau+\sigma)}\\
       X^{\mu}_{R}=x^{\mu}+\sqrt{\frac{\alpha'}{2}}&\Tilde{\alpha}^{\mu}_{0}(\tau-\sigma)+i\sqrt{\frac{\alpha'}{2}}\sum_{n\neq 0}\frac{1}{n}\Tilde{\alpha}^{\mu}_{n}e^{-in(\tau-\sigma)},\\ 
    \end{split}
\end{equation*}
where $\{\alpha,\Tilde{\alpha}\}$ satisfy the harmonic oscillator algebra. For compactified coordinates we use the same scaling we used before, namely \eqref{nakgol}. Consequently we have the new metric as in \eqref{BBS1}, along with new $B$ field given by
\eqref{ramgorur}. We also have new compactified boundary conditions exactly like in \eqref{scaled}. In terms of the scaled coordinates we now rewrite the compactified part of the gauge fixed action
\begin{align}
    S_{C}=\frac{T}{2}\int d^{2}\xi \Big(\eta^{\alpha\beta}\partial_{\alpha}\widetilde{X}^{I}\partial_{\beta}\widetilde{X}^{J}G_{IJ}+\epsilon^{\alpha\beta}\partial_{\alpha}\widetilde{X}^{I}\partial_{\beta}\widetilde{X}^{J}b_{IJ}\Big).
\end{align}
As we all know, the canonical momentum gets modified due to the presence of $B$ field as $\Pi_{I}=\frac{1}{2\pi\alpha'}\Big(\dot{\widetilde{X}^{J}}G_{IJ}+b_{IJ}\widetilde{X}'^{J}\Big)$,
and the internal momentum of the string becomes 
\begin{align}
    \pi_{I}=\int_{0}^{2\pi} d\sigma \Pi_{I}=\sum_{J}G_{IJ}K^{J}+\frac{1}{\alpha'}\sum_{J}b_{IJ}W^{J}.
\end{align}
Here again, following section \eqref{sec5}, we define the dimensionless field $Y^{I}$ as $X^{I}=\sqrt{\frac{\alpha'}{2}}Y^{I}$, where the mode expansion of $Y^{I}$ (splitting into left and right part) is given in terms of twisted tensile oscillators:
\begin{equation}
    \begin{split}
        Y^{I}_{L}&=y^{I}_{L}+\big(P^{I}\big)_{L}(\tau+\sigma)+i\sum_{n\neq 0}\frac{1}{n}\alpha^{\mu}_{n}e^{-in(\tau+\sigma)},\\
        Y^{I}_{R}&=y^{I}_{R}+\big(P^{I}\big)_{R}(\tau-\sigma)+i\sum_{n\neq 0}\frac{1}{n}\Tilde{\alpha}^{\mu}_{n}e^{-in(\tau-\sigma)}.
    \end{split}
\end{equation}

\subsubsection*{Quantization}
Classically, the residual gauge symmetry of this theory gives us two copies of the Virasoro algebra as well. We can also construct the same algebra from the classical constraints in the same way we did for usual string theory. However, the quantum theory for twisted strings is constructed on the flipped representation of the Virasoro algebra, which constitues an inner automorphism of the same. This makes it distinct from the usual tensile string theory, which is constructed on the highest weight representation of the Virasoro algebra. The physical state condition for this twisted theory turns out to be \cite{Bagchi:2020fpr}:
\begin{equation}\label{opo}
    \begin{split}
       \big(\mathcal{L}_{n}-a\delta_{n,0}\big)\ket{phys}=\big(\Bar{\mathcal{L}}_{-n}-\Bar{a}\delta_{n,0}\big)\ket{phys}=0\hspace{5mm}\forall n\geq 0 
    \end{split}
\end{equation}
where $\{\mathcal{L}_{n},\Bar{\mathcal{L}}_{n}\}$ are the (anti)holomorphic generators of the Virasoro algebra and  \{$a,\Bar{a}$\} are associated normal ordering constants. The vacuum in this theory is defined as a combination of highest weight and lowest weight conditions on the $\alpha$,
\begin{align}\label{opodartho}
    \alpha_{n}\ket{0}=\Tilde{\alpha}_{-n}\ket{0}=0,\hspace{5mm}\forall n>0
\end{align}
Since in the right modes the role of creation and annihilation operators are flipped, the definition of normal ordering for the right modes also change. The new definition of normal ordering for right modes is 
\begin{align}
    :\Tilde{\alpha}^{\mu}_{-m}\Tilde{\alpha}^{\nu}_{n}:=\Tilde{\alpha}^{\nu}_{n}\Tilde{\alpha}^{\mu}_{-m}.
\end{align}
The number operators in this theory are defined as 
\begin{align}
    N=\sum_{n=1}^{\infty}:\alpha_{-n}\cdot \alpha_{n}:\hspace{5mm}\widetilde N=-\sum_{n=1}^{\infty}:\Tilde{\alpha}_{-n}\cdot \Tilde{\alpha}_{n}:
\end{align}
The entire Hilbert space of this theory can be spanned by the eigenbasis of the number operators, and the basis states of the eigenbasis can be constructed by acting the oscillators $\{\alpha_{n},\Tilde{\alpha}_{-n}\}$ on the vacuum defined in \eqref{opodartho}. In order to make the wave function single valued, canonical momentum $\pi_{I}$ must be quantized i.e. $\pi_{I}=k_{I},\hspace{1mm}k_{I}\in \mathbb{Z}$. These result into the following, again familiar, expression for $\big(P_{I}\big)_{L,R}$:
\begin{align}
   \big(P_{I}\big)_{L,R}=\frac{1}{\sqrt{2}}\bigg\{\sqrt{\alpha'}\Big(k_{I}-\frac{1}{\alpha'}\sum_{J=1}^{d} b_{IJ}\omega^{J}\Big)\pm \frac{1}{\sqrt{\alpha'}}\sum_{J=1}^{d}G_{IJ}\omega^{J}\bigg\}.
\end{align}
Now, the zero modes $\mathcal{L}_{0}$ and $\Bar{\mathcal{L}_{0}}$ can be expressed in terms of the number operators, windings and momenta as
\begin{equation}
    \begin{split}
        \mathcal{L}_{0}&=\frac{\alpha'}{4}\sum_{I.J=1}^{d}\Big(k_{I}-\frac{1}{\alpha'}b_{IJ}\omega^{J}+\frac{1}{\alpha'}G_{IJ}\omega^{J}\Big)^2+\frac{\alpha'}{4}\sum_{\mu=0}^{25-d}k_{\mu}k^{\mu}+N\\
        \Bar{\mathcal{L}}_{0}&=\frac{\alpha'}{4}\sum_{I.J=1}^{d}\Big(k_{I}-\frac{1}{\alpha'}b_{IJ}\omega^{J}-\frac{1}{\alpha'}G_{IJ}\omega^{J}\Big)^2+\frac{\alpha'}{4}\sum_{\mu=0}^{25-d}k_{\mu}k^{\mu}-\widetilde{N}.
    \end{split}
\end{equation}
Using $\mathcal{L}_{0}$ and $\Bar{\mathcal{L}}_{0}$ conditions in \eqref{opo} with
$a=-\Bar{a}=1$ and adding and subtracting them, we have the following constraints on physical states of level $(r,s)$, where $r,s$ are eigenvalues of number operators $N$ and $\widetilde{N}$ respectively,
\begin{equation*}
\begin{split}
    &r+s+\frac{\alpha'}{4}\sum_{I.J=1}^{d}\Bigg[\Big(k_{I}-\frac{1}{\alpha'}b_{IJ}\omega^{J}+\frac{1}{\alpha'}G_{IJ}\omega^{J}\Big)^2-\Big(k_{I}-\frac{1}{\alpha'}b_{IJ}\omega^{J}-\frac{1}{\alpha'}G_{IJ}\omega^{J}\Big)^2\Bigg]-2=0\\
    r&-s+\frac{\alpha'}{4}\sum_{I.J=1}^{d}\Bigg[\Big(k_{I}-\frac{1}{\alpha'}b_{IJ}\omega^{J}+\frac{1}{\alpha'}G_{IJ}\omega^{J}\Big)^2+\Big(k_{I}-\frac{1}{\alpha'}b_{IJ}\omega^{J}-\frac{1}{\alpha'}G_{IJ}\omega^{J}\Big)^2\Bigg]-\frac{\alpha'}{2}m^2=0.
\end{split}
\end{equation*}
In the above, we have used $m^2=-\sum_{\mu=0}^{25-d}k_{\mu}k^{\mu}$, considering bosonic twisted tensile string theories are also consistent in $D=26$ .  Simplifying these constraints one gets the level matching condition and the mass formula:
\begin{subequations}
   \begin{equation}\label{beaver}
       r+s+\sum_{I=1}^{d}k_{I}\omega^{I}=2,
   \end{equation}
       \begin{equation}\label{raccoon}
            m^2=k^{T}G^{-1}k+\frac{1}{\alpha'^2}\omega^{T}(G-bG^{-1}b)\omega-\frac{2}{\alpha'}k^{T}bG^{-1}\omega+\frac{2}{\alpha'}(r-s).
       \end{equation}
\end{subequations}
There are immediately two important things to note as a consequence of these constraints: 
\begin{itemize}
    \item For non-compact target spacetime, the level matching condition implied that only states of level 2 could be physical. In \cite{Casali:2017mss,Lee:2017crr,Banerjee:2023ekd} it was found that single compact dimension allows more freedom: states of any level can be compactified for appropriate values of internal momentum and winding number. In \eqref{beaver} we see that for $d>1$, the freedom is extended even more, in this case for each level we will have infinite number of possible physical states.
    \item Comparing the mass spectrum \eqref{raccoon} with the mass spectrum for usual tensile string \eqref{ctb28}, one finds that the part of the expression arising due to compactification happens to be exactly same as that of tensile string. Hence, from this, one can conclude that this mass spectrum has all the discrete symmetries satisfied by the usual tensile mass spectrum. In other words the twisted mass spectrum will also have an explicit $O(d,d,\mathbb{Z})$ symmetry. 
\end{itemize}

\subsection{Tensionless flipped strings with $B$ field}

In this section we analyze the effect of constant $B$ field on the null flipped or ``\textit{ambitwistor}'' string theory, the tensionless progeny of the twisted theory we discussed in the last subsection. It is well known that this theory inherently possesses the highest weight representation of the BMS$_3$ algebra. The physical state conditions leading to this quantum theory are given by just the usual right action of the stress tensor constraints:
\begin{equation}\label{TSQR30}
    (L_{n}-a_{L}\delta_{n,0})\ket{phys}=0\hspace{5mm}\forall n\geq 0,~~~~
    (M_{n}-a_{M}\delta_{n,0})\ket{phys}=0\hspace{5mm}\forall n\geq 0.
\end{equation}

This theory also involves left and right moving oscillators like the tensile parent thereof, much like the theory defined on the oscillator vacuum. However, here the right moving oscillators are redefined, to comply with the parent vacuum conditions \cite{Bagchi:2020fpr}, as below:
\begin{align}
    \Tilde{\mathcal{C}}_{n}=\Tilde{C}_{-n}.
\end{align}
The flipped vacuum for a compactified theory, on which the ambitwistor string theory (defined by the subscript $A$) is based on, is defined somewhat like the condition \eqref{opodartho}, i.e.
\begin{align}\label{TSQR31}
C^{\mu}_{n}\ket{0,0,k^{\mu},k_{I},\omega^{I}}_{A}=\Tilde{\mathcal{C}}^{\mu}_{n}\ket{0,0,k^{\mu},k_{I},\omega^{I}}_{A}=0\hspace{5mm}\forall n>0.
\end{align}
As we can see, just like the tensile counterpart of the theory, here too, the role of the creation and annihilation operators have been flipped. The algebra satisfied by the newly defined oscillators turns out to be the following:
\begin{align}\label{TSQR33}
[C^{\mu}_{m},C^{\nu}_{n}]=m\eta^{\mu\nu}\delta_{m+n}\hspace{5mm}[\Tilde{\mathcal{C}}^{\mu}_{m},\Tilde{\mathcal{C}}^{\nu}_{n}]=-m\eta^{\mu\nu}\delta_{m+n},\hspace{5mm}[C^{\mu}_{m},\Tilde{\mathcal{C}}^{\nu}_{n}]=0.\end{align}
The generators $L_{n}$ and $M_{n}$ can be expressed in terms of $\{C,\Tilde{\mathcal{C}}\}$ as below
\begin{equation}\label{TSQR34}
\begin{split}
    L_{n}&=\frac{1}{2}\sum_{m}\left[C_{-m}\cdot C_{m+n}-\Tilde{\mathcal{C}}_{m}\cdot\Tilde{\mathcal{C}}_{-m+n}\right],\\
    M_{n}&=\frac{1}{2}\sum_{m}\left[C_{-m}\cdot C_{m+n}+\Tilde{\mathcal{C}}_{m}\cdot \Tilde{\mathcal{C}}_{-m+n}+2C_{-m}\cdot\Tilde{\mathcal{C}}_{m+n}\right].
    \end{split}
\end{equation}
Here again we use the equations \eqref{ctb24}-\eqref{chh49} and obtain the expression for the canonical momenta $(K_{I})_{L,R}$ which is exactly similar as we had in the case of oscillator vacuum, given by
\begin{align}
   (k_I)_{\hspace{.25mm}L,R}=\frac{1}{\sqrt{2}}\bigg\{\sqrt{c'}\Bigg(k_{I}-\sum_{J=1}^{d}\mathcal{B}_{IJ}\omega^{J}\Bigg)\pm \frac{1}{\sqrt{c'}}\sum_{J=1}^{d}G_{IJ}\omega^{J}\bigg\}.
\end{align}
The normal ordered form of the zero modes $L_{0}$ and $M_{0}$ in terms of $ (k_I)_{\hspace{.25mm}L,R}$ are given by
\begin{eqnarray}
\begin{split}
    L_{0}=N+\Bar N+\frac{1}{2}\sum_{I=1}^{d}\big((K^{I})_{L}(k_I)_{\hspace{.25mm}L}-(K^{I})_{R}(k_I)_{\hspace{.25mm}R}\big)
=N+\Bar N+\sum_{I=1}^{d}k_{I}\omega^{I}
\end{split}
\end{eqnarray}
and
\begin{equation}
\begin{split}
M_{0}&=\frac{1}{2c'}\sum_{I=1}^{d}(K^{I}_{L}+K^{I}_{R})(K_{I\hspace{.25mm}L}+K_{I\hspace{.25mm}R})+c'\sum_{\mu={0}}^{25-d}k_{\mu}k^{\mu}+N-\Bar N+X+Y \\
=&c'\sum_{I,J=1}^{d}\Big(k_{I}-\mathcal{B}_{IJ}\omega^{J}\Big)G^{IJ}\Big(k_{I}-\mathcal{B}_{IK}\omega^{K}\Big)+c'\sum_{\mu={0}}^{25-d}k_{\mu}k^{\mu}+N-\Bar N+X+Y.
\end{split}
\end{equation}
The physical state conditions for tensionless string on flipped vacuum is given in \eqref{TSQR30}.
As we have already seen before \cite{Bagchi:2021rfw}, this theory has $a_{L}=2$ and $a_{M}=0$. Using $a_L=2$ on the $L_{0}$ condition gives us the following physical state condition for a generic state of level $\{r,s\}$
\begin{align}\label{B10}
     r+s+\sum_{i=1}^{d}k_{I}\omega^{I}=2
\end{align}
Here we note that the level matching condition we obtained in \eqref{B10} is identical to that in \eqref{beaver}, indicating towards a one to one mapping from tensile and tensionless states through the limit. In \cite{Banerjee:2023ekd}, we have seen that when we compactify one dimension, we can have physical states in any possible level for suitable choice of internal momentum ($K$) and winding number ($W$). However, for states with $r+s\neq 2$, we could have only finite number of physical states at each level. For example, for level zero, ($r+s=0$), there would be only four possible states (since there are only four possible integer combinations of $K$ and $W$ which can ensure that $KW=2$). Similarly we can find only two states for level one. However, for more compact dimensions $d\geq 2$, even for $r+s\neq 2$, we will be able to get infinite number of possible combinations of $\{k_{I},\omega^{I}\}$ satisfying \eqref{B10}. As a result, for $d\geq 2$ we have infinite number of physical states at any level.  

\medskip

Now, we will apply the physical state conditions \eqref{TSQR30} (keeping in mind that $a_M=0$) along with \eqref{B10} on level 0, level 1 and level 2 states and study the constraints to be satisfied by them in order to become explicit physical states.

\subsection*{Level 0 states}
The level 0 states are essentially flipped vacuum states defined in \eqref{TSQR31}, i.e. $\ket{0,0,k^{\mu},k_{I},\omega^{I}}_{A}$. The $L_{0}$ and $M_{0}$ physical state conditions are the only non-trivial physical state conditions for this level. Since for vacuum both $r=s=0$, the $L_{0}$ condition leads us to:
\begin{align}\label{B11}
    \sum_{I=1}^{d}k_{I}\omega^{I}=2.
\end{align}
As we have already discussed, there are infinite number of states satisfying \eqref{B11}. Now let us look at the $M_{0}$ condition, which gives us the mass
\begin{equation}\label{FD14}
\begin{split}
    m^2&=-\sum_{\mu={0}}^{25-d}k_{\mu}k^{\mu} =\sum_{I,J,L=1}^{d}\Big(k_{I}-\mathcal{B}_{IJ}\omega^{J}\Big)G^{IL}\Big(k_{L}-\mathcal{B}_{LK}\omega^{K}\Big)\\&=k^{T}G^{-1}k-2k^{T}G^{-1}\mathbb{B}\omega-\omega^{T}\mathbb{B}G^{-1}\mathbb{B}\omega
\end{split}
\end{equation}

\subsection*{Level 1 states}
As discussed in \cite{Bagchi:2020fpr}, the physical state conditions given in \eqref{TSQR30} for $n>1$ will be trivially satisfied by a state of level 1. Hence we have only four nontrivial physical state conditions, namely the $L_{0}$, $M_{0}$, $L_{1}$ and $M_{1}$ conditions. A generic state of level 1 is given by
\begin{align}\label{chh35}
\ket{1,k_{I},\omega^{I}}_{A}=a_{\mu}C^{\mu}_{-1}\ket{0,0,k^{\mu},k_{I},\omega^{I}}_{A}+b_{\mu}\Tilde{\mathcal{C}}^{\mu}_{-1}\ket{0,0,k^{\mu},k_{I},\omega^{I}}_{A}.
\end{align}
The $L_{0}$ condition simply gives us $\sum_{I=1}^{d}k_{I}\omega^{I}=1$, which leads us to infinite number of level 1 states. The other non-trivial physical state conditions are then given by
\begin{align}
    L_{1}\ket{1,k_{I},\omega^{I}}_{A}=M_{1}\ket{1,k_{I},\omega^{I}}_{A}=M_{0}\ket{1,k_{I},\omega^{I}}_{A}=0
\end{align}
They lead us to the following constraints for $\{k^{\mu},k_{I},\omega^{I},a_{\mu},b_{\mu}\}$.
    \begin{align} \label{opodartho3}
        &\Bigg[\sum_{I,J,K=1}^{d}(k_{J}-\mathcal{B}_{JK}\omega^{k})G^{IJ}(a_{I}+b_{I})+\frac{1}{c'}\sum_{I=1}^{d}\omega^{I}(a_{I}-b_{I}) \nonumber \\ & \hspace{5.85cm}+\sum_{\mu=0}^{25-d}k^{\mu}(a_{\mu}+b_{\mu})\Bigg]\ket{0,0,k^{\mu},k_{I},\omega^{I}}_{A}=0 \nonumber \\
        &\Bigg[\sum_{I,J,K=1}^{d}(k_{J}-\mathcal{B}_{JK}\omega^{k})G^{IJ}(a_{I}-b_{I})+\sum_{\mu=0}^{25-d}k^{\mu}(a_{\mu}-b_{\mu})\Bigg]\ket{0,0,k^{\mu},k_{I},\omega^{I}}_{A}=0 \\
        &\Big[\big(A+1\big)a_{\mu}-b_{\mu}\Big]C^{\mu}_{-1}\ket{0,0,k^{\mu},k_{I},\omega^{I}}_{A}+\Big[\big(A-1\big)b_{\mu}+a_{\mu}\Big]\Tilde{\mathcal{C}}^{\mu}_{-1}\ket{0,0,k^{\mu},k_{I},\omega^{I}}_{A}=0. \nonumber
    \end{align}
where we have used:
\begin{equation}
    A=c'\sum_{\mu={0}}^{25-d}k_{\mu}k^{\mu}+c'\sum_{I,J,L=1}^{d}\Big(k_{I}-\mathcal{B}_{IJ}\omega^{J}\Big)G^{IL}\Big(k_{L}-\mathcal{B}_{LK}\omega^{K}\Big).
\end{equation}
Following the same line of argument provided in  \cite{Bagchi:2020fpr} it can be shown that $ A=0,\hspace{1mm} \text{and} \hspace{1mm}a_{\mu}=b_{\mu}, \hspace{1mm}\mu=\{0,1,...,25-d\}$. $A=0$ implies that the mass of the state has to be same as \eqref{FD14}. The norm of the state is then given by $\braket{1|1}=a_{\mu}a^{\mu}-b_{\mu}b^{\mu}=0$, implying that level 1 physical states are null states. The remaining physical state conditions give us the following additional constraints on the coefficients $a_{\mu}$ themselves:
\begin{align}
   \sum_{I,J,K=1}^{d}(k_{J}-\mathcal{B}_{JK}\omega^{k})G^{IJ}a_{I}+\sum_{\mu=0}^{25-d}k^{\mu}a_{\mu}=0.
\end{align}
\subsection*{Level 2 states}
We start from a generic state of level 2, which is given below in terms of all possible polarisations
\begin{align}\label{opodartho2}
    \ket{2,k_{I},\omega^{I}}_{A}&=a_{\mu}C^{\mu}_{-2}\ket{0}_{A}+e_{\mu\nu}C^{\mu}_{-1}C^{\nu}_{-1}\ket{0}_{A}+h_{\mu\nu}C^{\mu}_{-1}\Tilde{\mathcal{C}}^{\nu}_{-1}\ket{0}_{A}\nonumber\\
    &+b_{\mu}\Tilde{\mathcal{C}}^{\mu}_{-2}\ket{0}_{A}+f_{\mu\nu}\Tilde{\mathcal{C}}^{\mu}_{-1}\Tilde{\mathcal{C}}^{\nu}_{-1}\ket{0}_{A}+j_{\mu\nu}C^{\mu}_{-1}\Tilde{\mathcal{C}}^{\nu}_{-1}\ket{0}_{A}.
\end{align}
In the above expression for level 2, $\ket{0}_{A}$ is written as short form of $\ket{0,0,k^{\mu},k_{I},\omega^{I}}_A$, $e_{\mu\nu}$ and $f_{\mu\nu}$ are essentially symmetric, $h_{\mu\nu}$ and $j_{\mu\nu}$ are by definition  symmetric and antisymmetric respectively. Applying the $L_{0}$ condition on this state we inevitably get
\begin{align}\label{opodartho4}
    \sum_{i=1}^dk_{I}\omega^{I}=0.
\end{align}
For level 2, apart from $L_{0}$ condition there will be five more non-trivial physical state conditions. The $M_{0}$ condition is given by
\begin{equation}\label{chhagol6}
\begin{split}
     M_{0}\ket{2,k_{I},\omega^{I}}&=\left[(A+2)a_{\mu}-2b_{\mu}\right]C^{\mu}_{-2}\ket{0}_{A}+\left[(A-2)b_{\mu}-2a_{\mu}\right]\Tilde{\mathcal{C}}^{\mu}_{-2}\ket{0}_{A}\\
    &+\left[(A+2)e_{\mu\nu}-h_{\mu\nu}\right]C^{\mu}_{-1}C^{\nu}_{-1}\ket{0}_{A}+\left[(A-2)f_{\mu\nu}+h_{\mu\nu}\right]\Tilde{\mathcal{C}}^{\mu}_{-1}\Tilde{\mathcal{C}}^{\nu}_{-1}\ket{0}_{A}\\
    &+\left[A(h_{\mu\nu}+j_{\mu\nu})+2(e_{\mu\nu}-f_{\mu\nu})\right]C^{\mu}_{-1}\Tilde{\mathcal{C}}^{\nu}_{-1}\ket{0}_{A},
\end{split}
\end{equation}
where expression of $A$ is given in \eqref{opodartho3}. The remaining physical state conditions are given below.
\begin{equation}\label{chhagol7}
\begin{split}
L_{1}\ket{2,k_{I},\omega^{I}}&=\Bigg[2a_{\nu}+c'\sum_{I,J,K=1}^{d}(k_{J}-\mathcal{B}_{JK}\omega^{k})G^{IJ}(2e_{I\nu}+h_{I\nu}-j_{I\nu})\\
&+\sum_{I=1}^{d}\omega^{I}(2e_{I\nu}-h_{I\nu}+j_{I\nu})+c'\sum_{\mu=0}^{25-d}(2e_{\mu\nu}+h_{\mu\nu}-j_{\mu\nu})k^{\mu}\Bigg]C^{\nu}_{-1}\ket{0}_{A}\\&+\Bigg[2b_{\nu}+c'\sum_{I,J,K=1}^{d}(k_{J}-\mathcal{B}_{JK}\omega^{k})G^{IJ}(h_{I\nu}+j_{I\nu}+2f_{I\nu})\\&+\sum_{I=1}^{d}\omega^{I}(h_{I\nu}+j_{I\nu}-2f_{I\nu})+c'\sum_{\mu=0}^{25-d}(h_{\mu\nu}+j_{\mu\nu}+2f_{\mu\nu})k^{\mu}\Bigg]\Tilde{\mathcal{C}}^{\nu}_{-1}\ket{0}_{A}=0
\end{split}
\end{equation}
\begin{equation}\label{chhagol8}
\begin{split}
M_{1}\ket{2,k_{I},\omega^{I}}&=2\Bigg[(a_{\nu}-b_{\nu})+c'\sum_{I,J,K=1}^{d}(k_{J}-\mathcal{B}_{JK}\omega^{k})G^{IJ}(2e_{I\nu}+j_{I\nu}-h_{I\nu})\nonumber\\&+c'\sum_{\mu=0}^{25-d}(2e_{\mu\nu}+j_{\mu\nu}-h_{\mu\nu})k^{\mu}\Bigg]C^{\nu}_{-1}\ket{0}_{A}+2\Bigg[(a_{\nu}-b_{\nu})\\&+c'\sum_{I,J,K=1}^{d}(k_{J}-\mathcal{B}_{JK}\omega^{k})G^{IJ}(j_{I\nu}+h_{I\nu}-2f_{I\nu})\nonumber\\&+c'\sum_{\mu=0}^{25-d}(h_{\mu\nu}+j_{\mu\nu}-2f_{\mu\nu})k^{\mu}\Bigg]\Tilde{\mathcal{C}}^{\nu}_{-1}\ket{0}_{A}=0\nonumber
\end{split}
\end{equation}
\begin{equation}\label{chhagol9}
    \begin{split}
        L_{2}\ket{2,k_{I},\omega^{I}}=\Bigg[c'\sum_{I,J,K=1}^{d}(k_{J}-\mathcal{B}_{JK}\omega^{k})G^{IJ}(a_{I}+b_{I})+\sum_{I=1}^{d}\omega^I(a_{I}&-b_{I})+c'\sum_{\mu=0}^{25-d}(a_{\mu}+b_{\mu})k^{\mu}\\&+\frac{1}{2}(e^{\mu}_{\mu}-f^{\mu}_{\mu})\Bigg]\ket{0}_{A}=0
    \end{split}
\end{equation}
\begin{equation}\label{chhagol10}
    \begin{split}
M_{2}\ket{2,k_{I},\omega^{I}}=\Bigg[c'\sum_{I,J,K=1}^{d}(k_{J}-\mathcal{B}_{JK}\omega^{k})G^{IJ}(a_{I}-b_{I})&+c'\sum_{\mu=0}^{25-d}(a_{\mu}-b_{\mu})k^{\mu}\\&+\frac{1}{4}(e^{\mu}_{\mu}+f^{\mu}_{\mu}-h^{\mu}_{\mu})\Bigg]\ket{0}_{A}=0
     \end{split}
\end{equation}
The equations \eqref{chhagol6}-\eqref{chhagol10} yield the following constraints on the state 
\eqref{opodartho2}
\begin{subequations}
\begin{equation}\label{gadha1}
     a_{\mu}=b_{\mu}=0, \hspace{5mm} h_{\mu\nu}=e_{\mu\nu}=2h_{\mu\nu}
\end{equation}
  \begin{equation}\label{gadha3}
      A=0
  \end{equation}
   \begin{equation}\label{gadha4}
      \sum_{I,J,K=1}^{d}(k_{J}-\mathcal{B}_{JK}\omega^{k})G^{IJ}e_{I\nu}+\sum_{\nu=0}^{25-d}e_{\mu\nu}k^{\mu}= \sum_{I,J,K=1}^{d}(k_{J}-\mathcal{B}_{JK}\omega^{k})G^{IJ}j_{I\nu}+\sum_{\nu=0}^{25-d}j_{\mu\nu}k^{\mu}=0.
   \end{equation}
   \begin{equation}\label{gadha5}
       \sum_{I=1}^{d}j_{I\nu}W^{I}=0.
   \end{equation}
\end{subequations}
The constraint \eqref{gadha3} leads us to the mass spectrum identical to \eqref{FD14}. The only difference is that \{$k_{I},\omega^{I}$\} for level 2 has to satisfy the constraint \eqref{opodartho4}. The remaining constraints are on the coefficients and fixing those lead us to the following final form of the level 2 states, which decouple into symmetric and antisymmetric parts:
\begin{align}
    \ket{2,k_{I},\omega^{I}}=e_{\mu\nu}\Big[C^{\mu}_{-1}C^{\nu}_{-1}\ket{0}_{A}+2C^{\mu}_{-1}\Tilde{\mathcal{C}}^{\nu}_{-1}\ket{0}_{A}+\Tilde{\mathcal{C}}^{\mu}_{-1}\Tilde{\mathcal{C}}^{\nu}_{-1}\ket{0}_{A}\Big]+j_{\mu\nu}C^{\mu}_{-1}\Tilde{\mathcal{C}}^{\nu}_{-1}\ket{0}_{A}.  
\end{align}

\subsection*{Higher level states}
In this part we generalise our analysis presented above to an arbitrary level $l=r+s$. A generic state of this level is given by
\begin{align}
    \ket{r,s,k^\mu,k_I,\omega^I}=\sum_{j}\rho_{j}\Bigg(\prod_{i=1}^{p}C^{a_{i}}_{-m_{i}}\prod_{j=1}^{q}\Tilde{\mathcal{C}}^{b_{j}}_{-n_{j}}\Bigg)_{j}\ket{0,0,k^{\mu},k_I,\omega^I}_{c},
\end{align}
$a_{j}$  and $b_{j}$ being the powers of $C$ and $\Tilde{\mathcal{C}}$ respectively, and the level is given by
\begin{align}
    l=r+s=\sum_{i=1}^{p}a_{i}m_{i}+\sum_{i=1}^{q}b_{i}n_{i}.
\end{align}
The $L_{0}$ condition for this level will be
\begin{align}
    \sum_{I}k_{I}\omega^{I}=2-l.
\end{align}
After applying this condition, we shall be left with $(2l+1)$ number of physical state conditions. The $M_{0}$ condition will lead us to the same mass spectrum we have derived before for level 0,1 and 2 and the remaining conditions will restrict the coefficients $\rho_{j}$. We do not present all those details here for sake of brevity. 

\medskip

Another thing which is important to highlight here is that the mass spectrum obtained in this theory happens to be identical to the mass spectrum we previously obtained for oscillator vacuum and induced vacuum. Hence the symmetry structure of the mass spectrum too is identical to the other two theories discussed before. 

\subsection{Limit from tensile flipped string}

The level matching condition and the mass spectrum of the tensile counterpart of the flipped string theory has already been calculated. Let us rewrite the results
\begin{equation*}\label{flipspec}
    \begin{split}       &\hspace{2mm}r+s+\sum_{I=1}^{d}k_{I}\omega^{I}=2\\
        m^2=k^{T}G^{-1}k+\frac{1}{\alpha'^2}&\omega^{T}(G-bG^{-1}b)\omega-\frac{2}{\alpha'}k^{T}bG^{-1}\omega+\frac{2}{\alpha'}(r-s).
    \end{split}
\end{equation*}
As we have already highlighted, the level matching conditions for both tensile and tensionless flipped string theory are identical. From this we get an indication that the physical state with level $l$ of tensile flipped theory will reduce to the physical state of level $l$ in the tensionless theory. As we have seen in \cite{Banerjee:2023ekd}, that is indeed the case when we work in a background with only one dimension compactified. Working in a target space with $d$-dimensions compactified does not change anything: here too, the tensile physical states will reduce to tensionless physical states of the very same level. 
\medskip

We should also note here, taking direct tensionless limit of the mass spectrum in equation \eqref{raccoon} in the spirit of earlier computations, we can directly reproduce the mass spectrum in \eqref{FD14}. Hence the limiting analysis of tensionless twisted string theory also turns out to be consistent with the intrinsic flipped construction of the same.

\newpage 

\section{Conclusions and Discussions}\label{sec7}

 \subsection{Summary}

In this paper, we have extensively investigated the impact of a constant $B$ field on the three distinct quantum theories of tensionless strings, namely, those defined on Oscillator, Induced, and Flipped vacua. Our analysis, complemented with earlier works, revealed that while the $B$ field does not affect the level matching constraint of the states, it does modify the mass-shell condition when the target space is compactified (on a $T^d$ in this case). In our previous work, we have observed that the mass spectrum of all the three quantum theories seemingly violated T-duality as the string length became infinitely large in the tensionless limit. However, tensile string with constant $B$ field has  additional symmetries which turns out to be preserved even at tensionless limit, restoring a sense of T-duality to the spectrum. This has been a key point in our manuscript. 
\medskip

We also explicitly worked out the fate of $O(d,d)$ symmetries associated to tensile generalised metric in this case, and it turned out to be a degenerate one as well, leading to a broken version of the tensile self.
Notably, for all the vacua, introduction of a constant $B$ field leads to the appearance of winding number dependence in the mass spectrum, distinguishing it from the case without a $B$ field. This creates an important difference in the spectrum of states. We verified the consistency of our results by comparing them with the direct tensionless limit of the corresponding parent tensile theories for the Induced and Flipped cases. Furthermore, we discovered that in multi-dimensional compactification, as opposed to circle compactification, an infinite number of physical states emerge at each level for the flipped spectrum. 
Overall, our findings shed light on the intricate interplay between constant $B$ fields, compactification, and the tensionless limit, deepening our understanding of these fundamental aspects in string theory.

\subsection{Future directions}
As we mentioned at the very beginning, the study of null strings from a formalism viewpoint is still very nascent and requires close investigation. In this work, which is hopefully another one in a series of many, we are just starting to delve deeper into the theory of null string compactifications. With the inclusion of the antisymmetric field, we are one step closer to understanding tensionless sigma models and role of interactions therein. The road ahead is intriguing, and one can look at various related problems in close connection with the current one. We discuss some immediate avenues below.
\medskip

\textit{\textbf{Null duality and connection to other ``corners'':}}~As already discussed in the text, the manifestations of sectorised T-duality for our case takes the theory out of the null form of $\mathcal{G}$.  Although one could argue these two theories are related via the very same spectrum, more is required to show the equivalence of these theories. Fuelled by the recent observations \cite{Gomis:2023eav}, we also theorized that our formalism indicates sectorised T-duality transformations on Euclidean compact spaces for null strings may lead to tensile strings in target space equipped with a degenerate metric. By construction, our compact directions are spacelike, unlike the ones in \cite{Gomis:2023eav} giving rise to pure Carrollian target space structures. However this observation keeps open the possibility that something similar could also be done for our case, involving a timelike compact circle, so that a sectorised duality transforms our null string to some opposite unexplored corner with tensile strings on Carroll backgrounds.

\medskip

\textit{\textbf{Bose-Einstein condensation and $B$ fields:}} As we mentioned in the Induced spectrum discussion, the Induced vacuum state in the non-compact case simply boils down to a space-filling Neumann boundary state \cite{Bagchi:2019cay}. There has been hints that this appearance of what essentially are D-brane boundary states, are nothing but a Bose-Einstein like condensation of all closed perturbative degrees of freedom on open ones. For compact cases these new degrees of freedom are even more interesting, however they lack an inherent T-dual picture, which is explicitly remedied by the presence of a constant $B$ field. It is imperative to understand whether the physics of this phase transition can be directly linked to the more widely known phenomenon of open-closed duality \cite{Seiberg:1999vs} in presence of a $B$ field. 

\medskip

\textit{\textbf{Double field theory with degenerate metric:}} A nice observation coming out of our investigation was the presence of a degenerate DFT metric in the description of such tensionless strings. As we mentioned earlier, embedding of degenerate (non-Riemannian) target spaces in the DFT metric has been explored in recent literature. Although in these cases, the DFT metric itself is non-degenerate and $O(d,d)$ covariance in unbroken, it turns out novel connections to Carroll and Newton-Cartan geometries can be established \cite{Morand:2017fnv}. It would be important to explore how our setup fits into this picture, and how degenerate and non-degenrate DFT metrics encode non-Riemannian geometric data in different ways. It will also be interesting to find whether this degenerate theory has an explicit worldsheet embedding, in line with the $O(d,d)$ covariant doubled string sigma model formulation \cite{Aldazabal:2013sca}.

\medskip

\textit{\textbf{Target space complex structures:}}
Most intriguingly, a lot still remains to be understood about the target space of such tensionless strings. 
The absence of part of $O(2,2;\mathbb{Z})$ symmetry in the $2d$ compact case can be visualised in another way. As is well known, the $O(2,2;\mathbb{Z})$ has $SL(2,\mathbb{Z})\otimes SL(2,\mathbb{Z})$ as subgroup. In case of usual tensile string theory with a $B$ field $B\epsilon_{IJ}$ this symmetry can be observed by defining complexified Kähler modulus ($T$) and complex structure modulus ($U$) on the torus as below \cite{Blumenhagen:2013fgp}.
\begin{equation}\label{Kähler1}
\begin{split}
    T=\frac{1}{\alpha'}(BR_{1}R_{2}+i\sqrt{\det G})=(T_{1}+iT_{2}),~~
    U=\frac{1}{G_{11}}(G_{12}+i\sqrt{\det G})=U_{1}+iU_{2}
\end{split}
\end{equation}
The squares of the left and right moving momentum can be expressed in terms of these structure constants as
\begin{equation}\label{Kähler2}
            p^{2}_{L}=\frac{\alpha'}{4T_{2}U_{2}}|k_{2}-Uk_{1}+\overline{T}(\omega_{1}+U\omega_{2})|^2,~~
         p^{2}_{R}=\frac{\alpha'}{4T_{2}U_{2}}|k_{2}-Uk_{1}+T(\omega_{1}+U\omega_{2})|^2
\end{equation}
In the above $\overline{T}$ denotes the complex conjugate of $T$. Since the mass spectrum of the tensile theory is the sum of $p^{2}_{L}$ and $p^{2}_{R}$, the compactified part of mass spectrum was entirely expressed in terms of these two complex structures. The mass spectrum for tensile string had an $SL(2,\mathbb{Z})_{T}$ symmetry, with  the sectorised duality transformation leading to $T\leftrightarrow U$ \cite{Blumenhagen:2013fgp}. The implication is that the mass spectrum will also have an $SL(2,\mathbb{Z})_{U}$ symmetry. 
\medskip

In case of tensionless strings too, the momenta $K^{2}_{L}$ and $K^{2}_{R}$ can be expressed in terms of appropriately defined complex structures\footnote{$\alpha'$ in \eqref{Kähler1} and \eqref{Kähler2} in this case has to be replaced by $c'$.}. However, the mass spectrum of the tensionless string, as we can see in \eqref{gadha7} also has a $2K^{I}_{L}K_{I\hspace{.5mm}R}$ part, which cannot be expressed as a function of the complex structure constants. As we have already discussed, the $g_{D_{i}}$ transformation does not lead us to a definite $G$ and $B$, which means we won't get any transformed values of $T$ and $U$ either, so $T\leftrightarrow U$ cannot happen here. As a result, we cannot observe the $SL(2,\mathbb{Z})_{U}$ symmetry in the mass spectrum. However, the $SL(2,\mathbb{Z})_{T}$ symmetry can still be observed. The direct tensionless limit of the complex Kähler modulus ($T$) evolves to a real number:
\begin{align}
  \lim_{\alpha'\to\infty}T_{1}=bR_{1}R_{2}\hspace{5mm}\lim_{\alpha'\to\infty}T_{2}=0.  
\end{align}
 We know for tensile strings, the shift symmetry implies symmetry under the transformation $T\to T+1$ i.e. $T_{1}\to T_{1}+1$, $T_{2}\to T_{2}$. Since $T\to T_{1}$ at tensionless limit, here the mass spectrum is invariant under the $T_{1}\to T_{1}+1$ transformation (also called the modular T-transformation). Also the inversion symmetry leads to the transformation $T_{1}\to-1/T_{1}$ (also called the modular S-transformation). It is well known that the modular T-transformation and modular S-transformation together generate the entire $SL(2,\mathbb{Z})$ group, which means $SL(2,\mathbb{Z})_{T}$ still survives in the tensionless limit. In earlier sections we found indication that for $T^2$ compactification, individual $g_{D_{i}}$ transformations connect tensionless theory to other limits of string theory involving a degenerate torus. Our expectation is that the lost $SL(2,\mathbb{Z})_{U}$ symmetry probably resides on those yet to be explored corners. We plan to come back to these pressing questions in future communications. 

\section*{Acknowledgements}
It is a pleasure to thank Arjun Bagchi for numerous illuminating discussions and comments on the manuscript. The authors would also like to thank Stephane Detournay, Sabyasachi Dhar, Daniel Grumiller, Jelle Hartong, Johannes Lahnsteiner, Shailesh Lal,  Hisayoshi Muraki, Niels Obers, Shahin Sheikh-Jabbari, Stefan Vandoren, and Ziqi Yan for many useful discussions. Initial version of this work was presented at OIST Okinawa and BITS Goa. ArB is supported in part by an OPERA grant and a seed grant NFSG/PIL/2023/P3816 from BITS-Pilani.  RC would like to acknowledge the partial support of CSIR grant File No: 09/092(0991)/2018-EMR-I and PHY/SERB/2019525. RC would also like to thank TU Wien, University of Utrecht, University of Amsterdam, NORDITA, DESY Hamburg, ULB Brussels, I.I.T. Bombay and BITS Goa for hospitality during various stages of this work. PP is partially supported by SPO/SERB/PHY/2019525 and by Institute Post Doctoral Fellowship (IPDF) from IIT kanpur. PP would also like to thank Kyoto University, OIST Okinawa and BITS Goa for kind hospitality during various stages of this work.
 
\appendix

\section{Symmetries of tensile string spectrum over two compact directions}\label{apA}
Let us consider Polyakov action for the tensile string theory compactified on $T^2$. The mass spectrum for this theory is given in \eqref{tensile}. Considering the $B$ field along the torus to be $B_{IJ}=b\epsilon_{IJ}$ the mass spectrum of tensile string is as given in \eqref{beltola}. For finite tension one can always take $\alpha'=1$. In that case the compactfied part of the mass spectrum will be
\begin{align}
     m^2=\frac{k_{1}^2}{R_{1}^2}+\frac{k_{2}^2}{R_{2}^2}+2b\Bigg(\frac{R_{1}}{R_{2}}k_2\omega_1-\frac{R_{2}}{R_{1}}k_1\omega_2\Bigg)+(1+b^2)(R_1^2\omega_1^2+R_2^2\omega_2^2)
\end{align}
This mass spectrum can be rewritten in the following form. 
 \begin{equation}\label{Bombagarh}
 \begin{split}
 m^2= \big[\omega^{T} \hspace{3mm} k^{T}\big]
\hspace{2mm}\mathcal{G}^{-1}_{\alpha'}
\begin{bmatrix}
    \omega \\ k
\end{bmatrix}, \hspace{5mm}
    \mathcal{G}^{-1}_{\alpha'}=\begin{bmatrix}
        (1+b^2)R^{2}_{1} & 0 & 0 & \frac{bR_{1}}{R_{2}}
       \\  0 & (1+b^2)R^{2}_{2} & -\frac{bR_{2}}{R_{1}} & 0\\
       0 & -\frac{bR_{2}}{R_{1}} & \frac{1}{R^{2}_{1}} & 0\\
       \frac{bR_{1}}{R_{2}} & 0 & 0 & \frac{1}{R^{2}_{2}}
    \end{bmatrix}   
 \end{split}
   \end{equation}
This mass spectrum shows complete $O(2,2;\mathbb{Z})$ symmetry. For $T^2$, there can be only two sectorised duality transformation 
\begin{align}
   g_{D_{1}}=\begin{bmatrix}
        0 & 0 & 1 & 0
       \\  0 & 1 & 0 & 0\\
       1 & 0 & 0 & 0\\
       0 & 0 & 0 & 1
    \end{bmatrix},\hspace{5mm} g_{D_{2}=}\begin{bmatrix}
        1 & 0 & 0 & 0
       \\  0 & 0 & 0 & 1\\
       0 & 0 & 1 & 0\\
       0 & 1 & 0 & 0
    \end{bmatrix}
\end{align}
This transformation leads to the following transformations on $\mathcal{G}^{-1}_{\alpha'}$
\begin{equation}
\begin{split}
   \mathcal{G}^{-1}_{\alpha'} \xrightarrow{g_{D_1}}{} g_{D_{1}}\mathcal{G}^{-1}_{\alpha'}g^T_{D_{1}}&=\begin{bmatrix}
        \frac{1}{R^{2}_{1}} & -\frac{bR_{2}}{R_{1}} & 0 & 0
       \\  -\frac{bR_{2}}{R_{1}} & (1+b^2)R^{2}_{2} & 0 & 0\\
       0 & 0 & (1+b^2)R^{2}_{1} & \frac{bR_{1}}{R_{2}}\\
       0 & 0 & \frac{bR_{1}}{R_{2}} & \frac{1}{R^{2}_{2}}
    \end{bmatrix}
\\  \mathcal{G}^{-1}_{\alpha'} \xrightarrow{g_{D_2}}{} g_{D_{2}}\mathcal{G}^{-1}_{\alpha'}g^T_{D_{2}}&=\begin{bmatrix}
        (1+b^2)R^{2}_{1} & \frac{bR_{1}}{R_{2}} & 0 & 0
       \\  \frac{bR_{1}}{R_{2}} & \frac{1}{R^{2}_{2}} & 0 & 0\\
       0 & 0 & \frac{1}{R^{2}_{1}} & -\frac{bR_{2}}{R_{1}}\\
       0 & 0 & -\frac{bR_{2}}{R_{1}} & (1+b^2)R^{2}_{2}
    \end{bmatrix}.
\end{split}
\end{equation}
By definition $g_{D_{1}}$ and $g_{D_{2}}$ flips $k_{1}\leftrightarrow\omega_{1}$ and $k_{2}\leftrightarrow\omega_{2}$ and the transformed inverse generalised metric along with the transformed $\{\omega,k\}$ vector gives us the same mass spectrum. As highlighted earlier in \eqref{bhishmalochan}, the transformed inverse generalised metric correspond to a new $G_{IJ}$ with off diagonal terms and $B_{IJ}=0$. This spectrum also obeys the shift symmetry. As one can see, the mass spectrum is invariant under the transformation \eqref{babui}, i.e.
\begin{align*}
    b\to b+\frac{n}{R_{1}R_{2}}\hspace{5mm}k_{I}\to k_{I}+n\epsilon_{IJ}\omega^J\hspace{5mm}n\in\mathbb{Z}.
\end{align*}
We recall that \eqref{babui} belongs to the category defined in \eqref{alu}. This is also invariant under the basis change transformation defined in \eqref{potol}. It can be shown that under the basis change transformation the metric of the torus $G_{IJ}$ and the $B$ field $B_{IJ}$ will transform as below
\begin{align}
    \mathbb{B}'=C\mathbb{B}(C^T)\hspace{5mm}G'=CG(C^T),
\end{align}
which is the same as \eqref{ghenchu}.
\section{Double sectorised duality calculation for $d=3$}\label{factorised}
Let us recall $\mathbb{B}_{IJ}^1$ from \eqref{level}
\begin{align}\label{a1}
    \mathbb{B}_{IJ}^1=\left(
\begin{array}{ccc}
 0 & a {R_1} {R_2} & 0 \\
 -a {R_1} {R_2} & 0 & 0 \\
 0 & 0 & 0 \\
\end{array}
\right)
\end{align}
Using $\eqref{a1}$ along with $G_{IJ}=R_I^2\delta_{IJ}$, we compute $\mathcal{G}$ as defined in \eqref{Buscher1} and obtain the following
\begin{align}
  \mathcal{G}=  \left(
\begin{array}{cccccc}
 a^2 R_1^2 & 0 & 0 & 0 & -\frac{a R_1}{R_{2}} & 0
   \\
 0 & a^2 R_{2}^2 & 0 & \frac{a R_{2}}{R_{1}} & 0 & 0 \\
 0 & 0 & 0 & 0 & 0 & 0 \\
 0 & -\frac{a R_{2}}{R_{1}} & 0 & \frac{1}{R_{1}^2} & 0
   & 0 \\
 \frac{a R_{1}}{R_{2}} & 0 & 0 & 0 & \frac{1}{R_{2}^2}
   & 0 \\
 0 & 0 & 0 & 0 & 0 & \frac{1}{R_{3}^2} \\
\end{array}
\right)
\end{align}
From the definition of $g_{D_i}$ given in \eqref{hotol}, it can be easily seen that 
\begin{align}
   g_{D_1}g_{D_2}= \left(
\begin{array}{cccccc}
 0 & 0 & 0 & 1 & 0 & 0 \\
 0 & 0 & 0 & 0 & 1 & 0 \\
 0 & 0 & 1 & 0 & 0 & 0 \\
 1 & 0 & 0 & 0 & 0 & 0 \\
 0 & 1 & 0 & 0 & 0 & 0 \\
 0 & 0 & 0 & 0 & 0 & 1 \\
\end{array}
\right)
\end{align}
Now let us recall that, under any transformation $A\in O(d,d;\mathbb{Z})$, $\mathcal{G}$ will transform like 
\begin{equation*}
    \mathcal{G}\to \mathcal{G}'= A\mathcal{G}A^T
\end{equation*}
For $A=g_{D_1}g_{D_2}$, $\mathcal{G}'$ will be given by
\begin{align}
\mathcal{G}'=\left(
\begin{array}{cccccc}
 \frac{1}{R_{1}^2} & 0 & 0 & 0 & -\frac{a R_{2}}{R_{1}}
   & 0 \\
 0 & \frac{1}{R_{2}^2} & 0 & \frac{a R_{1}}{R_{2}} & 0
   & 0 \\
 0 & 0 & 0 & 0 & 0 & 0 \\
 0 & -\frac{a R_{1}}{R_{2}} & 0 & a^2 R_{1}^2 & 0 & 0
   \\
 \frac{a R_{2}}{R_{1}} & 0 & 0 & 0 & a^2 R_{2}^2 & 0 \\
 0 & 0 & 0 & 0 & 0 & \frac{1}{R_{3}^2} \\
\end{array}
\right)
\end{align}
This $\mathcal{G}'$ can be computed from $B$ field $\mathbb{B}_{IJ}'$ and metric $G_{IJ}'$ using the same expression given in \eqref{Buscher1}, where $\mathbb{B}_{IJ}'$ and $G_{IJ}'$ are given by
\begin{align*}
    \mathbb{B}_{IJ}'= &\begin{bmatrix}
        0 && -\frac{1}{aR_1R_2}  && 0 \\
        \frac{1}{aR_1R_2} && 0 && 0 \\
        0  && 0  && 0\\
    \end{bmatrix},\hspace{5mm} G_{IJ}'=\begin{bmatrix}
        \frac{1}{a^2R_1^2} && 0  && 0 \\
        0 && \frac{1}{a^2R_2^2} && 0 \\
        0  && 0  && R_3^2\\
     \end{bmatrix}.
     \end{align*}

\section{Tensionless closed strings from limit}\label{apB}

The  generic mode expansion for the tensile bosonic string is given by
\begin{equation}\label{tensilemodeexp}
     X^{\mu}(\tau,\sigma)=x^{\mu}+\sqrt{\frac{\alpha'}{2}}\big[(\alpha_0^\mu+\tilde{\alpha}_0^\mu)\tau+(\alpha_0^\mu-\tilde{\alpha}_0^\mu)\sigma\big]+i\sqrt{\frac{\alpha'}{2}}\sum_{n\neq0}\frac{1}{n}\left[\alpha^{\mu}_{n}e^{-in(\tau+\sigma)}+\Tilde\alpha^{\mu}_{n}e^{-in(\tau-\sigma)}\right].
\end{equation}
Here $\alpha_n^\mu$ and $\tilde{\alpha}_n^\mu$ are left moving and right moving oscillators. For closed bosonic string, periodicity condition $(X^\mu(\tau,\sigma+2\pi)=X^\mu(\tau,\sigma))$ demands, $\alpha_0^\mu =\tilde{\alpha}_0^\mu$, which reduces the above mode expansion to,
\begin{equation}\label{tensilemodexp}
     X^{\mu}(\tau,\sigma)=x^{\mu}+\sqrt{2\alpha'} \alpha_0^\mu\tau+i\sqrt{\frac{\alpha'}{2}}\sum_{n\neq0}\frac{1}{n}\left[\alpha^{\mu}_{n}e^{-in(\tau+\sigma)}+\Tilde\alpha^{\mu}_{n}e^{-in(\tau-\sigma)}\right].
\end{equation}
Now in order to obtain closed bosonic tensionless string mode expansion, we consider the following limit on the worldsheet coordinates,
\begin{equation}
    \tau\to\epsilon\tau,~~~~~\sigma\to\sigma ~~~ \text{and}~~~\alpha'\to c'/\epsilon,~~\epsilon\to 0,
\end{equation}
where $c'$ is a finite parameter. The mode expansion \eqref{tensilemodexp} in this limit takes the following form
\begin{equation}\label{tsc4mode}
     X^{\mu}(\tau,\sigma)=x^{\mu}+\sqrt{2\epsilon c'} \alpha_0^\mu\tau+i\sqrt{\frac{c'}{2}}\sum_{n\neq0}\frac{1}{n}\left[\frac{\alpha_n^\mu-\tilde\alpha_{-n}^\mu}{\sqrt{\epsilon}}-in\tau\sqrt{\epsilon}(\alpha_n^\mu+\tilde\alpha_{-n}^\mu)\right]e^{-in\sigma}.
\end{equation}
We now compare \eqref{tsc4mode} with the tensionless mode expansion \eqref{fcb1} given by
\begin{equation}
    X^{\mu}(\tau,\sigma)=x^{\mu}+\sqrt{\frac{c'}{2}}B^{\mu}_{0}\tau+i\sqrt{\frac{c'}{2}}\sum_{n\neq 0}\frac{1}{n}(A^{\mu}_{n}-in\tau B^{\mu}_{n})e^{-in\sigma}
\end{equation}
and obtain the relation between ($\alpha,\tilde\alpha$) with $A$'s and $B$'s as:
\begin{equation}\label{tsc8}
    A_n^\mu=\frac{1}{\sqrt{\epsilon}}(\alpha_n^\mu-\tilde\alpha_{-n}^\mu),~~~ B_n^\mu=\sqrt{\epsilon}(\alpha_n^\mu+\tilde\alpha_{-n}^\mu).
\end{equation}
\section{More oscillator states from $T^2$ compactification}\label{states}
Here we will consider the states with \{$r=1,s=0$\} and {$\{r=0,s=1\}$} for the oscillator spectrum for $T^2$ compactification as given in \eqref{twomass}.
We begin with the constraint satisfied by \{$k_{I},\omega^{I}$\} for the state \{$r=1,s=0$\} which is given by
\begin{align}
    k_{1}\omega_{1}+k_{2}\omega_{2}=-1
\end{align}

\begin{table}[ht]\label{masstable}
\small
\centering
\begin{tabular}{ |c|c|c|c| }
\hline
$k_{1},\omega_{1}$ & $k_{2},\omega_{2}$ & $m^2$ & Condition for $m^2<0$ \\
\hline
\multirow{2}{6em}{$k_{1}=\omega_{1}=0$} & $k_{2}=1, \omega_{2}=-1$ & $m^2=\frac{1}{R^2}+b^2R^2-\frac{1}{c'}$ & $\frac{1}{R^2}+b^2R^2<\frac{1}{c'}$ \\
& $k_{2}=-1,\omega_{2}=1$ & & \\

\hline
\multirow{2}{7em}{$k_{1}=0,~\omega_{1}\neq0$} & $k_{2}=1, \omega_{2}=-1$ & $m^2=\left(\frac{1}{R}-b R\omega_1\right)^2+b^2R^2-\frac{1}{c'}$ &  $\left(\frac{1}{R}-b R\omega_1\right)^2+b^2R^2<-\frac{1}{c'}$\\ 

& $k_{2}=-1,~\omega_{2}=1$ & $m^2=\left(\frac{1}{R}+b R\omega_1\right)^2+b^2R^2-\frac{1}{c'}$ & $\left(\frac{1}{R}+b R\omega_1\right)^2+b^2R^2<-\frac{1}{c'}$\\

\hline
\multirow{2}{7em}{$k_{1}\neq0,~\omega_{1}=0$} & $k_{2}=1, \omega_{2}=-1$ & $m^2=\left(\frac{k_1}{R}-b R\right)^2+\frac{1}{R^2}-\frac{1}{c'}$ &  $\left(\frac{k_1}{R}-b R\right)^2+\frac{1}{R^2}<-\frac{1}{c'}$\\ 

& $k_{2}=-1,~\omega_{2}=1$ & $m^2=\left(\frac{k_1}{R}+b R\right)^2+\frac{1}{R^2}-\frac{1}{c'}$ & $\left(\frac{k_1}{R}+b R\right)^2+\frac{1}{R^2}<-\frac{1}{c'}$\\ 

\hline

\multirow{2}{7em}{$k_{1}\neq0,~\omega_{1}\neq0$} & $k_{2}\neq0,~\omega_{2}\neq0$ & $m^2=(\omega_1^2+\omega^2_{2})\left(\frac{k_{2}}{R\omega_{1}}-bR\right)^2$ & $k_{2}=\omega_{1}bR^2$ \\
& & $-\frac{1}{c'}$ & $(\omega_1^2+\omega^2_{2})\left(\frac{k_{2}}{R\omega_{1}}-bR\right)^2<\frac{1}{c'}$ \\
\hline
\end{tabular}
\captionof{table}{Mass square of states belonging to level $r=1,s=0$ for different $k_I,\omega^I$. \label{Tab4}}
\end{table}
The Table \eqref{Tab4} provides us with the expression of $m^2$ for the state \{$r=1,s=0$\} with the relevant constraint and the condition for getting tachyons for given values of $k_1,k_2,\omega_1$ and $\omega_2$. Similarly, we write the constraint satisfied by \{$k_{I},\omega^{I}$\} for the state \{$r=0,s=1$\} which is given by
\begin{align}
    k_{1}\omega_{1}+k_{2}\omega_{2}=1.
\end{align}
The corresponding Table \eqref{Tab5} with expression for $m^2$ for different values of $k_1,k_2,\omega_1$ and $\omega_2$ is given below.
\begin{center}\label{mass2table}
\small
\begin{tabular}{ |c|c|c|c| } 
\hline
$k_{1},\omega_{1}$ & $k_{2},\omega_{2}$ & $m^2$ & Condition for $m^2<0$ \\
\hline
\multirow{2}{6em}{$k_{1}=\omega_{1}=0$} & $k_{2}=1, \omega_{2}=1$ & $m^2=\frac{1}{R^2}+b^2R^2-\frac{1}{c'}$ & $\frac{1}{R^2}+b^2R^2<\frac{1}{c'}$ \\
& $k_{2}=-1,\omega_{2}=-1$ & & \\

\hline
\multirow{2}{7em}{$k_{1}=0,~\omega_{1}\neq0$} & $k_{2}=1, \omega_{2}=1$ & $m^2=\left(\frac{1}{R}-b R\omega_1\right)^2+b^2R^2-\frac{1}{c'}$ &  $\left(\frac{1}{R}-b R\omega_1\right)^2+b^2R^2<-\frac{1}{c'}$\\ 

& $k_{2}=-1,~\omega_{2}=-1$ & $m^2=\left(\frac{1}{R}+b R\omega_1\right)^2+b^2R^2-\frac{1}{c'}$ & $\left(\frac{1}{R}+b R\omega_1\right)^2+b^2R^2<-\frac{1}{c'}$\\

\hline
\multirow{2}{7em}{$k_{1}\neq0,~\omega_{1}=0$} & $k_{2}=1, \omega_{2}=1$ & $m^2=\left(\frac{k_1}{R}+b R\right)^2+\frac{1}{R^2}-\frac{1}{c'}$ &  $\left(\frac{k_1}{R}+b R\right)^2+\frac{1}{R^2}<-\frac{1}{c'}$\\ 

& $k_{2}=-1,~\omega_{2}=-1$ & $m^2=\left(\frac{k_1}{R}-b R\right)^2+\frac{1}{R^2}-\frac{1}{c'}$ & $\left(\frac{k_1}{R}-b R\right)^2+\frac{1}{R^2}<-\frac{1}{c'}$\\ 

\hline

\multirow{2}{7em}{$k_{1}\neq0,~\omega_{1}\neq0$} & $k_{2}\neq0,~\omega_{2}\neq0$ & $m^2=(\omega_1^2+\omega^2_{2})\left(\frac{k_{2}}{R\omega_{1}}-bR\right)^2$ & $k_{2}=\omega_{1}bR^2$ \\
& & $-\frac{1}{c'}$ & $(\omega_1^2+\omega^2_{2})\left(\frac{k_{2}}{R\omega_{1}}-bR\right)^2<\frac{1}{c'}$ \\
\hline
\end{tabular}
\captionof{table}{Mass square of states belonging to level $r=0,s=1$ for different $k_I,\omega^I$. \label{Tab5}}
\end{center}

\newpage

\bibliographystyle{ieeetr}
\bibliography{ref}

\begin{thebibliography}{10}

\bibitem{Schild:1976vq}
A.~Schild, ``{Classical Null Strings},'' {\em Phys. Rev. D}, vol.~16, p.~1722,
  1977.

\bibitem{Gross:1987kza}
D.~J. Gross and P.~F. Mende, ``{The High-Energy Behavior of String Scattering
  Amplitudes},'' {\em Phys. Lett. B}, vol.~197, pp.~129--134, 1987.

\bibitem{Gross:1987ar}
D.~J. Gross and P.~F. Mende, ``{String Theory Beyond the Planck Scale},'' {\em
  Nucl. Phys. B}, vol.~303, pp.~407--454, 1988.

\bibitem{Gross:1988ue}
D.~J. Gross, ``{High-Energy Symmetries of String Theory},'' {\em Phys. Rev.
  Lett.}, vol.~60, p.~1229, 1988.

\bibitem{Vasiliev:2003cph}
M.~A. Vasiliev, ``{Higher spin gauge theories in various dimensions},'' {\em
  PoS}, vol.~JHW2003, p.~003, 2003.

\bibitem{Sagnotti:2003qa}
A.~Sagnotti and M.~Tsulaia, ``{On higher spins and the tensionless limit of
  string theory},'' {\em Nucl. Phys. B}, vol.~682, pp.~83--116, 2004.

\bibitem{Bonelli:2003kh}
G.~Bonelli, ``{On the tensionless limit of bosonic strings, infinite symmetries
  and higher spins},'' {\em Nucl. Phys. B}, vol.~669, pp.~159--172, 2003.

\bibitem{Sezgin:2002rt}
E.~Sezgin and P.~Sundell, ``{Massless higher spins and holography},'' {\em
  Nucl. Phys. B}, vol.~644, pp.~303--370, 2002.
\newblock [Erratum: Nucl.Phys.B 660, 403--403 (2003)].

\bibitem{Gaberdiel:2014cha}
M.~R. Gaberdiel and R.~Gopakumar, ``{Higher Spins \& Strings},'' {\em JHEP},
  vol.~11, p.~044, 2014.

\bibitem{Bagchi:2021ban}
A.~Bagchi, A.~Banerjee, S.~Chakrabortty, and R.~Chatterjee, ``{A Rindler road
  to Carrollian worldsheets},'' {\em JHEP}, vol.~04, p.~082, 2022.

\bibitem{Bagchi:2020ats}
A.~Bagchi, A.~Banerjee, and S.~Chakrabortty, ``{Rindler Physics on the String
  Worldsheet},'' {\em Phys. Rev. Lett.}, vol.~126, no.~3, p.~031601, 2021.

\bibitem{PhysRevD.26.3735}
R.~D. Pisarski and O.~Alvarez, ``Strings at finite temperature and
  deconfinement,'' {\em Phys. Rev. D}, vol.~26, pp.~3735--3737, Dec 1982.

\bibitem{Olesen:1985ej}
P.~Olesen, ``{Strings, Tachyons and Deconfinement},'' {\em Phys. Lett. B},
  vol.~160, pp.~408--410, 1985.

\bibitem{Atick:1988si}
J.~J. Atick and E.~Witten, ``{The Hagedorn Transition and the Number of Degrees
  of Freedom of String Theory},'' {\em Nucl. Phys. B}, vol.~310, pp.~291--334,
  1988.

\bibitem{Bagchi:2019cay}
A.~Bagchi, A.~Banerjee, and P.~Parekh, ``{Tensionless Path from Closed to Open
  Strings},'' {\em Phys. Rev. Lett.}, vol.~123, no.~11, p.~111601, 2019.

\bibitem{Bagchi:2022iqb}
A.~Bagchi, D.~Grumiller, and M.~M. Sheikh-Jabbari, ``{Horizon Strings as 3d
  Black Hole Microstates},'' 10 2022.

\bibitem{Karlhede:1986wb}
A.~Karlhede and U.~Lindstrom, ``{The Classical Bosonic String in the Zero
  Tension Limit},'' {\em Class. Quant. Grav.}, vol.~3, pp.~L73--L75, 1986.

\bibitem{Lizzi:1986nv}
F.~Lizzi, B.~Rai, G.~Sparano, and A.~Srivastava, ``{Quantization of the Null
  String and Absence of Critical Dimensions},'' {\em Phys. Lett. B}, vol.~182,
  pp.~326--330, 1986.

\bibitem{Gamboa:1989zc}
J.~Gamboa, C.~Ramirez, and M.~Ruiz-Altaba, ``{NULL SPINNING STRINGS},'' {\em
  Nucl. Phys. B}, vol.~338, pp.~143--187, 1990.

\bibitem{Gamboa:1989px}
J.~Gamboa, C.~Ramirez, and M.~Ruiz-Altaba, ``{QUANTUM NULL (SUPER)STRINGS},''
  {\em Phys. Lett. B}, vol.~225, pp.~335--339, 1989.

\bibitem{Gustafsson:1994kr}
H.~Gustafsson, U.~Lindstrom, P.~Saltsidis, B.~Sundborg, and R.~van Unge,
  ``{Hamiltonian BRST quantization of the conformal string},'' {\em Nucl. Phys.
  B}, vol.~440, pp.~495--520, 1995.

\bibitem{Lindstrom:2003mg}
U.~Lindstrom and M.~Zabzine, ``{Tensionless strings, WZW models at critical
  level and massless higher spin fields},'' {\em Phys. Lett. B}, vol.~584,
  pp.~178--185, 2004.

\bibitem{Isberg:1993av}
J.~Isberg, U.~Lindstrom, B.~Sundborg, and G.~Theodoridis, ``{Classical and
  quantized tensionless strings},'' {\em Nucl. Phys. B}, vol.~411,
  pp.~122--156, 1994.

\bibitem{Bagchi:2013bga}
A.~Bagchi, ``{Tensionless Strings and Galilean Conformal Algebra},'' {\em
  JHEP}, vol.~05, p.~141, 2013.

\bibitem{Bagchi:2015nca}
A.~Bagchi, S.~Chakrabortty, and P.~Parekh, ``{Tensionless Strings from
  Worldsheet Symmetries},'' {\em JHEP}, vol.~01, p.~158, 2016.

\bibitem{Bondi:1962px}
H.~Bondi, M.~G.~J. van~der Burg, and A.~W.~K. Metzner, ``{Gravitational waves
  in general relativity. 7. Waves from axisymmetric isolated systems},'' {\em
  Proc. Roy. Soc. Lond. A}, vol.~269, pp.~21--52, 1962.

\bibitem{Sachs:1962wk}
R.~K. Sachs, ``{Gravitational waves in general relativity. 8. Waves in
  asymptotically flat space-times},'' {\em Proc. Roy. Soc. Lond. A}, vol.~270,
  pp.~103--126, 1962.

\bibitem{Ashtekar:1996cd}
A.~Ashtekar, J.~Bicak, and B.~G. Schmidt, ``{Asymptotic structure of symmetry
  reduced general relativity},'' {\em Phys. Rev. D}, vol.~55, pp.~669--686,
  1997.

\bibitem{Lévy1965}
J.-M. Lévy-Leblond, ``Une nouvelle limite non-relativiste du groupe de
  poincaré,'' {\em Annales de l'I.H.P. Physique théorique}, vol.~3, no.~1,
  pp.~1--12, 1965.

\bibitem{NDS}
N.~Sen~Gupta, ``{On an Analogue of the Galileo Group},'' {\em Nuovo Cim. 54
  (1966) 512 • DOI: 10.1007/BF02740871}.

\bibitem{Henneaux:1979vn}
M.~Henneaux, ``{Geometry of Zero Signature Space-times},'' {\em Bull. Soc.
  Math. Belg.}, vol.~31, pp.~47--63, 1979.

\bibitem{Bagchi:2010zz}
A.~Bagchi, ``{Correspondence between Asymptotically Flat Spacetimes and
  Nonrelativistic Conformal Field Theories},'' {\em Phys. Rev. Lett.},
  vol.~105, p.~171601, 2010.

\bibitem{Bagchi:2012xr}
A.~Bagchi, S.~Detournay, R.~Fareghbal, and J.~Sim\'on, ``{Holography of 3D Flat
  Cosmological Horizons},'' {\em Phys. Rev. Lett.}, vol.~110, no.~14,
  p.~141302, 2013.

\bibitem{Bagchi:2012cy}
A.~Bagchi and R.~Fareghbal, ``{BMS/GCA Redux: Towards Flatspace Holography from
  Non-Relativistic Symmetries},'' {\em JHEP}, vol.~10, p.~092, 2012.

\bibitem{Bagchi:2012yk}
A.~Bagchi, S.~Detournay, and D.~Grumiller, ``{Flat-Space Chiral Gravity},''
  {\em Phys. Rev. Lett.}, vol.~109, p.~151301, 2012.

\bibitem{Bagchi:2014iea}
A.~Bagchi, R.~Basu, D.~Grumiller, and M.~Riegler, ``{Entanglement entropy in
  Galilean conformal field theories and flat holography},'' {\em Phys. Rev.
  Lett.}, vol.~114, no.~11, p.~111602, 2015.

\bibitem{Bagchi:2016bcd}
A.~Bagchi, R.~Basu, A.~Kakkar, and A.~Mehra, ``{Flat Holography: Aspects of the
  dual field theory},'' {\em JHEP}, vol.~12, p.~147, 2016.

\bibitem{Asadi:2016plj}
M.~Asadi, O.~Baghchesaraei, and R.~Fareghbal, ``{Stress tensor correlators of
  CCFT$_2$ using flat-space holography},'' {\em Eur. Phys. J. C}, vol.~77,
  no.~11, p.~737, 2017.

\bibitem{Jiang:2017ecm}
H.~Jiang, W.~Song, and Q.~Wen, ``{Entanglement Entropy in Flat Holography},''
  {\em JHEP}, vol.~07, p.~142, 2017.

\bibitem{Donnay:2019jiz}
L.~Donnay and C.~Marteau, ``{Carrollian Physics at the Black Hole Horizon},''
  {\em Class. Quant. Grav.}, vol.~36, no.~16, p.~165002, 2019.

\bibitem{Bagchi:2023ysc}
A.~Bagchi, K.~S. Kolekar, and A.~Shukla, ``{Carrollian Origins of Bjorken
  Flow},'' 2 2023.

\bibitem{Bagchi:2023rwd}
A.~Bagchi, K.~S. Kolekar, T.~Mandal, and A.~Shukla, ``{Heavy-ion collisions,
  Gubser flow, and Carroll hydrodynamics},'' {\em Phys. Rev. D}, vol.~109,
  no.~5, p.~056004, 2024.

\bibitem{Armas:2023dcz}
J.~Armas and E.~Have, ``{Carrollian fluids and spontaneous breaking of boost
  symmetry},'' 8 2023.

\bibitem{Bidussi:2021nmp}
L.~Bidussi, J.~Hartong, E.~Have, J.~Musaeus, and S.~Prohazka, ``{Fractons,
  dipole symmetries and curved spacetime},'' {\em SciPost Phys.}, vol.~12,
  no.~6, p.~205, 2022.

\bibitem{Figueroa-OFarrill:2023vbj}
J.~Figueroa-O'Farrill, A.~P\'erez, and S.~Prohazka, ``{Carroll/fracton
  particles and their correspondence},'' {\em JHEP}, vol.~06, p.~207, 2023.

\bibitem{Bagchi:2022eui}
A.~Bagchi, A.~Banerjee, R.~Basu, M.~Islam, and S.~Mondal, ``{Magic fermions:
  Carroll and flat bands},'' {\em JHEP}, vol.~03, p.~227, 2023.

\bibitem{Marsot:2022imf}
L.~Marsot, P.~M. Zhang, M.~Chernodub, and P.~A. Horvathy, ``{Hall effects in
  Carroll dynamics},'' 12 2022.

\bibitem{Duval:2014lpa}
C.~Duval, G.~W. Gibbons, and P.~A. Horvathy, ``{Conformal Carroll groups},''
  {\em J. Phys. A}, vol.~47, no.~33, p.~335204, 2014.

\bibitem{Bagchi:2016yyf}
A.~Bagchi, S.~Chakrabortty, and P.~Parekh, ``{Tensionless Superstrings: View
  from the Worldsheet},'' {\em JHEP}, vol.~10, p.~113, 2016.

\bibitem{Bagchi:2017cte}
A.~Bagchi, A.~Banerjee, S.~Chakrabortty, and P.~Parekh, ``{Inhomogeneous
  Tensionless Superstrings},'' {\em JHEP}, vol.~02, p.~065, 2018.

\bibitem{Bagchi:2018wsn}
A.~Bagchi, A.~Banerjee, S.~Chakrabortty, and P.~Parekh, ``{Exotic Origins of
  Tensionless Superstrings},'' {\em Phys. Lett. B}, vol.~801, p.~135139, 2020.

\bibitem{Bagchi:2020fpr}
A.~Bagchi, A.~Banerjee, S.~Chakrabortty, S.~Dutta, and P.~Parekh, ``{A tale of
  three \textemdash{} tensionless strings and vacuum structure},'' {\em JHEP},
  vol.~04, p.~061, 2020.

\bibitem{Bagchi:2021rfw}
A.~Bagchi, M.~Mandlik, and P.~Sharma, ``{Tensionless tales: vacua and critical
  dimensions},'' {\em JHEP}, vol.~08, p.~054, 2021.

\bibitem{Casali:2016atr}
E.~Casali and P.~Tourkine, ``{On the null origin of the ambitwistor string},''
  {\em JHEP}, vol.~11, p.~036, 2016.

\bibitem{Lee:2017utr}
K.~Lee, S.-J. Rey, and J.~A. Rosabal, ``{A string theory which
  isn\textquoteright{}t about strings},'' {\em JHEP}, vol.~11, p.~172, 2017.

\bibitem{Chen:2023esw}
B.~Chen, Z.~Hu, Z.-f. Yu, and Y.-f. Zheng, ``{Path-integral quantization of
  tensionless (super) string},'' 2 2023.

\bibitem{Polchinski:1998rq}
J.~Polchinski, {\em {String theory. Vol. 1: An introduction to the bosonic
  string}}.
\newblock Cambridge Monographs on Mathematical Physics, Cambridge University
  Press, 12 2007.

\bibitem{Blumenhagen:2013fgp}
R.~Blumenhagen, D.~L\"ust, and S.~Theisen, {\em {Basic concepts of string
  theory}}.
\newblock Theoretical and Mathematical Physics, Heidelberg, Germany: Springer,
  2013.

\bibitem{Banerjee:2023ekd}
A.~Banerjee, R.~Chatterjee, and P.~Pandit, ``{Tensionless Tales of
  Compactification},'' 7 2023.

\bibitem{Becker:2006dvp}
K.~Becker, M.~Becker, and J.~H. Schwarz, {\em {String theory and M-theory: A
  modern introduction}}.
\newblock Cambridge University Press, 12 2006.

\bibitem{Gangopadhyay:2006gmx}
S.~Gangopadhyay, A.~G. Hazra, and A.~Saha, ``{Noncommutativity in interpolating
  string: A Study of gauge symmetries in noncommutative framework},'' {\em
  Phys. Rev. D}, vol.~74, p.~125023, 2006.

\bibitem{Gomis:2023eav}
J.~Gomis and Z.~Yan, ``{Worldsheet Formalism for Decoupling Limits in String
  Theory},'' 11 2023.

\bibitem{Bagchi:2024unl}
A.~Bagchi, A.~Banerjee, S.~Mondal, D.~Mukherjee, and H.~Muraki, ``{Beyond
  Wilson? Carroll from current deformations},'' 1 2024.

\bibitem{Hartong:2015xda}
J.~Hartong, ``{Gauging the Carroll Algebra and Ultra-Relativistic Gravity},''
  {\em JHEP}, vol.~08, p.~069, 2015.

\bibitem{Bergshoeff:2017btm}
E.~Bergshoeff, J.~Gomis, B.~Rollier, J.~Rosseel, and T.~ter Veldhuis,
  ``{Carroll versus Galilei Gravity},'' {\em JHEP}, vol.~03, p.~165, 2017.

\bibitem{Bagchi:2022eav}
A.~Bagchi, A.~Banerjee, S.~Dutta, K.~S. Kolekar, and P.~Sharma, ``{Carroll
  covariant scalar fields in two dimensions},'' {\em JHEP}, vol.~01, p.~072,
  2023.

\bibitem{Aldazabal:2013sca}
G.~Aldazabal, D.~Marques, and C.~Nunez, ``{Double Field Theory: A Pedagogical
  Review},'' {\em Class. Quant. Grav.}, vol.~30, p.~163001, 2013.

\bibitem{Morand:2017fnv}
K.~Morand and J.-H. Park, ``{Classification of non-Riemannian
  doubled-yet-gauged spacetime},'' {\em Eur. Phys. J. C}, vol.~77, no.~10,
  p.~685, 2017.
\newblock [Erratum: Eur.Phys.J.C 78, 901 (2018)].

\bibitem{Giveon:1994fu}
A.~Giveon, M.~Porrati, and E.~Rabinovici, ``{Target space duality in string
  theory},'' {\em Phys. Rept.}, vol.~244, pp.~77--202, 1994.

\bibitem{Gomis:2000bd}
J.~Gomis and H.~Ooguri, ``{Nonrelativistic closed string theory},'' {\em J.
  Math. Phys.}, vol.~42, pp.~3127--3151, 2001.

\bibitem{Blair:2023noj}
C.~D.~A. Blair, J.~Lahnsteiner, N.~A.~J. Obers, and Z.~Yan, ``{Unification of
  Decoupling Limits in String and M-theory},'' 11 2023.

\bibitem{Casali:2017mss}
E.~Casali and P.~Tourkine, ``{Windings of twisted strings},'' {\em Phys. Rev.
  D}, vol.~97, no.~6, p.~061902, 2018.

\bibitem{Lee:2017crr}
K.~Lee and J.~A. Rosabal, ``{A Note on Circle Compactification of Tensile
  Ambitwistor String},'' {\em Nucl. Phys. B}, vol.~933, pp.~482--510, 2018.

\bibitem{Seiberg:1999vs}
N.~Seiberg and E.~Witten, ``{String theory and noncommutative geometry},'' {\em
  JHEP}, vol.~09, p.~032, 1999.

\end{thebibliography}

\end{document}